\documentclass{aa}

\usepackage{natbib}
\usepackage{epsfig}
\usepackage{txfonts}

\begin{document} 

\newcommand{\chisq}{\mbox{$\chi^2$}}
\newcommand{\es}{erg s$^{-1}$}                          
\newcommand{\ecms}{erg cm$^{-2}$ s$^{-1}$}              
\newcommand{\halpha}{H$\alpha$}                   
\newcommand{\hbeta}{H$\beta$}
\newcommand{\kms}{km~s$^{-1}$}       
\newcommand{\cmthree}{cm$^{-3}$}
\newcommand{\msun}{M$_{\odot}$} 
\newcommand{\xmm}{XMM-\emph{Newton}} 
\newcommand{\chandra}{\emph{Chandra}} 
\newcommand{\spitzer}{\emph{Spitzer}} 
\newcommand{\nh}{\mbox{$N({\rm H})$}}
\newcommand{\rhoph}{$\rho$~Oph}
\newcommand{\eltn}{Elias 29}
\newcommand{\eltf}{Elias 24}
\newcommand{\hlt}{HL~Tau}

\title{The onset of X-ray emission in young stellar objects}
\subtitle{A \chandra\ observation of the Serpens star-forming region}

\author{G. Giardino\inst{1} \and F. Favata\inst{1} \and G.
  Micela\inst{2} \and S.  Sciortino\inst{2} \and E. Winston\inst{3}}

\institute{Astrophysics Division -- Research and Science Support
  Department of ESA, ESTEC, 
  Postbus 299, NL-2200 AG Noordwijk, The Netherlands
\and
INAF -- Osservatorio Astronomico di Palermo, 
Piazza del Parlamento 1, I-90134 Palermo, Italy 
\and
Harvard-Smithsonian Center for Astrophysics -- 60 Garden Street Cambridge, MA 02138, USA
}

\offprints{G. Giardino}

\date{Received date / Accepted date}

\abstract 
  {}
  {To study the properties of X-ray emission from young
  stellar objects (YSOs) through their evolution from Class I to Class
  III and determine whether Class 0 protostars emit X-rays.}  
  {A deep
  \chandra\ X-ray observation of the Serpens star-forming region was
  obtained. The Serpens Cloud Core is ideally suited for this type of
  investigation, being populated by a dense and extremely young
  cluster whose members are found in all evolutionary
  stages, including six well-studied Class 0 sources.}  
  {None of the six Class 0 protostars is detected in our
  observations, excluding the presence of sources with typical
  X-ray luminosities $\ga 0.4 \times 10^{30}$ erg s$^{-1}$ (for column
  densities of the order of 4 $\times 10^{23}$ cm$^{-2}$, or $A_V \sim
  200$). A total of 85 X-ray sources are detected and the light curves
  and spectra of 35 YSOs are derived. There is a clear trend of
  decreasing absorbing column densities as one moves from Class I to
  Class III sources, and some evidence of decreasing plasma
  temperatures, too.  We observe a strong, long-duration, flare from a
  Class II low-mass star, for which we derive a flaring loop length of
  the order of 20 stellar radii. We interpret the flaring event as
  originating from a magnetic flux tube connecting the star to its
  circumstellar disk. The presence of such a disk is supported by the
  detection, in the spectrum of this star, of 6.4 keV Fe fluorescent
  emission.}
 {}


\maketitle

\section{Introduction}
\label{sec:intro}


Young stellar objects (YSOs) are well-known X-ray sources, and X-ray
observations have become a routine tool for the study of star-forming
regions. Beside its intrinsic astrophysics interest, X-ray emission in
the early stages of stellar evolution may have a significant influence
on the circumstellar environment. As a ionizing source for the
circumstellar accretion disk, X-ray emission may play an important
role in regulating the coupling between the disk and the magnetic
field and affect the chemistry of the disk itself. X-rays are an
important component in the complex feedback processes regulating
star-formation: \cite{ntr2006} speculate on the possibility that the
observed spread in accretion rates observed in classical T Tauri stars
(CCTS) of any given age (which is too large to be explained by
``classical'' accretion disk theory) might be explained by the
large spread in X-ray luminosity among otherwise similar
pre-main sequence stars, if the main source of disk ionization is
indeed provided by the stellar X-ray emission. Additionally, as the
hot X-ray emitting plasma needs to be confined by strong magnetic
fields ($B\ge 100$ G), tracing the X-ray emission and its spatial
extent provides a mean of locating strong magnetic fields,
which (in the magnetospheric accretion model of low-mass YSOs) have a
key role in channeling the accreting plasma and thus in regulating the
accretion.

One open question regards the onset of X-ray emission: at which
evolutionary phase do YSOs start to emit X-rays? Current evidence
shows that X-ray emission is common (although perhaps not universal)
in Class I sources (e.g \citealp{ogm05}) -- Class I sources are
protostars which already have an accretion disk but are still supplied
with a relatively massive circumstellar envelope. In Class II sources
(or CTTS, i.e.\ young stars actively accreting from a circumstellar
disk, but without a circumstellar envelope) X-ray emission is an
universal feature, as shown, for instance, by the deep \chandra\
observations of the Orion Nebula Cluster (\citealp{fdm+03};
\citealp{pkf+05}). In more evolved sources, that are no longer
accreting (Class III, or weak-lined T Tauri stars -- WTTS), X-ray
emission is also an universal occurrence, with characteristics and
luminosity similar to active main-sequence stars
(e.g. \citealp{fdm+03}; \citealp{pkf+05}).

Class 0 sources are young protostars at the beginning of the main
accretion phase. Observationally, they are characterized by strong,
centrally-condensed dust continuum emission at submillimeter
wavelengths, powerful jet-like outflows, and very little emission
shortward of 10 $\mu$m. According to the defintion given in
\citet{awb2000}, their submillimeter luminosity (measured longward of
350 $\mu$m), should be greater than 0.5\% of their bolometric
luminosity -- thereby suggesting that the envelope mass exceeds the
central stellar mass. The evidence for X-ray emission from Class 0
protostars is still fragmentary. \citet{tkh+2001} claimed the
discovery of X-rays from two highly embedded sources in the OMC-2/3
clouds, within a few arcsecs from the Class 0 candidates MMS 2 and MMS
3. The two detected sources show some Class 0 characteristics: a very
large absorption ($N({\rm H}) = 1-3 \times 10^{23}$ cm$^{-2}$), no
near-IR counterparts, and associations with millimeter radio clumps.
However, follow-up radio and near-IR observations by \citet{tkk+2004}
did not unambiguously classify either source as Class 0, and
associated one of them with emission from a proto-stellar jet
originating in a Class I source.

Another claim for X-ray emission from Class 0 sources was made by
\citet{hcp+2005}, who detected X-ray emission from 2 embedded sources
in the R Coronae Australis star-forming core. The two sources are
associated with VLA centimeter radio sources and their X-ray spectra
imply absorbing column densities of the order of $3 \times 10^{23}$
cm$^{-2}$, or $A_V \simeq 180$ mag. One of the two sources, IRS 7W,
does not appear to be a Class 0 source: according to \citet{fpm06} the
emerging picture is that of IRS 7W being an infrared-detected,
deeply-embedded protostar, probably a Class I or II source.  The other
source, IRS 7E, is proposed as a Class 0 source (\citealp{fpm06}). The
X-ray emission from this source displays strong variability (its
luminosity varying between $0.2 \times 10^{31}$ and $2\times 10^{31}$
\es on timescales of 3 to 30 months) and it is characterized by a high
plasma temperature, $kT \simeq 3-4$ keV.

Finally, \citet{gfg+06} have recently reported X-ray emission associated with
the luminous Class 0/I protostar IRAS 21391+5802 (in IC 1396N). Their
\chandra\ data reveal a faint, extremely hard, X-ray source within
$0.5''$ of BIMA 2, the millimeter counterpart of the IRAS source. From
the 8 extracted photons they derive a median energy of the source of
6.0 keV and infer an absorbing column density of $\nh \sim 10^{24}$
cm$^{-2}$ ($A_V \sim 500$). The inferred intrinsic hard band
luminosity corrected for absorption is $L_{\rm X} \sim 2\times
10^{31}$ \es, a high luminosity consistent with IRAS 21391+5802 being
an intermediate mass protostar with circumstellar mass $\sim
5~M_{\sun}$. The analysis of the photon arrival time suggests that 
IRAS 21391+5802 was seen during the decay of a magnetic reconnection
flare (\citealp{gfg+06}).

Thus, while some evidence of Class 0 X-ray emission is gathering, this
is still patchy; the Serpens cloud is an ideal environment to try to
investigate X-ray emission from protostars further since it contains a
set of well-studied, well-identified Class 0 sources and is nearby.

Here we present a 90 ks \chandra\ observation of the Serpens Cloud
Core.  The Serpens cloud has been previously observed in X-rays both
with the ROSAT HRI and with \xmm. The ROSAT observation was relatively
short ($\simeq 19$ ks) and resulted in the detection of 7 sources
(\citealp{pre98}), while a total of 45 X-ray sources were detected in
the shorter \xmm\ observation ($\simeq 12$ ks,
\citealp{pre2003}). This last observation was combined by
\cite{pre2004} with two other subsequent \xmm\ observations for a
total exposure time of $\sim 52$~ks.  The present \chandra\
observation is therefore the deepest X-ray observation conducted to
date of the Serpens region, allowing us to detect fainter sources (for
a total number of 85) and to study the spectrum and the temporal
variability of a significant number of YSOs in the region.

\subsection{Characteristics of the Serpens cloud}

 The Serpens Cloud Core (diameter $\sim 6$ arcmin) is one of the more
 active, nearby star-forming regions and has been the subject of many
 observational studies during the last 15 years. Its distance is
 estimated at 260 pc by \citet{scb96}, using photometry of the
 brightest stars in the region.  In the IR, the Serpens Cloud Core is
 dominated by two conical regions of diffuse emission extending out of
 a large disk-like absorption feature along the northeast-southwest
 direction (centered on the young star SVS 2); to the south, a
 filamentary, eye-shaped structure of intense emission is centered
 around SVS 20. The region is populated by a deeply-embedded and
 extremely young cluster whose members are found in many different
 evolutionary stages. Near-IR surveys identified more than 150
 sources embedded in the cloud core (\citealp{ec92}; \citealp{stg+97};
 \citealp{gcn+98}; \citealp{kob+04};). Many of these sources were
 classified as Class II YSOs, by means of mid-IR Infrared Space
 Observatory (ISO) observations (\citealp{kob+04}); these observations
 also revealed a significant number of flat-spectrum sources and Class I
 protostars. Sub-millimeter, millimeter, and far-IR observations show
 that the cloud core is also populated by one of the richest
 collection of Class 0 objects and pre-stellar condensations
 (\citealp{ced+93}; \citealp{hb96}; \citealp{ts98};
 \citealp{etc+2005}).  The age of the Class II population of the
 cluster is estimated at 2-3 Myr (\citealp{gcn+98}; \citealp{kob+04}),
 however on-going star formation is evident from the presence of
 protostars and pre-stellar condensations as well as several molecular
 outflows (\citealp{bl83}; \citealp{wce+95}; \citealp{hbb97};
 \citealp{dmr+99}) and the age of the Class I population is believed
 to be less than a few $10^{5}$ yr (\citealp{gcn+98};
 \citealp{kob+04}). Serpens was also the target of a
 \emph{Spitzer} observation, as described in Sect.~\ref{sec:spi}. We
 made use of the \emph{Spitzer} data to provide a classification
 of the X-ray sources detected in the \emph{Chandra} observation. In
 particular, we used a \emph{Spitzer} color-color diagram to
 determine whether a given YSO is Class I, II or III.

\section{Observations and data analysis}
\label{sec:obs}

\begin{figure}[!htbp]
  \begin{center} \leavevmode 
     
     \epsfig{file=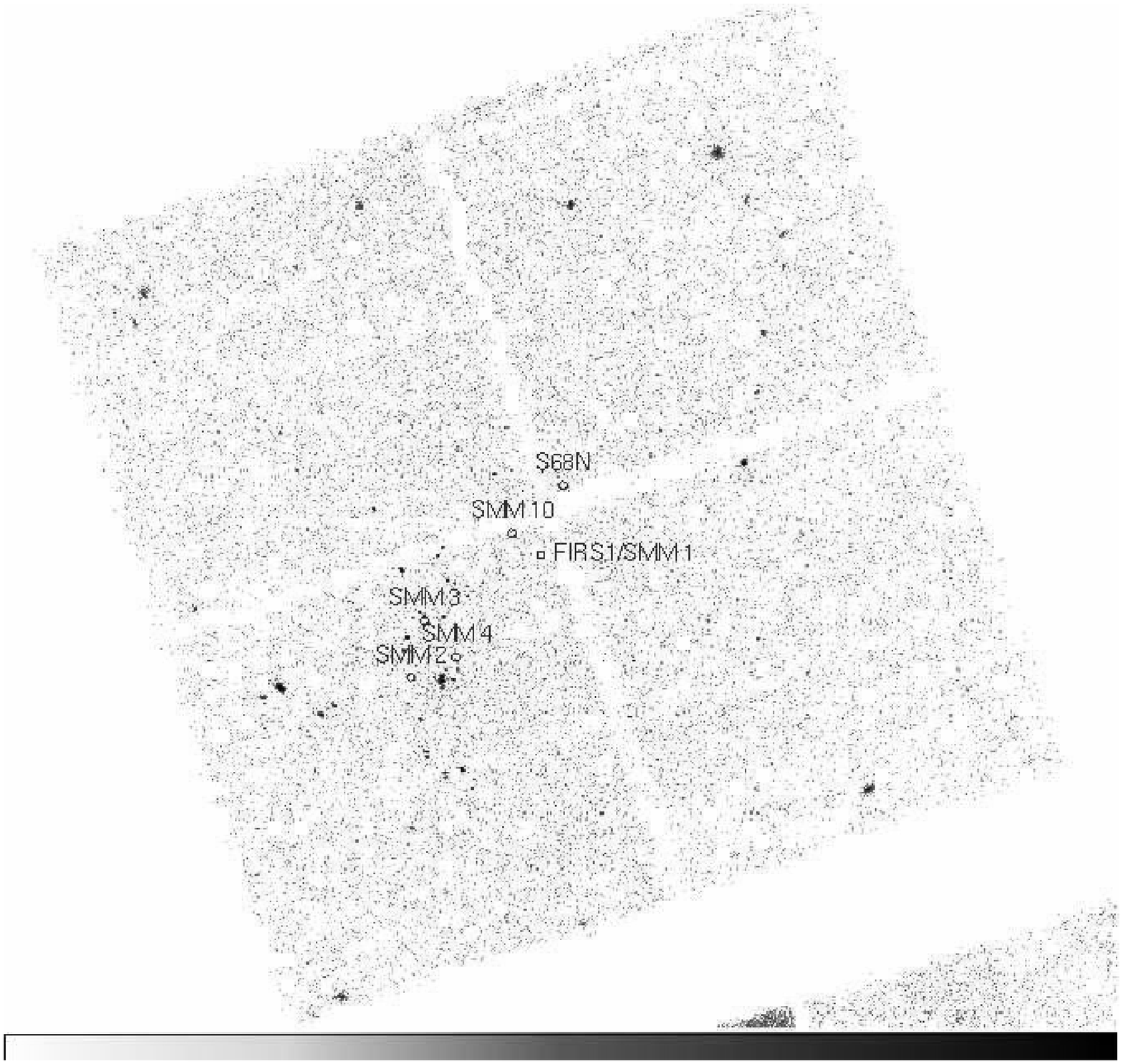, height=7.9cm, bbllx=37, bblly=160,bburx=576, bbury=660, clip=}	
\epsfig{file=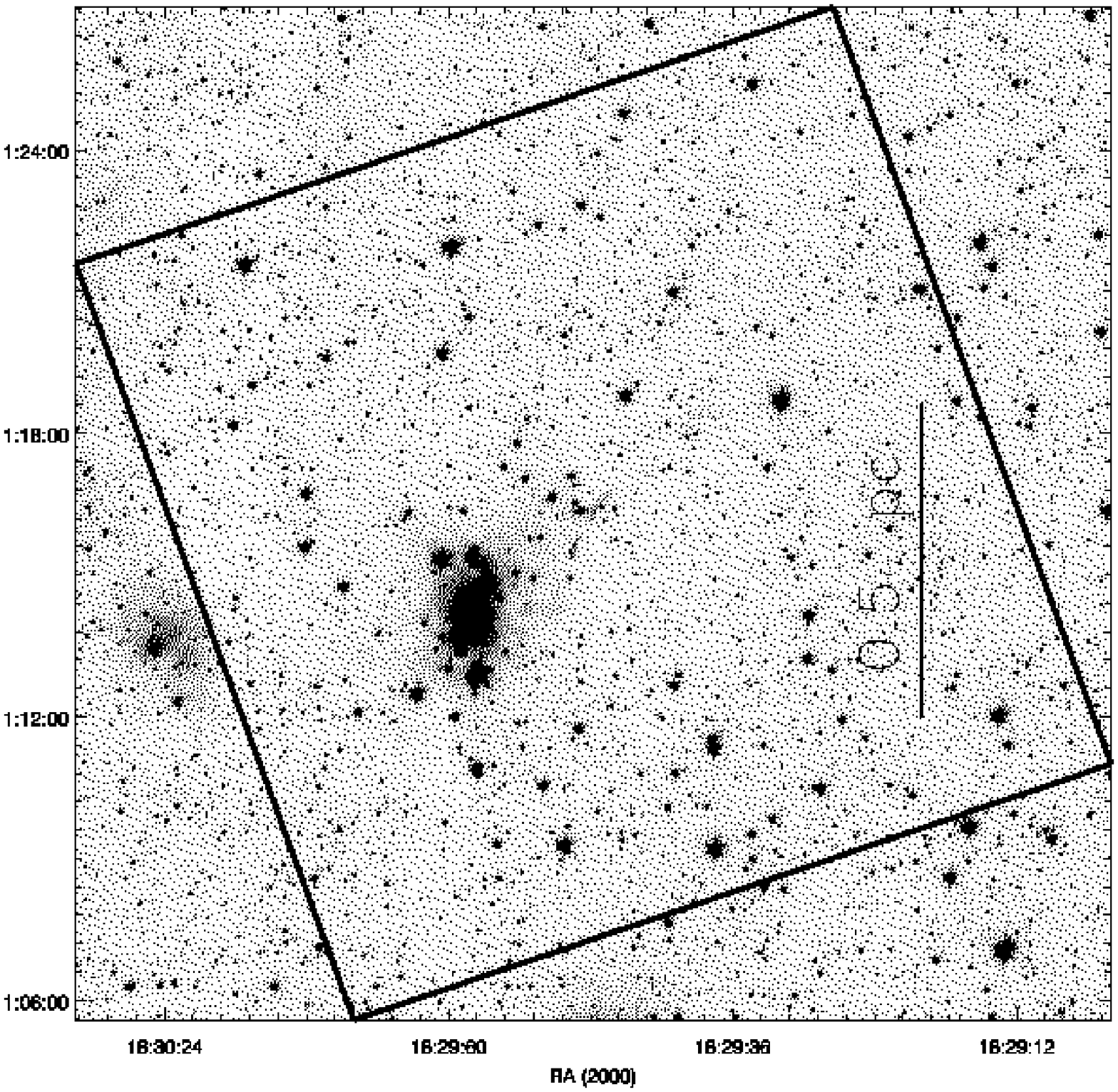, height=10.5cm}	     

    \caption{{\it Top} --
    \chandra\ X-ray image of the Serpens Cloud Core; the positions of Class 0 sources are
    indicated. {\it Bottom} -- \spitzer\ IRAC band 1 view of the
    Serpens (the outline gives the Chandra FOV). }.
  
 \label{fig:pic}	
  \end{center}
\end{figure}

The core of the Serpens cloud was observed with \chandra\ starting at
21:43 UT on June 19, 2004. The observation lasted for 91.4 ks
(although the scheduled duration was 100 ks). The data quality as
received from the CXC was satisfactory, and we did not reprocess
the data set.  The image of the X-ray observation is shown in
Fig.\ref{fig:pic}, together with an IR image derived from the \spitzer\
observation in IRAC band 1 (3.6 ${\mu}$m).
 
We performed the source
detection on the unfiltered event list,
using the Wavelet Transform detection algorithm developed at Palermo
Astronomical Observatory (\textsc{Pwdetect}, available at
http://oapa.astropa.unipa.it/progetti\_ricerca/PWDetect).  A total of
85 X-ray sources were detected, of which we expect at most 1 to
be spurious, based on the source significance limit with which
\textsc{pwdetect} was run. The coordinates and ACIS-I count rate
of the detected sources are listed in Table~\ref{tab:src}. Note
that sources 49 and 51 are unresolved from source 48, as this is
characterized by an elongated shape in the
NE direction. Source 48 is the X-ray counterpart of SVS 20,
the deeply-embedded IR double source located inside the ``eye''
structure at the center of the Serpens nebula (see
Sect.\,\ref{sec:src48} for more details). 

Table~\ref{tab:cross-id} provides cross-identification of sources as
obtained from the \textsc{Simbad} database (search radius of 5
arcsec).

For sufficiently bright X-ray sources (with count rate $\ge 0.5$
cts/ks) individual light curves and spectra were extracted using
source and background regions defined in \textsc{ds9} and \textsc{ciao
3.3} threads, which were also used for the generation of the relative
response matrices. For these sources, spectral fits were
performed using \textsc{xspec 11.3} (see Sect.~\ref{sec:res}).

\begin{figure}[!tbp]
  \begin{center} \
        \leavevmode 
        \epsfig{file=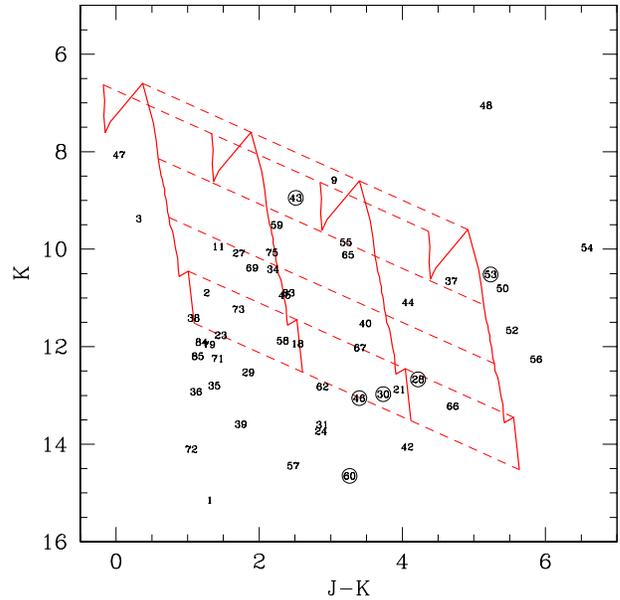, width=9cm}

        \caption{Color magnitude diagram for the X-ray sources with
          2MASS counterparts (Table 5).  Theoretical 2 Myr isochrones
          are also shown for different extinction coefficients (from
          left to right $A_K={\rm 0, 1, 2, 3}$).  The dashed lines
          indicate the reddening vector for stars of constant mass
          (from top to bottom 3, 7, 2.2, 1, 0.2, and 0.1 $M_{\sun}$).
          Circled numbers indicates Class I sources: their position in
          the diagram should not be compared to the 2 My isochrones, as
          their ages is estimated to be less than $\sim 10^{5}$ yr.}

          \label{fig:cmd_2myr} 
	\end{center}
\end{figure}

Fig.\,\ref{fig:cmd_2myr} shows the $K$ vs. $J-K$ color-magnitude
diagram for the X-ray sources listed in Tables~\ref{tab:src} which
have a 2MASS counterpart. In the diagram 2 Myr isochrones are shown
for 4 values of extinction ($A_K={\rm 0, 1, 2, 3}$). Reddening vectors
have been computed following the relation $A_K = R_K \times E(J-K)$,
where $E(J-K)$ is the colour excess ($E(J-K) = (J-K) - (J-K)_0$) and
$R_K = 0.66$ (\citealp{rl85}). The isochrones are from \cite{sdf00},
for a metal abundance $Z=0.02$ plus overshooting, shifted to the
distance of the Serpens (260 pc). Although, various simplifying
assumptions apply to the model isochrones (e.g. they include neither
rotation nor accretion) and evolutionary models for pre-main sequence
stars are not yet well established, they allow a mass estimate for
most of our sources to be derived. The X-ray sample comprises a
significant number of very low-mass stars ($M < 0.1 M_{\sun}$), 6
intermediate mass stars and the highly reddened double source SVS 20
(source 48).

\begin{figure}[!tbp]
  \begin{center} \
        \leavevmode 
        \epsfig{file=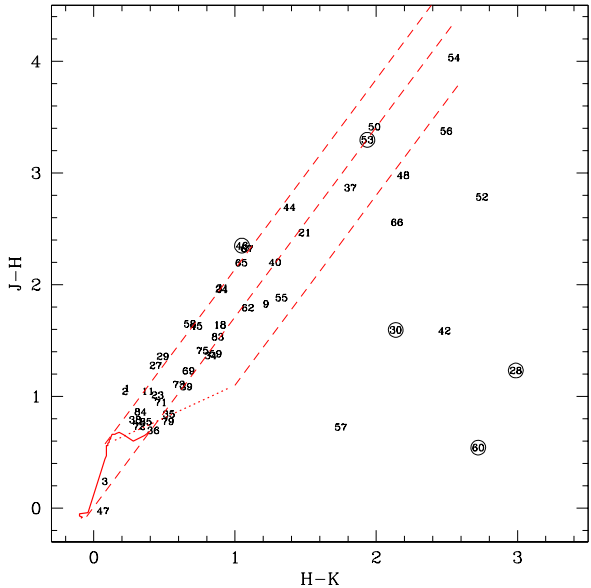, width=9cm}

        \caption{Color-color diagram for the X-ray sources in Table 5
          with 2MASS counterpart.  The solid line at lower-left is the
          theoretical 2 Myr isochrone.  The dotted line yields the
          locus of dereddened colours of classical T Tauri stars
          according to Meyer et al. (1997). The dashed lines mark the
          reddening band for normally reddened stars.  Class I sources
          are identified by a circle.}

  \label{fig:ccd_2myr}
  \end{center}
\end{figure}

The $J-H$ versus $H-K$ color-color diagram in Fig.~\ref{fig:ccd_2myr}
allows normally reddened stars to be discriminated from stars with IR
excess (indicative of warm circumstellar dust in addition to a
reddened photosphere). The majority of stars with high IR excess are
indeed Class I sources or Flat spectrum sources (Class I/II).

\subsection{Spitzer data}
\label{sec:spi}

A detailed description of the processing and data reduction of the 
{\it Spitzer} infrared data can be found in \citet{wmw+06}.

The {\it Spitzer} Space Telescope observed the Serpens region at
infrared wavelengths from 3.6 - 70 ${\mu}$m as part of the Guaranteed
Time Observations (GTO) program PID 6.  The InfraRed Array Camera
(IRAC; \citealp{fha+04}) observed the Serpens Cloud Core at 3.6, 4.5,
5.8, 8.0 ${\mu}$m, while the Multiband Imaging Photometer for {\it
Spitzer} (MIPS; \citealp{rye+04}) provided observations at 24 and
70 ${\mu}$m.  The IRAC bands detected sources below the hydrogen
burning limit in Serpens.  The IRAC and MIPS data were combined with
near-IR $J$, $H$, and $K$ band data from 2MASS. The overlap region for these
eight bands covered the entire \chandra\ field of view.

The classification of the sources was carried out using the slope of
the Spectral Energy Distributions (SEDs) between the wavelengths
3.6 ${\mu}$m and 8.0 ${\mu}$m, as a measure of the excess IR emission
from circumstellar material.  Sources were classified as follows:
Class I are sources with a slope greater than 0.  Flat spectrum (Class
I/II) sources have 'flat' slopes between 0.0 and $-0.5$.  Class II YSOs
show excess emission from a circumstellar disk, with a slope between
$-0.5$ and $\sim -3.5$. Class III sources exhibit photospheric emission, with
slopes of $\sim -3.5$. Transition Disk sources (Class II/III) show
photospheric emission shortwards of 8.0 ${\mu}$m, with excess emission
at 24 ${\mu}$m. In this case we only observe disk emission from the
outer regions of the disk, indicative of disks with large inner
holes. These may be due to grain growth or disk clearing by
(proto)planets.  A more detailed description of the classification is
given in \cite{wmw+06}.

The X-ray and infrared lists were merged with a positional accuracy of
$\sim 1$ arcsec. Of the 85 X-ray sources, 72 were detected in at least one
of the IR bands.  Of these 72, 56 could be classified by our infrared
data.  The remaining 16 sources were detected in too few \spitzer\
bands to construct their SEDs and were faint ($>15$ mag) at
3.6 ${\mu}$m.  We consider these objects likely to be background
galaxies.  The 56 brighter objects were classified as follows: 9 Class
I, 9 Flat Spectrum, 19 Class II, 2 Transition Disks, and 17 Class III
sources.

\begin{figure}[!htbp]
  \begin{center} \leavevmode

\epsfig{file=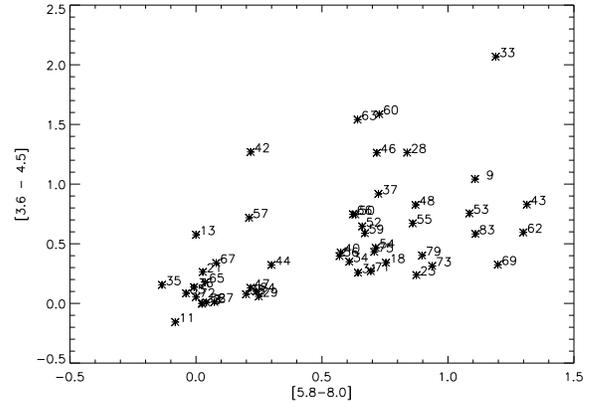, width=8.0cm}

\caption{\spitzer\ color-color diagram for the X-ray sources detected in all four \spitzer\ bands.}

    \label{fig:spitzer}
  \end{center}
\end{figure}

\section{Results}
\label{sec:res}

\subsection{The lack of X-ray emission from Class 0 sources}
\label{sec:c0}

The Serpens cloud has been the subject of many observational studies
in the submm and far IR, which have identified a number of Class 0
sources. \citet{ced+93} published the first list of submm sources
without near-IR counterparts in the Serpens (FIRS1/SMM1, SMM2, SMM4,
SMM3) and \citet{hb96} indicated these four objects plus S68N (SMM 9)
to be Class 0. These objects were also detected in a later survey of 3
mm radio emission by \citet{ts98}, where a total of 32 sources in a
$5.5 \times 5.5$-arcmin area, were identified. \citet{wbw+98} also
identified S68N as a Class 0 object and reported a more accurate
position for it. \citet{hds+99} confirmed the Class 0 status of FIRS1,
SMM3 and SMM4 but suggested that SMM2 may not contain a central
protostar. \citet{dmr+99} obtained wide-field submm continuum and CO
J=2-1 observations of the Serpens Cloud Core. They detected all the
above mentioned sources plus SMM8, SMM10 and SMM11, however for SMM8
and SMM11 they were unable to derive a spectral index\footnote{Due to
the lack of 3-mm fluxes from Testi \& Sargent (1998) for these two
sources.}. \citet{dmr+99} also suggested that these Serpens submm
source may be ``warm'', late Class 0 or early Class I objects.

\citet{etc+2005} searched for 3.5 cm VLA sources in Serpens and,
together with eighteen more evolved sources, they detected FIRS1,
SMM4, SMM9/S68N, and SMM10. They also searched for ISO counterparts
and suggested that these four sources had an ISO counterpart; they
noted that the Class 0 classification given in the literature, for
these four sources, is based on (sub-)millimetric observations
(\citealp{hb96}; \citealp{wbw+98}; \citealp{hds+99}), while the Class
I classification is taken from ISO mid-IR observations
(\citealp{kob+04}). In the case of FIRS1, however, the
cross-identification with ISO objects 258a and 258b is uncertain,
since ISO objects 258a and 258b lie $\ga$ 9 arcsec away from the
position of the mm/submm source and, as discussed by \citet{kob+04},
they could be scattered light from the far-IR source FIRS1.
\citet{kob+04} also exclude ISOCAM detections for SMM2 and SMM3 within
the positional uncertainties.

Table~\ref{tab:c0} provides the positions for this sample of
Class 0 or 0/I sources which fall in the ACIS field of view of our
Serpens observation. 
None of these 6 sources was detected in the recent \spitzer\
observation confirming the high ratio of sub-millimiter to mid-IR
luminosity for these sources, and thereby their Class 0 nature.

\begin{table}[!thbp]
  \begin{center} 

\caption{Class 0 and Class 0/I sources in the field of view of our
\chandra\ ACIS observation. Their ISO counterpart is also
indicated. Coordinates are from: Hogerheijde et al. (1999) for FIRS1,
SMM2, SMM3, and SMM4; Wolf-Chase et al. (1998) for S68N and Davis et
al. (1999) for SMM10. Hogerheijde et al. (1999) indicated SMM2 as
star-less.}

\footnotesize
\leavevmode
\begin{tabular}{l|cccl}
Name & RA & Dec &  ISO  & Cl. \\
     &    &     &  id   & \\
\hline
\footnotesize{FIRS1/SMM1} & 18 29 49.7 &  $+$01 15 21 & -        & 0 \\
\footnotesize{SMM2} &        18 30 00.4 &  $+$01 12 50 & -        &  0? \\
\footnotesize{SMM3} &        18 29 59.3 &  $+$01 14 00 & -        &  0 \\
\footnotesize{SMM4} &        18 29 56.7 &  $+$01 13 15 & 308  &  0/I\\
\footnotesize{S68N/SMM9} &   18 29 47.9 &  $+$01 16 47 & 241  &  0/I \\
\footnotesize{SMM10} &       18 29 52.1 &  $+$01 15 48 & 270  &  0/I\\
\end{tabular}
    \label{tab:c0}
  \end{center}
\end{table}

An important result from the present study is that none of the six
Class 0 sources in Table \ref{tab:c0} are individually detected as
X-ray sources. This non-detection can be translated in an estimate of
the absorbing column density necessary to absorb a source's emission
to the level where it would become undetectable in our data (below
$\sim$ 0.1 cts/ks, as per Table~\ref{tab:src}), assuming a spectrum
and luminosity for the source. We used as proxy the X-ray spectrum of
a Class I source in our field, source 60 (with 0.7 cts/ks),
and increased the absorbing column density of its best-fit model to
the value at which the model predicted source counts would be $\sim$ 0.1
ks$^{-1}$, i.e. our detection limit. Source 60 would become
undetectable in our data behind an absorbing column density of $\nh \sim
40\times 10^{22}$ cm$^{-2}$ or $A_V \sim 200$.  Source 60 has a best
fit $kT$ value of 2.4 keV and an (intrinsic) luminosity of $10^{30}$
\es, which are typical for Class I sources.  In the sample of five
Class I sources of $\rho$ Oph compiled by \citet{ogm05}, the average
values of $kT$ and $L_{\rm X}$ are 3.2 keV and $2.9\times 10^{30}$ erg
s$^{-1}$. Therefore, in the assumption that Class 0 sources have X-ray
characteristics similar to Class I objects, source 60 is a good proxy.

\begin{figure}[!htbp]
  \begin{center} \leavevmode 

\epsfig{file=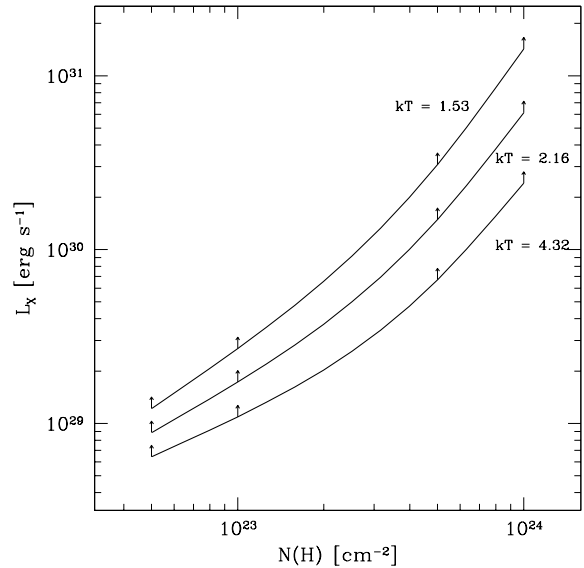, height=8.0cm}

\caption{Minimum X-ray luminosity required for a source to be
detectable in our observations, as a function of the absorbing column
density, for different temperatures of the emitting plasma. The curves
were obtained using the {\sc pimms} software at {\sc Heasarc},
assuming for the emitting plasma a Raymond-Smith model with 
$Z=0.2~Z_{\sun}$.}

    \label{fig:limLx}
  \end{center}
\end{figure}

As already mentioned, \citet{tkh+2001} and \citet{hcp+2005} derive
column densities of the order of $(10-30)\times 10^{22}$ cm$^{-2}$ in
their X-ray observations of highly embedded YSOs. Since these objects
appear to have higher plasma temperatures than source 60 ($kT \sim
3-4$~keV) and similar luminosity, than we can exclude the possibility
that X-ray sources similar to the one reported by the above authors
are embedded within the mm/submm sources of Table ~\ref{tab:c0}.  As
shown in Fig.\,\ref{fig:limLx}, a source with a higher plasma
temperature than source 60 and the same luminosity would have to be
screened by an absorbing column density higher than $40\times 10^{22}$
cm$^{-2}$ for its count rate to be below our detection threshold. For
comparison, a source with $kT = 4.3$ keV and $L_X = 10^{30}$ \es\,
would be undetectable in our data only if screened by an absorbing
column density $\ga 60\times 10^{22}$ cm$^{-2}$ ($A_V \ga 300$ mag).

A source with characteristics similar to IRAS 21391+5802, reported by
\citet{gfg+06}, would have also been well visible in our observation,
its properties ($\nh \sim 10^{24}$, $kT = 6.0$ keV and $L_{\rm X} \sim
2\times 10^{31}$ \es) implying in ACIS a count rate of 1.0
ks$^{-1}$. If the high luminosity and plasma temperature of IRAS
21391+5802 are, however, the result of a flare and the more common
situation is the one in which a lower luminosity, lower plasma
temperature, source (e.g. with $L_{\rm X} = 10^{30}$ \es\ and $kT = 2
- 4$ keV) is hidden behind 500 mag of extinction, then our observation
is not sensitive enough to detect such a source (see
Fig.\,\ref{fig:limLx}).

\begin{figure*}[!htbp]
  \begin{center} \leavevmode 

\epsfig{file=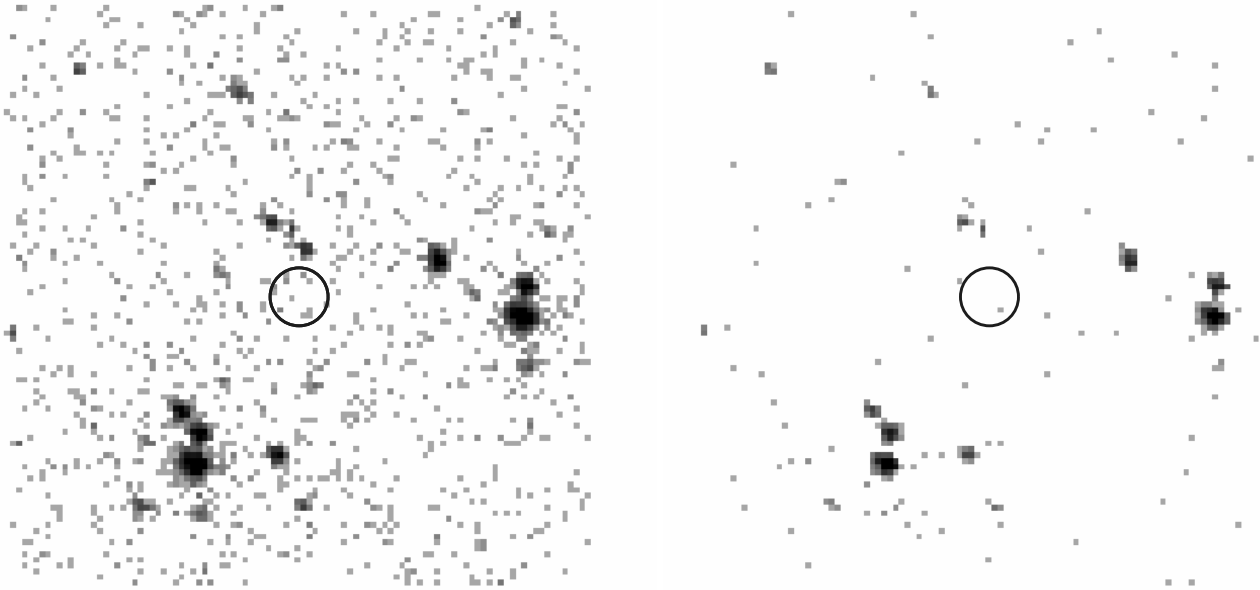, height=8.0cm, bbllx=37, bblly=284, bburx=576, bbury=520, clip= }

\caption{The coadded image obtained by adding ACIS events from six
regions of $200 \times 200$ pixels centered on the position of the
Class 0 sources, for the energy intervals $\Delta E = 0.5 - 8.0$ keV
({\it left}) and $\Delta E = 4.0 - 8.0$ keV ({\it right}). The
circular region in the center is $5''$, corresponding to positional
uncertainties of the mm/submm sources. No X-ray source is present
within this area indicating that the Class 0 sources in our sample are
unlikely to be X-ray sources with intensities just below our detection
threshold. The sources present in the image are genuine sources
falling by chance near the position of one of the Class 0 sources,
which have been ``coadded''.}

    \label{fig:coadd}
  \end{center}
\end{figure*}

Of course, we cannot exclude the possibility that an X-ray emitting
source with properties similar to the ones of source 60 is hiding
beneath a column of absorbing material higher than $40\times 10^{22}$
cm$^{-2}$\footnote{In $\rho$ Oph, \citet{mon98} derives, for compact
cores, molecular hydrogen column densities of $N({\rm H2}) = (10-80)
\times 10^{22}$ cm$^{-2}$, corresponding to $A_V = 100-800$.} or that
these Class 0 sources are weaker sources. In order to derive a
more sensitive upper limit for the typical Class 0 source in the
Serpens Cloud Core, we registered and co-added the X-ray photons from
$200 \times 200$ pixels ($98\times 98$ arcsec) regions around each of
the sources, centered on the coordinates given in
Table\,\ref{tab:c0}. Since there are six such regions, the result of
this operation is equivalent (under the assumption that the six Class
0 sources are all similar to each other) to an observation of a single
Class 0 source for an integration time six times longer than our
original observation, i.e.\ $\sim 540$ ks. The image of the
``coadded'' event list, resulting from this operation, is shown in
Fig.\,\ref{fig:coadd} for the two energy intervals $\Delta E = 0.5 -
8.0$ keV and $\Delta E = 4.0 - 8.0$ keV\footnote{If Class 0 or Class
0/I sources have high plasma temperatures, as in the case reported by
Getman et al. (2006), then they could be more easily detectable in an
hard-band image.}. No source appears to be present in the centre of
the coadded image, within a $5''$ radius, corresponding to the
positional uncertainties of the mm/submm objects. The total number of
counts within the $3''$-radius central area is 15 (a $3''$ radius
being typical of source extraction regions for ACIS data), completely
consistent with the background level for the coadded event list.  The
measured average background level in the coadded event file is $0.5$
cts arcsec$^{-2}$, i.e.\ $14\pm 4$ counts in a circular area with a
$3''$ radius. This agrees well with the level of background in the
original image, 0.083 cts arcsec$^{-2}$, indicating that the
``coadded'' image does not suffer from significant systematics.

In the coadded image, a source at 6$\sigma$ level\footnote{5$-$6
$\sigma$ above background being the significance level of the weaker
sources in our list.} above the background would have implied 41
counts within a $3''$-radius area, corresponding to a source count
rate of 0.04 cts/ks. The fact that there is no such a source sets a
more stringent upper limit on the typical level of X-ray
emission from these Class 0 sources of $L_{\rm X} < 0.4 \times
10^{30}$\es (in the above assumption of $\nh = 40\times 10^{22}$
cm$^{-2}$ and $kT = 2.3$~keV).

The median luminosity of the YSOs for which spectra have been studied
in the present work (32 out of 35 -- as three were undergoing a strong
flare) is $0.4 \times 10^{30}$\es, the same as the upper limit on the
typical X-ray luminosity of the Class 0 sources, implying that if the
Class 0 sources in the Serpens cloud are indeed X-ray sources, then
they have luminosities below the level typical of the other YSO in the
region (or they are hidden behind an absorbing column density higher
than $40 \times 10^{22}$ cm$^{-2}$).

\subsection{Spectral characteristics of the X-ray-bright YSOs}

A spectral and timing analysis was carried out on all sufficiently
bright sources (count rate $> 0.5$ cts/ks). The spectra and light
curves of all the sources brighter than 2.0 cts/ks are shown in
Fig.\,\ref{fig:lc} and Fig.\,\ref{fig:lc2}, the results of the
spectral analysis for these source are summarised in
Table\,\ref{tab:psfit}. The results of the spectral analysis for some
of the sources with count rate between 0.5 and 2.0 cts/ks are also
summarised in Table\,\ref{tab:psfit}. For these weaker sources, only
the results for the sources with \spitzer\ classification are
reported.  In the sample of sources stronger than 0.5 cts/ks, there
are no sources with a ISO counterpart but no \spitzer\ data, while
there is a number of sources observed by \spitzer\ for which there are
no ISO counterparts, so in Table\,\ref{tab:psfit} we used only the
classification derived from the \spitzer\ observation. The sources are
subdivided in the table according to their classification.

The foreground absorption toward the Serpens complex can be estimated
using the galactic extinction law of \citet{bs80},
$$
A_V = \alpha_{\rm inf} (1 - exp(-z/100))
$$
where $\alpha_{\rm inf} = 0.15/\sin b$, $z = 10^{((DM+5)/5)}\sin b$,
$b$ being the galactic latitude and $DM$ the distance modulus. Using
the distance to Serpens of 260 pc ($DM = 7$) and its latitude
($b=5.39$) one derives a minimum value of $A_V = 0.3$ or $N({\rm H}) =
0.06 \times 10^{22}$ cm$^{-2}$. The source with the lowest value of
absorbing column density is source 38 for which $N({\rm H}) = 0.05 \pm
0.06 \times 10^{22}$ cm$^{-2}$, consistent with all the YSOs for which
spectra and light curves have been studied here being part of the
Serpens complex (rather than being foreground objects).

\begin{figure}[!htbp]
  \begin{center} \leavevmode 

\epsfig{file=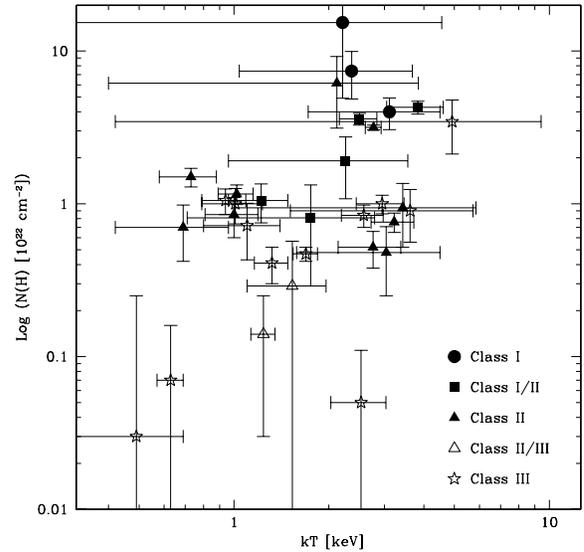, height=8.0cm}

\caption{Values of absorbing column densities vs plasma temperatures
for the sources in Table\,\ref{tab:psfit}. Sources undergoing a strong flare
(79, 66, 44) or likely non-stellar (6, 13 and 70) have been
omitted. Error bars are at 1$\sigma$.}

    \label{fig:spe_table}
  \end{center}
\end{figure}

Fig.\,\ref{fig:spe_table} shows the values of the absorbing column
densities vs. the plasma temperatures of the sources in
Table\,\ref{tab:psfit}, except for the sources undergoing a strong
flare (66, 79 and 44) or unlikely to be stars (6, 13 and 70), which have
been omitted. A trend of decreasing absorbing column densities from
Class I to Class III is clearly present. There is a lack of
low-plasma-temperature high-absorption sources, which is expected, as
with low plasma temperature higher luminosities are necessary in order
to see a source through higher column densities. There is also,
possibly, a lack of sources in the lower-right region of the diagram
(where sources have a high plasma temperature and small absorption,
and are thus easy to detect), hinting at an evolutionary effect from
the Class I (high X-ray temperature, high extinction) to the Class III
stage (lower X-ray temperature and
extinction). Table\,\ref{tab:medians} lists the median values of $\nh$
and $kT$ for the different classes from which the trend of decreasing
absorbing column density from Class I to Class III is clear and a
trend of decreasing plasma temperature is also somewhat present
(although less clear). Note also that, since these median values are
based on detected sources only (we did not include upper limits,
because of the lack of a parent catalogue), the trend in $kT$ could be due
to a selection effect (i.e. Class I/II soft sources, being very
absorbed, are not detected).

The diagram in Fig.\,\ref{fig:spe_table} is similar to the one derived
by \citet{ogm05} for the $\rho$ Oph cloud core. They also note a
lack of sources in the high-plasma-temperature small-absorption region
of the diagram and interpret this as evidence of an evolutionary
effect. A similar effect is noted by \cite{fms2006} in their study of
the NGC\,2264 star-forming complex.  

As shown in Table\,\ref{tab:medians}, no evolutionary pattern is seen
in the sources' luminosities. The higher X-ray luminosities of Class I
YSOs likely being due to a selection effect, since only the brightest
sources are detected behind higher column densities.

\begin{table*}
  \begin{center}

  \caption{Best-fit spectral parameters for 38 X-ray sources with
  count rate greater than 0.5 cts/ks. Results for all sources with
  count rate greater than 2 cts/ks are given; for the sources with rate
  between $0.5 - 2.0 {\rm cts/ks}$, only results for sources which
  have a counterpart with a \spitzer\-derived YSO class are
  reported. $E\!M$ is the emission measure, $P$ the null-hypothesis
  probability of the fit, and $L_{\rm X}$ is the X-ray luminosity
  derived from the ``unabsorbed'' X-ray flux. Units are $N_{22} =
  10^{22}~{\rm cm^{-2}}$, $E\!M_{53} = 10^{53}$~cm$^{-3}$, $F_{-13}=
  10^{-13}$~\ecms, and $L_{30} = 10^{30}$~\es. The spectral fits were
  carried out in the energy range 0.3--7.5~keV unless otherwise
  indicated. $C_{\rm KS}$ gives the probability of constancy from the
  Kolmogorov-Smirnov test for the sources' light curves.}

    \leavevmode
     \begin{tabular}{l|cccccccc|c}
Src &  $N({\rm H})$ & $kT$ & $Z$ & $E\!M$ & $\chi^2$ & $P$ & $F_{\rm X}$ & $L_{\rm X}$ & $C_{\rm KS}$ \\
\hline
~ & $N_{22}$ & keV &  $Z_{\odot}$ & $E\!M_{53}$ & ~ & ~ & $F_{-13}$  & $L_{30}$ & \\
\hline
Class I &  &  &  &  &  &  &  &  & \\

32 & 15.38 $\pm$ 10.47 & 2.21 $\pm$ 2.35 & 0.30 (froz.) & 2.01 $\pm$ 9.01 & 1.13 & 0.34 & 0.27 & 2.20
& 0.03\\

60 & 7.4 $\pm$ 2.54 & 2.36 $\pm$ 1.32 & 0.3 (froz.) & 0.93 $\pm$ 1.67 & 0.26 & 1 & 0.21 & 1.01 & 0.11 \\

61 & 4.00 $\pm$ 0.94 & 3.11 $\pm$ 1.39 & 0.30 (froz.) & 0.54 $\pm$ 0.56 & 1.23 & 0.20 & 0.24 & 0.68 & 0.05\\

Class I/II &  &  &  &  &  &  \\

13* & 4.34 $\pm$ 0.81 & 21.82 $\pm$ 32.48 & 0.3 (froz.) & 0.42 $\pm$ 0.45 & 0.8 & 0.91 & 1.07 & 1.7 & 0.1e-5 \\

37$^{\dag}$ & 1.91 $\pm$ 0.83 & 2.26 $\pm$ 1.3 & 0.3 (froz.) & 0.14 $\pm$ 0.19 & 1.2 & 0.3 & 0.06 & 0.15 & 0.9e-4 \\

48 & 3.6 $\pm$ 0.32 & 2.5 $\pm$ 0.34 & 0.3 (froz.) & 3.5 $\pm$ 1.3 & 0.84 & 0.91 & 1.26 & 3.9 & 0.1e-4 \\

53 & 4.28 $\pm$ 0.41 & 3.83 $\pm$ 0.78 & 0.3 (froz.) & 2.65 $\pm$ 1.03 & 0.63 & 1 & 1.39 & 3.47 & 0 \\

55$^{\dag}$ & 1.05 $\pm$ 0.3 & 1.22 $\pm$ 0.26 & 0.3 (froz.) & 0.46 $\pm$ 0.34 & 0.47 & 0.98 & 0.14 & 0.44 & 0.45 \\

66$^{\dag}$ f & 1.75 $\pm$ 0.75 & 4.33 $\pm$ 4.25 & 0.3 (froz.) & 0.16 $\pm$ 0.15 & 1.01 & 0.44 & 0.13 & 0.22 & 0 \\

83$^{\dag}$ & 0.81 $\pm$ 0.52 & 1.75 $\pm$ 1.04 & 0.3 (froz.) & 0.11 $\pm$ 0.14 & 1.02 & 0.43 & 0.05 & 0.11 & 0.02 \\

Class II &  &  &  &  &  &  &  &  \\

6* & 3.36 $\pm$ 1.60 & 28.02 $\pm$ 96.82 & 0.30 (froz.) & 0.19 $\pm$ 0.08 & 0.77 & 0.73 & 0.18 & 0.20 & 0.15\\

18$^{\dag}$ & 0.48 $\pm$ 0.23 & 3.04 $\pm$ 1.46 & 0.3 (froz.) & 0.09 $\pm$ 0.06 & 0.99 & 0.46 & 0.08 & 0.11 & 0.08 \\

23$^{\dag}$ & 0.7 $\pm$ 0.28 & 0.69 $\pm$ 0.27 & 0.3 (froz.) & 0.3 $\pm$ 0.56 & 1.08 & 0.37 & 0.07 & 0.31 & 0.01 \\

34 & 0.76 $\pm$ 0.11 & 3.22 $\pm$ 0.5 & 0.3 (froz.) & 0.78 $\pm$ 0.21 & 0.7 & 0.99 & 0.65 & 0.96 & 0 \\

44 f & 1.41 $\pm$ 0.34 & 6.7 $\pm$ 3.41 & 0.3 (froz.) & 0.41 $\pm$ 0.2 & 0.73 & 0.91 & 0.43 & 0.62 & 0 \\

50$^{\dag}$ & 0.94 $\pm$ 0.42 & 3.43 $\pm$ 2.42 & 0.3 (froz.) & 0.25 $\pm$ 0.24 & 0.32 & 0.96 & 0.21 & 0.32 & 0.11 \\

54 & 3.16 $\pm$ 0.11 & 2.77 $\pm$ 0.16 & 0.15 $\pm$ 0.06 & 17.68 $\pm$ 2.02 & 1.08 & 0.17 & 7.12 & 19.2 & 0.6e-2 \\

59 & 1.5 $\pm$ 0.21 & 0.73 $\pm$ 0.15 & 0.3 (froz.) & 0.74 $\pm$ 0.7 & 0.65 & 0.87 & 0.08 & 0.76 & 0.6e-3 \\

63$^{\dag}$ & 6.17 $\pm$ 3.04 & 2.12 $\pm$ 1.72 & 0.3 (froz.) & 0.46 $\pm$ 1.34 & 0.8 & 0.57 & 0.1 & 0.48 & 0.05 \\

69 & 1.16 $\pm$ 0.17 & 1.02 $\pm$ 0.13 & 0.3 (froz.) & 0.69 $\pm$ 0.37 & 0.98 & 0.51 & 0.16 & 0.69 & 0.7e-5 \\

73 & 0.85 $\pm$ 0.25 & 1 $\pm$ 0.19 & 0.3 (froz.) & 0.42 $\pm$ 0.23 & 0.83 & 0.72 & 0.12 & 0.42 & 0.8e-3 \\

75 & 0.52 $\pm$ 0.14 & 2.76 $\pm$ 0.62 & 0.3 (froz.) & 0.35 $\pm$ 0.14 & 0.65 & 0.98 & 0.29 & 0.41 & 0.02 \\

79$^{\ddag}$ f & 0.52 $\pm$ 0.04 & 3.71 $\pm$ 0.34 & 0.3 (froz.) & 2.45 $\pm$ 0.26 & 1.00 & 0.50 & 2.44 & 3.16 & 0 \\

Class II/III &  &  &  &  &  &  &  &   & \\

84 & 0.14 $\pm$ 0.11 & 1.24 $\pm$ 0.11 & 0.3 (froz.) & 0.13 $\pm$ 0.07 & 1.1 & 0.33 & 0.1 & 0.12 & 0.18 \\

85 & 0.29 $\pm$ 0.28 & 1.53 $\pm$ 0.43 & 0.3 (froz.) & 0.11 $\pm$ 0.08 & 1.11 & 0.34 & 0.07 & 0.1 & 0.32 \\

Class III &  &  &  &  &  &  &  \\

2 & 0.41 $\pm$ 0.11 & 1.32 $\pm$ 0.16 & 0.11 $\pm$ 0.07 & 0.71 $\pm$ 0.36 & 0.86 & 0.78 & 0.29 & 0.53 & 0.1e-4 \\

3$^{\dag}$ & 0.03 $\pm$ 0.22 & 0.49 $\pm$ 0.2 & 0.3 (froz.) & 0.06 $\pm$ 0.14 & 0.54 & 0.85 & 0.06 & 0.05 & 0.26 \\

11 & 0.47 $\pm$ 0.05 & 1.69 $\pm$ 0.15 & 0.03 $\pm$ 0.04 & 2.65 $\pm$ 0.72 & 0.96 & 0.62 & 1.2 & 2.02 & 0.02 \\

21$^{\dag}$ & 0.9 $\pm$ 0.34 & 3.62 $\pm$ 2.11 & 0.3 (froz.) & 0.19 $\pm$ 0.14 & 1.15 & 0.29 & 0.16 & 0.24 & 0 \\

27 & 0.72 $\pm$ 0.29 & 1.1 $\pm$ 0.3 & 0.04 $\pm$ 0.06 & 0.82 $\pm$ 0.96 & 0.68 & 0.96 & 0.18 & 0.49 & 0.07 \\

29$^{\dag}$ & 1 $\pm$ 0.26 & 1.01 $\pm$ 0.22 & 0.3 (froz.) & 0.26 $\pm$ 0.22 & 0.48 & 0.94 & 0.07 & 0.26 & 0.4e-3 \\

38 & 0.05 $\pm$ 0.06 & 2.53 $\pm$ 0.5 & 0.43 $\pm$ 0.36 & 0.26 $\pm$ 0.13 & 0.98 & 0.52 & 0.34 & 0.31 & 0.3e-6 \\

45 & 1.05 $\pm$ 0.20 & 0.94 $\pm$ 0.15 & 0.30 (froz.)  & 0.35 $\pm$ 0.24 & 0.85 & 0.65 & 0.08 & 0.37 & 0.09\\

47 & 0.07 $\pm$ 0.09 & 0.63 $\pm$ 0.06 & 0.3 (froz.) & 0.14 $\pm$ 0.09 & 0.88 & 0.66 & 0.14 & 0.14 & 0.32 \\

57$^{\dag}$ & 3.45 $\pm$ 1.33 & 4.92 $\pm$ 4.5 & 0.3 (froz.) & 0.17 $\pm$ 0.23 & 0.56 & 0.85 & 0.12 & 0.25 & 0.9e-3 \\

65 & 1 $\pm$ 0.14 & 2.95 $\pm$ 0.51 & 0.3 (froz.) & 0.91 $\pm$ 0.28 & 0.69 & 0.99 & 0.65 & 1.09 & 0 \\

67 & 0.84 $\pm$ 0.14 & 2.58 $\pm$ 0.39 & 0.3 (froz.) & 0.91 $\pm$ 0.29 & 0.75 & 0.95 & 0.61 & 1.03 & 0.5e-7 \\

Not a star &  &  &  &  &  \\
70 & 1.1 $\pm$ 0.52 & 63.96 $\pm$ 230.2 & 0.3 (froz.) & 0.05 $\pm$ 0.05 & 0.83 & 0.77 & 0.35 & 0.41 & 0.1e-3 \\
   
    \end{tabular}
    \label{tab:psfit}
  \end{center}

f -- Source undergoing an intense flare

* -- Unlikely a stellar source

\dag Spectral fit $\Delta E = 0.1-7.5$~keV
\ddag Spectral fit $\Delta E = 0.3-8.5$~keV

\end{table*}

\begin{table}
  \begin{center}

  \caption{Median values of absorbing column density, plasma
temperature, and X-ray luminosity for the different classes. Units as
from Table 2. Flaring sources and sources likely non-stellar are not included.}
\leavevmode
     \begin{tabular}{l|ccc}
Class &  $N({\rm H})$ & $kT$ &  $L_{\rm X}$ \\
\hline
~ & $N_{22}$ & keV & $L_{30}$ \\
\hline
 I & 7.4 & 2.4 & 1.7\\
I/II & 1.9 & 2.3 & 0.3\\
 II & 0.9 & 2.8 & 0.4\\
 II/III & 0.3 & 1.5 & 0.1\\
III & 0.8 & 1.7 & 0.2\\
\hline
I/II $+$ II & 1.0 & 2.3 & 0.3\\
II/III $+$ III & 0.7 & 1.5 & 0.2\\
\end{tabular}
    \label{tab:medians}
  \end{center}
\end{table}

\subsection{Temporal variability of the the X-ray bright YSOs}

The majority of the YSOs with count rate $> 0.5$ cts/ks, i.e. bright
enough for their light curves to be analysed, shows significant
variability, although with different characteristics. Three sources
(44, 66 and 79) undergo a flare, with that of source 79 being the
longest and most intense. This flare is analysed and discussed in detail
in Sec.\,\ref{sec:src79}.

To evaluate the presence of X-ray variability, the Kolmogorov-Smirnov
test, which measures the maximum deviation of the integral photon
arrival times from a constant source model, was applied to all the
brighter X-ray sources (count rate $> 0.5$ cts/ks). The results, in
terms of constancy probability, are given in
Table\,\ref{tab:psfit}. The majority of sources, 21 out of 38, has a
probability of constancy below 2\%. Of these 21, 19 are classified as
YSOs (sources 13 and 70 are unlikely to be stars).

In order to quantify the X-ray variability of these 19 YSOs we
computed the normalized cumulative distribution of the amplitude of
their variability. This represents the fraction of time that a source
spends in a state with the flux larger than a given value, expressed
in terms of a given normalization value, which can be the minimum
count rate, the median count rate, etc. For the present sample we took
as normalization value the count rate above which a source spends 90\%
of the time; this is less sensitive to noise fluctuations than the
real minimum.

\begin{figure}[!tbp]
  \begin{center} \leavevmode \epsfig{file=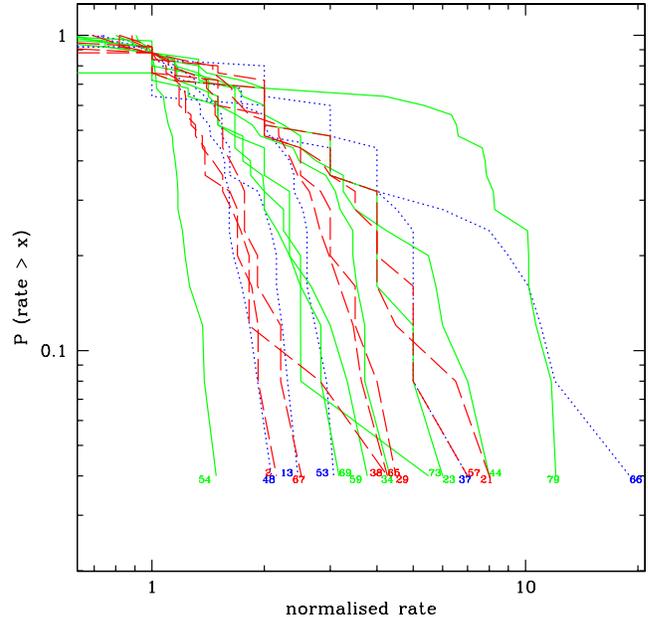, width=9.0cm}

    \caption{Normalized cumulative count-rate distribution for all the
      YSOs in Table 2 with Kolmogorov-Smirnov probability of constancy
      ($C_{\rm KS}$) lower than 2\%. Dotted line indicates Class I/II
      objects, continuous line Class II, and short-dashed Class III
      objects.}

    \label{fig:intAll}
  \end{center}
\end{figure}

Figure~\ref{fig:intAll} shows the distribution for the stars in our
sample. Some sources (e.g. 2, 48 and 54) show mostly low-amplitude
variability, in which less than 30\% of the time is spent in a state
1.3$-$2 times above the minimum. Other sources (e.g. 21, 37, and 65)
on the other hand, spend more than 50\% of the time in such a
state. Kolmogorov-Smirnov tests for the cumulative distributions of
various pairs of sources also indicate the presence of different
behaviors. However sources of the same class are present in
groups with different variability behavior, so that an interpretation
in terms of source evolution is not possible. This lack of trends in
the variability of different classes of YSOs was also observed for a
sample of WTTS and CTTS in L1551 (\citealp{gfs+06}).

\section{Interesting individual sources}
\label{sec:peculiar}

As already mentioned, we extracted the light curves and spectra
of all the sources with count rate $\ge 2$ cts/ks. These are shown in
Fig.~\ref{fig:lc} and Fig.~\ref{fig:lc2} and the results of the spectral
fits for these sources are summarised in Table~\ref{tab:psfit}. In
this section we comment on those sources which have a peculiarity or show
an interesting behavior.

\subsection{Source 13}

A relatively bright X-ray source, it has no 2MASS counterpart but it
has a \spitzer\ counterpart with colors apparently compatible with a
Flat Spectrum YSO. Its X-ray spectrum is however very hard and
strongly absorbed. If it were a thermal spectrum, the temperature
would be in excess of 20 keV. A $\alpha = 1.8$ power law provides as
good a fit as the thermal model; in both cases a significant soft
excess remains. The absorbing column density is 3 to $4 \times
10^{22}$ cm$^{-2}$, depending on the spectral model used. Such
spectrum is very unlikely to originate in a stellar source, and
therefore we consider this source most likely an active background
extragalactic object (an AGN). The source is present in the \xmm\
observation of \cite{pre2003}, who identified an IR counterpart (but
does not extract an X-ray spectrum).

\subsection{Source 44}

Classified as a Class II source, on the basis of the ISO and its
\spitzer\ colors, the source is observed during an intense (a factor
of $\simeq 4$) flare, as reflected in the high X-ray temperature, $kT
= 6.7$ keV. The large absorption $\nh = 1.4\times 10^{22}$ cm$^{-2}$
($A_V=7.3$) points to a source deep in the cloud. The source was first
detected at infrared wavelengths by \citet{ec89} and \citet{ec92} and
later observed with the VLA (\citealp{bont96}).

\subsection{Source 48}
\label{sec:src48}

\begin{figure}[!htbp]
  \begin{center} \leavevmode 
     
     \epsfig{file=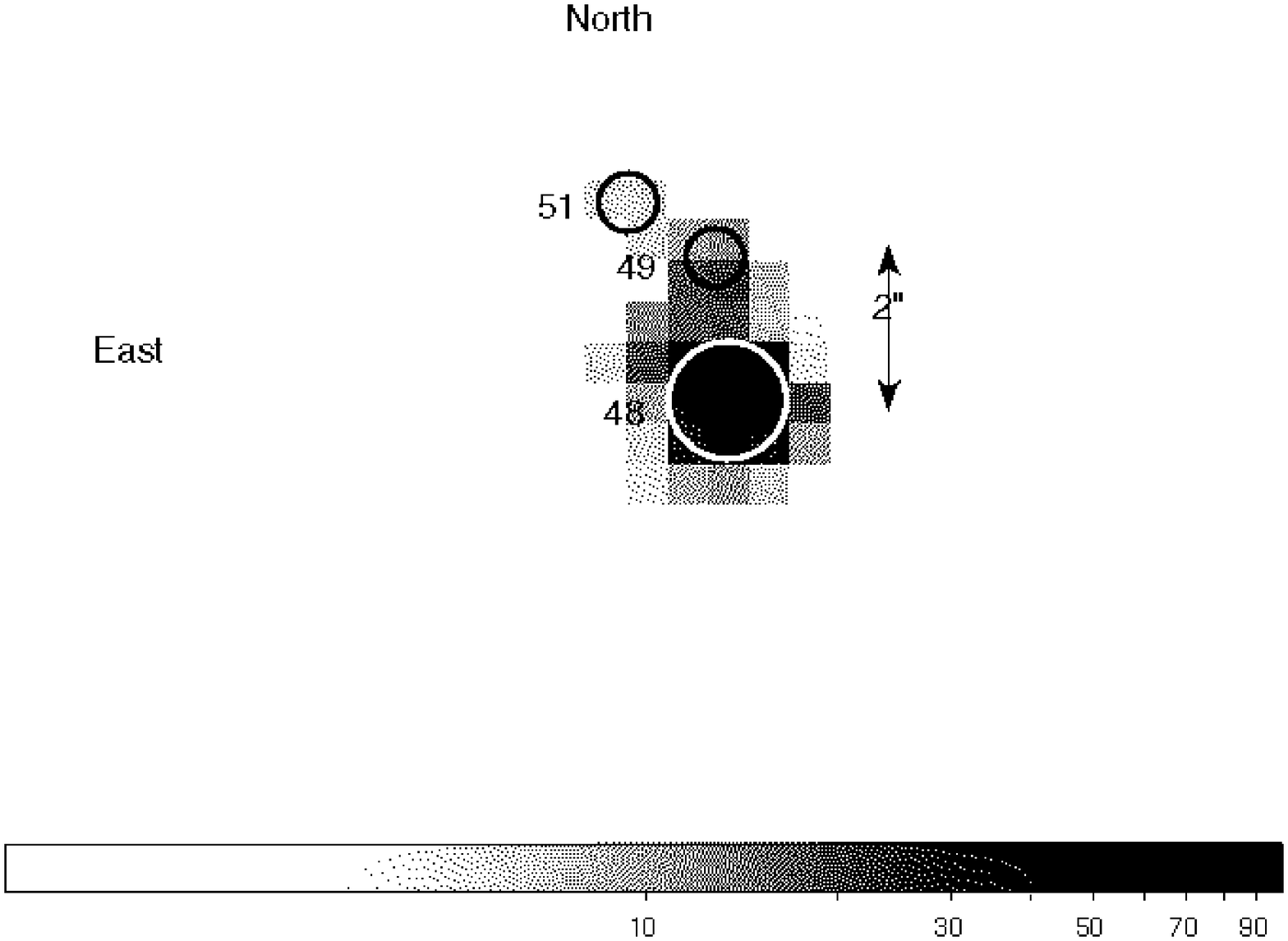, width=8cm, bbllx=37, bblly=236,bburx=576, bbury=600, clip=}	
\caption{X-ray image of SVS 20 (source 48). Source 48 and 49 are at
the right distance and orientation to be the X-ray counterpart of SVS
20-S and SVS20-N respectively. Source 51 could be a very absorbed
stellar source or a background AGN (see text.)}
  
 \label{fig:img48}	
  \end{center}
\end{figure}

Source 48 is the X-ray counterpart of SVS 20, the deeply embedded IR
double source located inside the ``eye'' structure at the center of
the Serpens nebula.  \citet{ell+87} derived a separation of $1.6$
arcsec between the two components and estimated a total luminosity in
the range $7-45~L_{\sun}$; \citet{ctp+05} found that the temperatures
and luminosities of the two protostellar objects are quite different,
for SVS-S they derived $L \approx 20-80 L_{\sun}$ and for SVS-N $L
\approx 0.9 L_{\sun}$. \citet{dgc+05} estimated for SVS-S a luminosity
of $24.7~L{\sun}$. As can be seen from Fig.\,\ref{fig:img48}, in the
\chandra\ image the PSF of this source is elongated in the North-East
direction and \textsc{Pwdetect} places three sources (48, 49 and 51)
within it.  At about 3 arcmin off-axis the two components of SVS 20,
cannot be fully resolved by \chandra. Source 48 and 49, however, are at
the right distance and orientation to be the X-ray counterparts of SVS
20-S and N respectively. The presence of emission to the NE of Source
49 is unexpected and points to a more complex situation, perhaps the
presence of a third source (without IR counterpart) or of an X-ray
emitting jet. Using near-IR polarimetry combined with narrow-band
imaging of molecular hydrogen emission, \citet{hwk97} detected outflow
activity in the surroundings of SVS 20, and in particular they observed
a knot of emission at $\sim 20''$ distance from SVS 20, in the
NE direction.

We extracted the photons from a small ($\sim 1''$ radius) region
around source 51 and obtained roughly 20 photons with a median energy
of $\simeq$ 5 keV. This is inconsistent with X-ray emission from a
stellar outflow whose typical energy are below 1 keV and would point
to a very absorbed stellar source or an extragalactic background
source. The source, however, is not fully resolved from sources 49 and
48, which will likely contaminate it (more so at high energies,
because of the energy dependence of the PSF), so that one cannot draw
definite conclusions on the spectral energy distribution of this
source.

ISO and \spitzer\ colors for the unresolved system indicate a Flat
Spectrum source. In the X-ray the star shows significant variability,
with variations of a factor of two on a 20 ks time scale. With $kT =
2.5$ keV and $\nh = 3.6\times 10^{22}$ cm$^{-2}$, equivalent to $A_V =
19$ mag, the source is relatively hot and deeply embedded in the
cloud.

\subsection{Source 54 -- [EC92] 95}

[EC92] 95 is a heavily absorbed intermediate-mass ($\sim 4 M_{\sun}$)
YSO, for which \cite{pre99} derived, from IR spectroscopy, a spectral
type K2, with $A_V \sim 36$ mag and $L_{\rm bol} \sim
60\,L_\odot$. Assuming that the X-ray absorption was the same as the
optical one, \cite{pre98} attributed an extremely large X-ray
luminosity to this star ($L_{\rm X} \sim 10^{33}$), much higher than
any known stellar source. \xmm\ observations \citep{pre2003} allowed
the X-ray absorption to be directly determined, showing it to be over
a factor of two lower than the optical estimate, thus bringing the
X-ray luminosity to a value within the range observed in YSOs. The
reason why the two absorption determinations (optical and X-ray)
differ is, however, still unknown.

The X-ray spectral parameters determined from this \chandra\
observation are $kT=2.7\pm0.2$ keV and $L_{\rm X} = 1.9\times 10^{31}$
erg s$^{-1}$, with $N_{\rm H} = (3.2\pm 0.1) \times 10^{22}$ cm$^{-2}$
and $Z = 0.2\,Z_\odot$, remarkably consistent with the parameters
derived by \cite{pre2003} for the \xmm\ observation, showing that the
source has little long-term variability in its X-ray emission.

\cite{pre2003} speculated about a
number of possible scenarios for the discrepancy, including some
unidentified peculiarity of flat-spectrum YSOs, by analogy with IRS 5
in L1551. However, the line of reasoning of \cite{pre2003} is based on
the assumption that X-ray emission is observed from the IRS
5 protostar (which has $A_V \sim 150$ mag). As demonstrated by
\cite{ffm+02} and \cite{bfr2003} the X-ray emission in this case comes
from the associated Herbig-Haro object HH~154, which has, given that
the jet penetrates through the absorbing cloud, a much lower
absorption, with $A_V \sim 7$ mag, compatible with the absorption
measured for the X-ray source.

One possible explanation for the peculiar discrepancy between the
X-ray and optically determined absorption columns is that the source
is surrounded by a thick accreting disk, and is seen nearly edge-on
(explaining the large optically determined $A_V$). If at the same time
the X-ray emission comes from a region somewhat displaced above or
below the disk (as it could be the case for a large polar corona),
this could explain the lower absorbing column density observed in
X-rays. However, this implies a corona with a significant scale height
above the photosphere, and this is in conflict with recent evidence,
for example, with the observed occurrence of rotational modulation in
YSOs in Orion (\citealp{fms+05A}), which implys that most coronal
material is located in compact structures. While large \emph{flaring}
magnetic loops have been observed, also in Orion YSOs
(\citealp{ffr+2005}), the consistency of the X-ray emission from
[EC92] 95 over a number of years appears to rule out the hypothesis
that one may be seeing a large flaring loop extending at a large
distance from the star's photosphere. Indeed, if the observed source
variability (Fig.\,\ref{fig:lc}), is due to a flare, then the
timescales of such variability ($\la 20$ ks) would imply a relatively
compact structure ($L_{\rm loop} \la 5 R_{\sun}$).

Another possible explanation for the discrepancy between the X-ray and
optically determined absorption columns would be the presence of a
lower mass companion. This companion star would have to be displaced
from the disk of the main star, and be the main source of X-ray
luminosity while contributing only a small fraction of the optical
flux. Nevertheless, for such a companion star to have an X-ray
luminosity of $1.9 \times 10^{31}$ \es, its mass should be around $2-3
M_{\sun}$, as can be inferred by looking at the scatter plot of X-ray
luminosity versus star mass in e.g. \citet{fdm03} and therefore would
provide significant contribution to the luminosity of the system.

\begin{figure*}[!htbp]
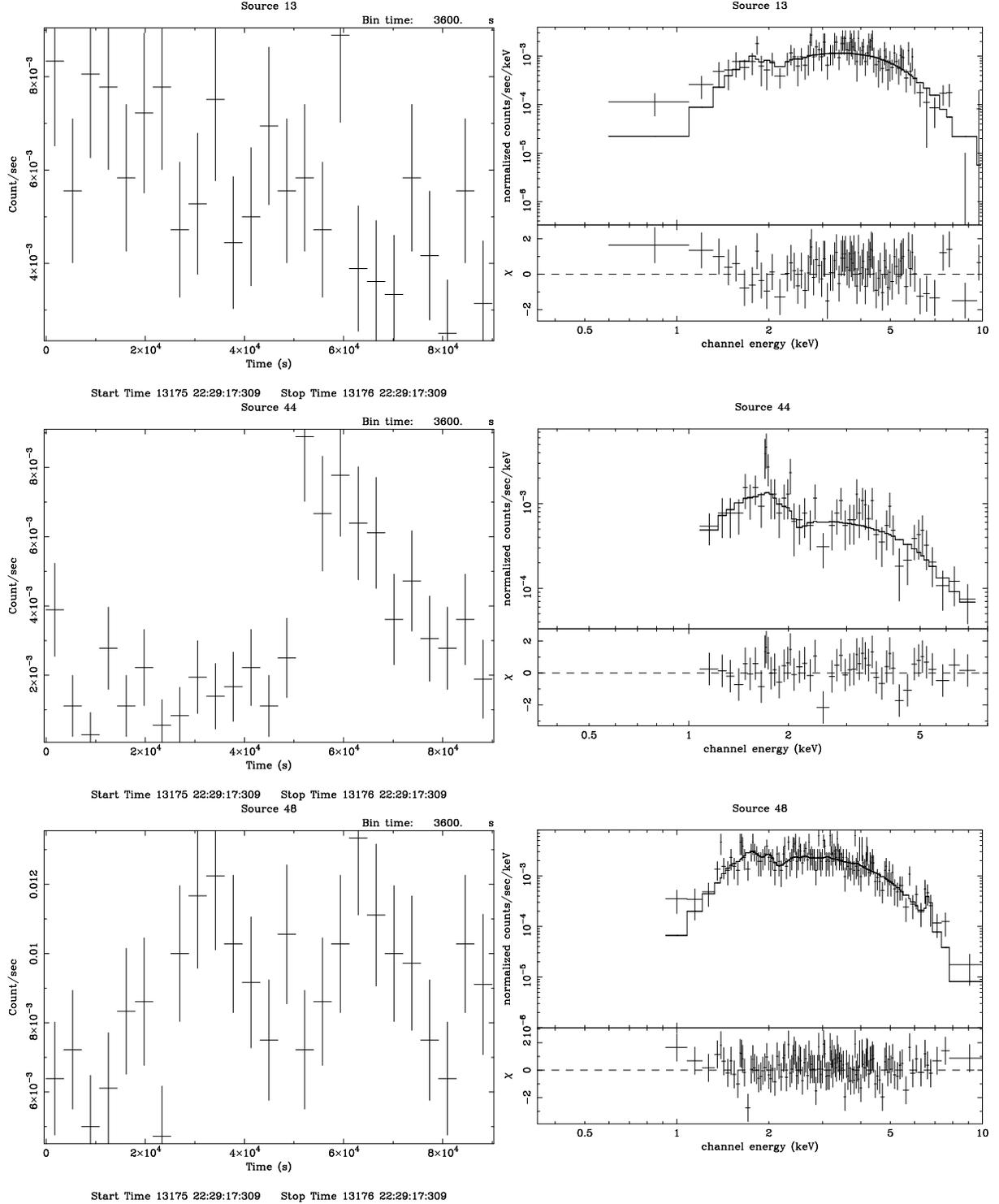

  \begin{center} \leavevmode 

        \epsfig{file=6424fg11.ps, height=8.0cm, angle=270}
        \epsfig{file=6424fg12.ps, height=8.0cm, angle=270}
        \epsfig{file=6424fg13.ps, height=8.0cm, angle=270}
        \epsfig{file=6424fg14.ps, height=8.0cm, angle=270}
	\epsfig{file=6424fg15.ps, height=8.0cm, angle=270}	
        \epsfig{file=6424fg16.ps, height=8.0cm, angle=270}

\caption{ Light curves and spectra with spectral fits of sources
discussed in the text. See Sect. 4.}

    \label{fig:lc}
  \end{center}
\end{figure*}

\addtocounter{figure}{-1}
\begin{figure*}[!htbp]
  \begin{center} \leavevmode

     \epsfig{file=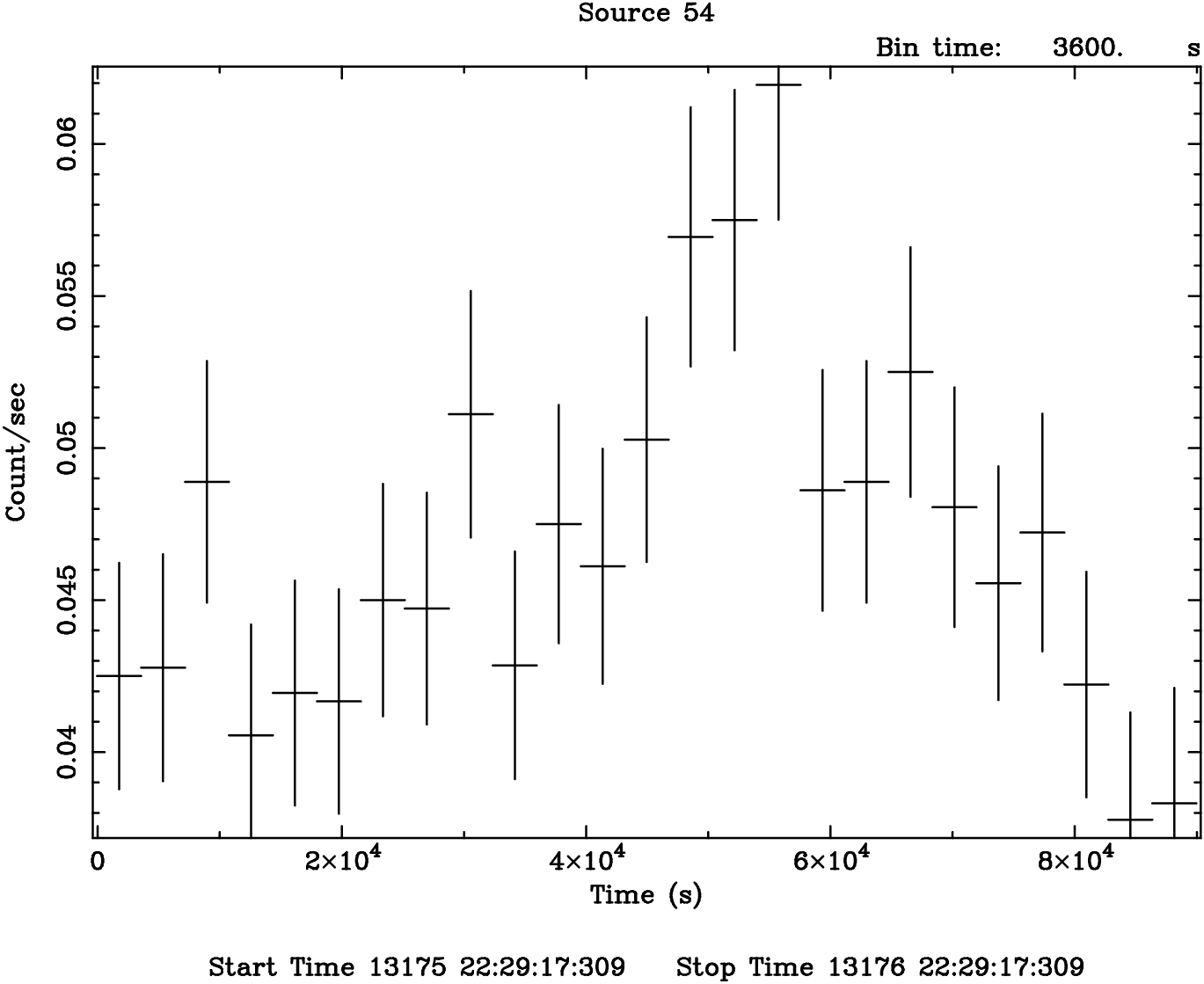, height=8.0cm, angle=270}
     \epsfig{file=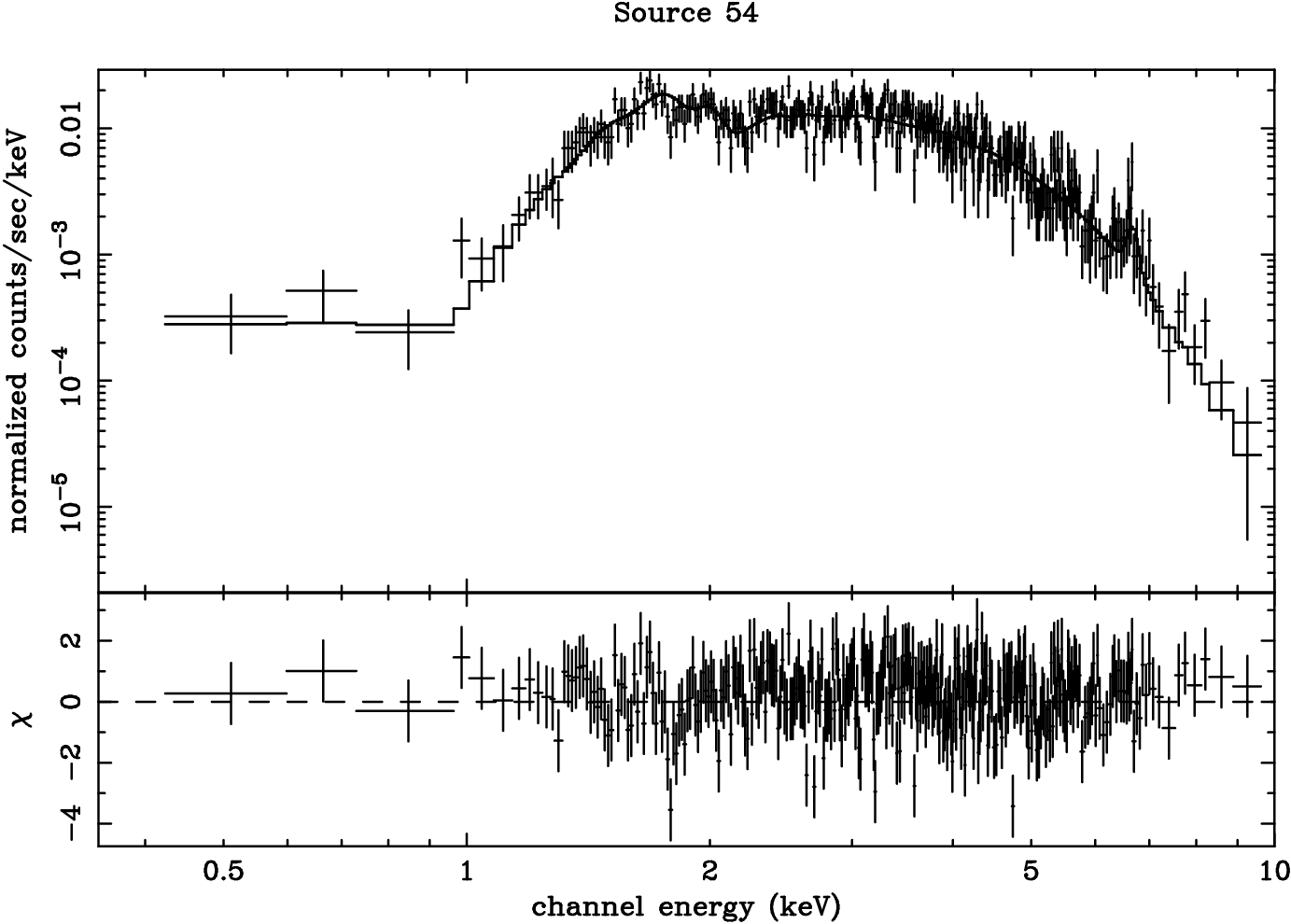, height=8.0cm, angle=270}
     \epsfig{file=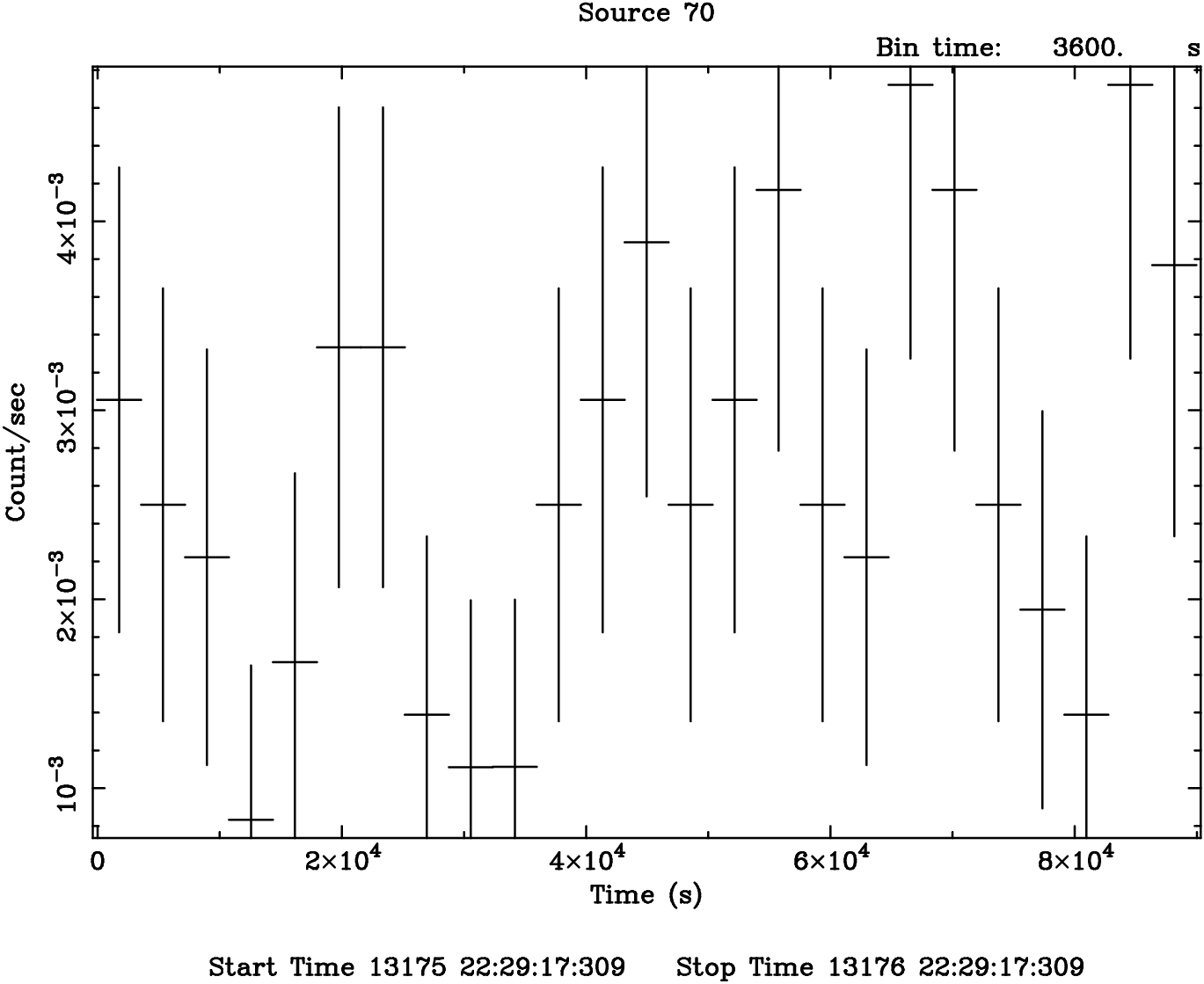, height=8.0cm, angle=270}
     \epsfig{file=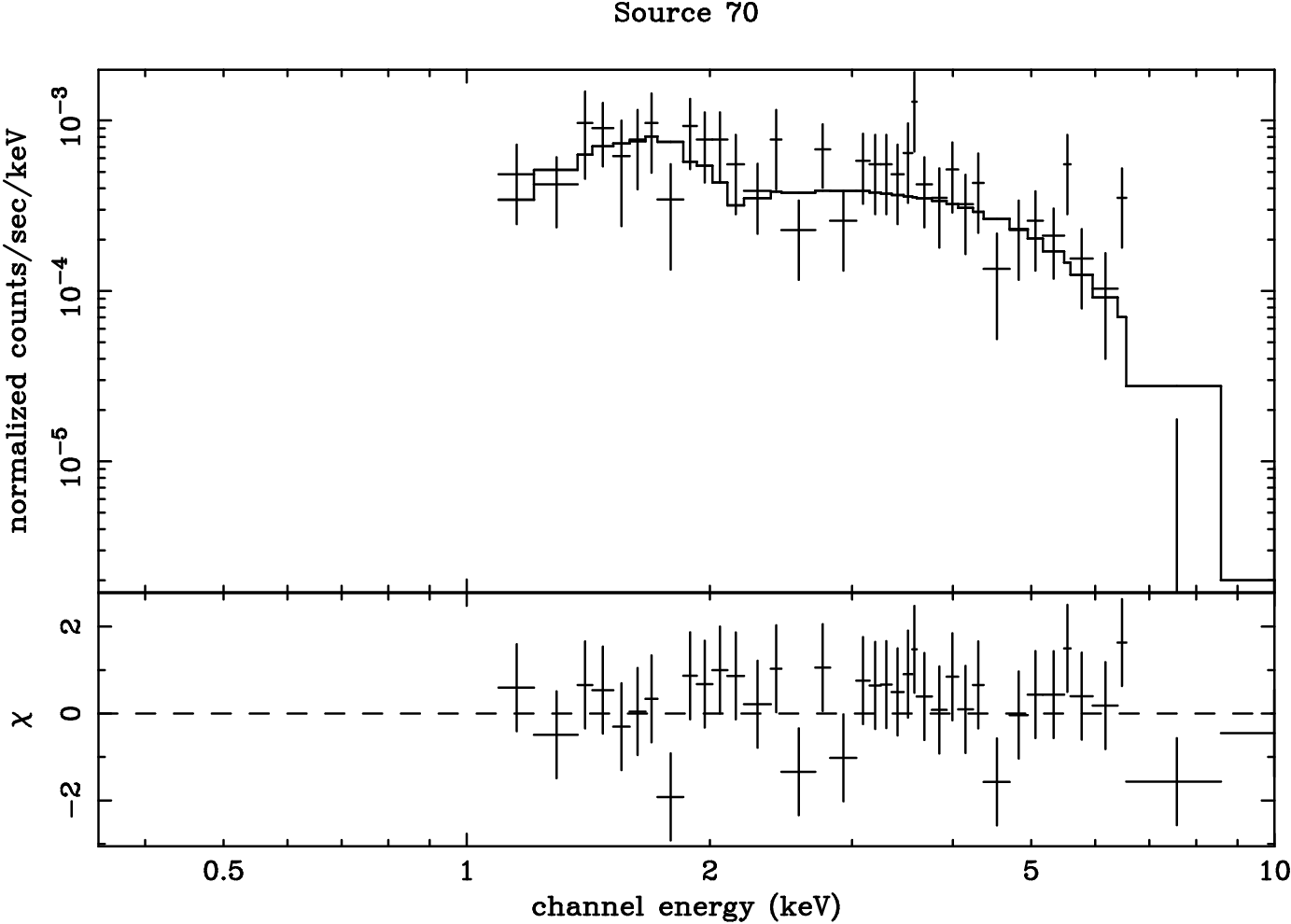, height=8.0cm, angle=270}
     \epsfig{file=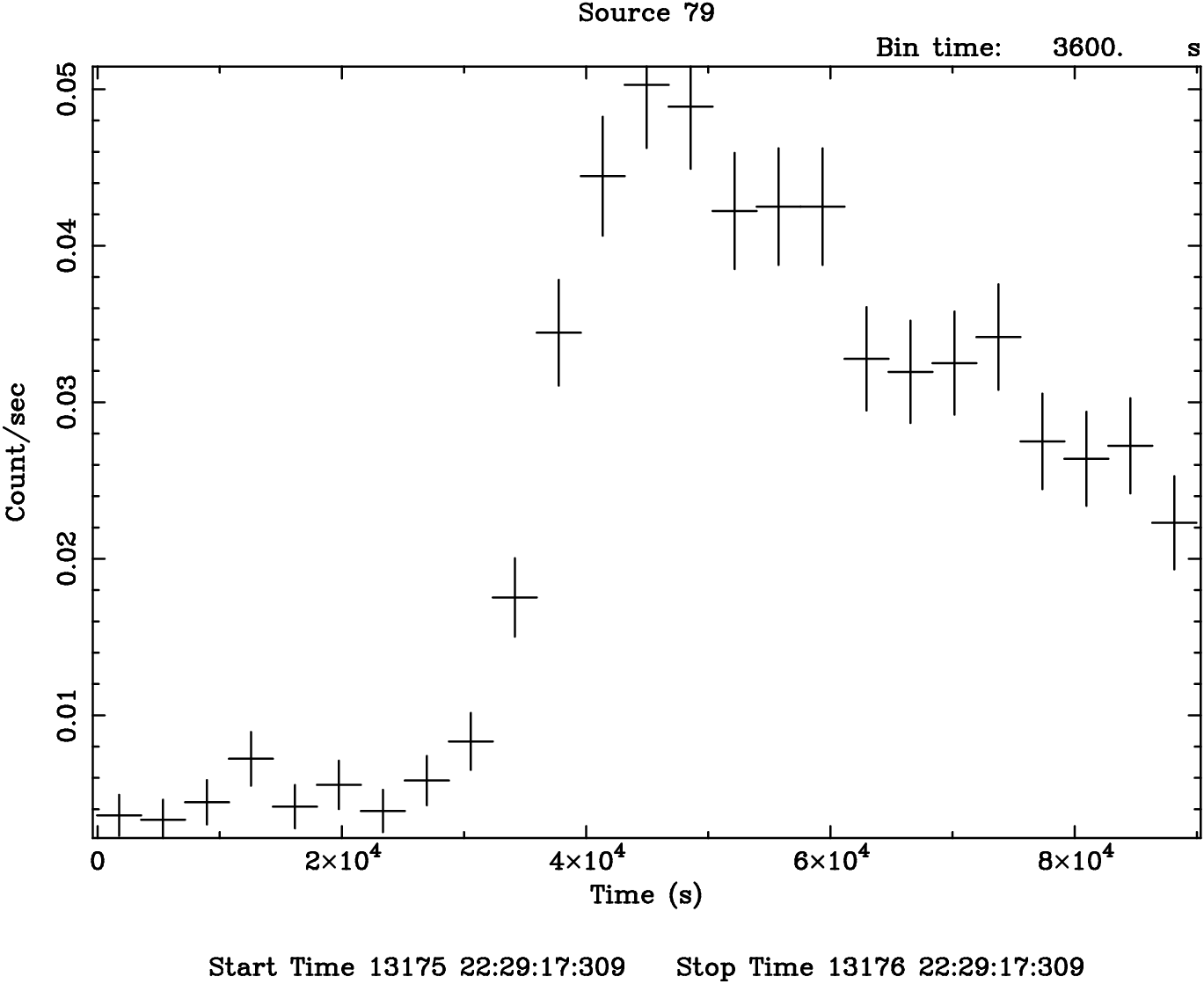, height=8.0cm, angle=270}
     \epsfig{file=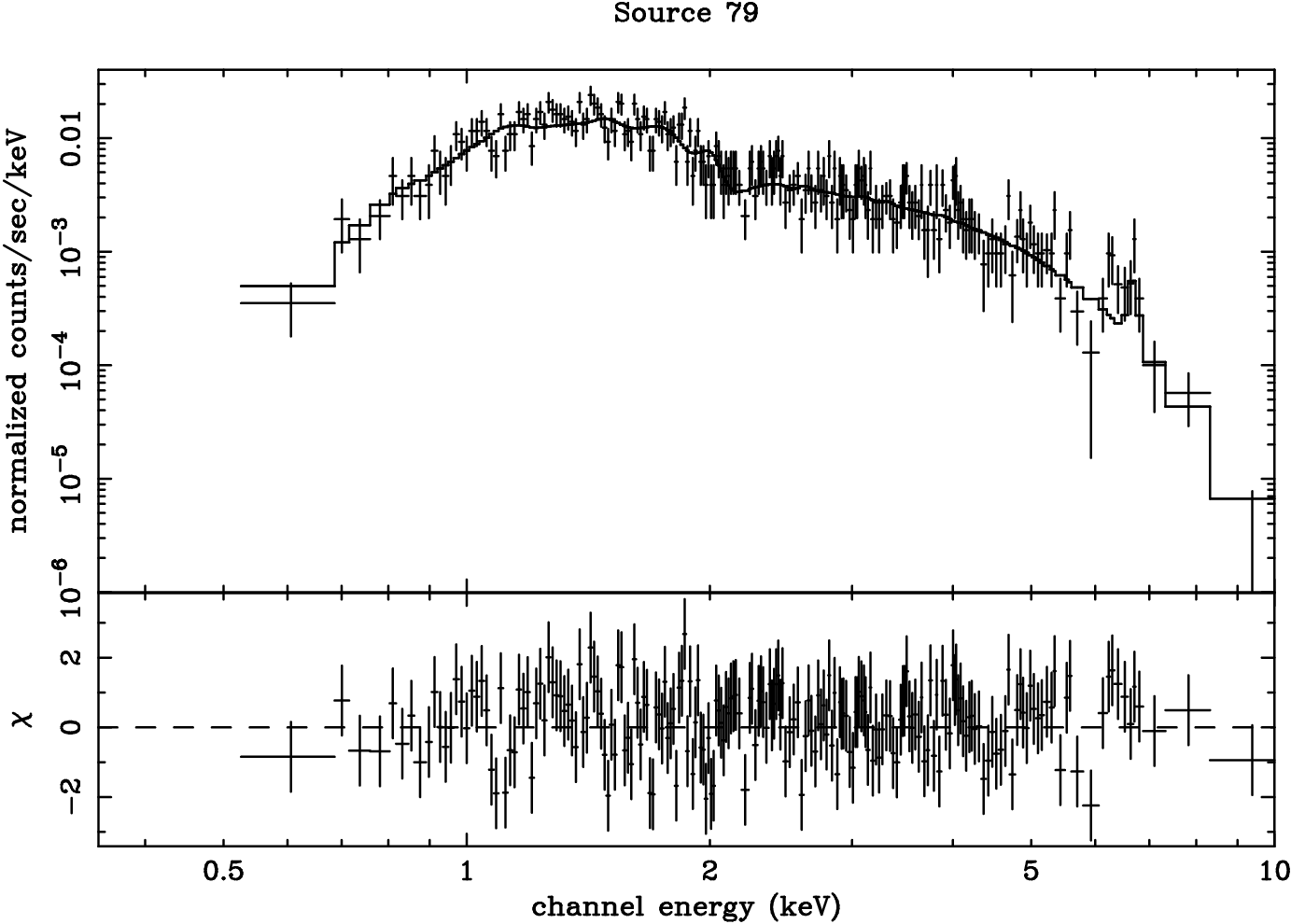, height=8.0cm, angle=270}

    \caption{({\it Continued.}) Light curves and spectra with spectral
fits of sources of sources discussed in the text. See Sect. 4.}

  \end{center}
\end{figure*}

\subsection{Source 57}

This source lies at 3.7 arcsec from the position of the Herbig-Haro
flow HH\,456 identified by \citet{dmr+99}. We believe it
unlikely that the X-ray emission originates from shocks generated by
the Herbig-Haro flow (as reported by e.g.\ \citet{pfg+2001} for HH\,2
or \citet{ffm+02} for HH\,154). The source is weak so its spectral
characteristics cannot be well constrained, however it appears to be
harder than generally observed in Herbig-Haro flows.

The source has a 2MASS counterpart at
only 0.2 arcsec distance and a corresponding \emph{Spitzer} source
classified as a Class III star.  A spectral fit
with an absorbed 1T plasma gives $N({\rm H}) = (3.5 \pm 1.3) \times
10^{22}$ cm$^{-2}$ and $kT = 4.9 \pm 4.5$ ($P = 0.85$), consistent
with the X-ray emission from the stellar corona of a CTTS or WTTS
star.

\subsection{Source 70}

This source has no 2MASS nor optical counterpart. It has no
counterpart within the ISO catalogue by \citet{kob+04} and it was
detected by \spitzer\ only in three of the IRAC channels (4.5, 5.8,
and 8.0 $\mu$m), so the source could not be classified.  Its X-ray
spectrum is very hard. A fit with a thermal spectrum gives a
temperature of 60 keV, but it is badly constrained. A $\alpha = 1.2$
power law provides as good a fit as the thermal model. Such a spectrum
is unlikely to originate in a stellar source, and therefore we
consider this most likely an X-ray bright background extragalactic
object. The source is present in the \xmm\ observation of
\cite{pre2003}, who also does not detect any near IR counterpart.

\subsection{Source 79}
\label{sec:src79}

During our observation, this Class II object, identified both in the
ISO and \spitzer\ data, undergoes a strong, long-duration, flare: the
intensity of the source increases by a factor of 10, impulsively, and
then slowly decays over a period of more than 50 ks.  To determine the
spectral parameters of the flaring emission we subdivided the data for
this source in four time intervals: the quiescent phase, the rise
phase of the flare, its peak, and the flare decay. The parameters of
the spectral fits derived for these four time intervals are summarised
in Table\,\ref{tab:flare}.

To derive the flare's physical parameters we used the approach
initially discussed by \citet{rbp+97} and since then applied to a
variety of stellar flares. This approach uses the slope $\zeta$ of the
flare decay in the $\log T$ versus $\log \sqrt{E\!M}$ diagram to
account properly for the presence of sustained heating during the
flare decay. The calibration of the method for \chandra\ ACIS, and a
detailed explanation of the physics behind it, can be found in
\citet{ffr+2005}, to which the reader is referred. In this
formulation, the semi-length of the flaring loop is given by
\begin{equation}
L = \frac{\tau_{\rm LC} \sqrt{T_{\rm max}}}{\alpha F(\zeta)}~~~~~~~0.32 <
\zeta \la 1.5
\label{eq:loop}
\end{equation}
where $\alpha = 3.7 \times 10^{-4} {\rm cm^{-1} s^{-1} K^{1/2}}$,
$\tau_{\rm LC}$ is the $1/e$ folding time of the light curve decay,
and $T_{\rm max}$ is the peak temperature of the plasma in the flaring
loop. The limits of applicability of Eq.\,\ref{eq:loop} correspond, on one
side ($\zeta \simeq 1.5$), to a freely decaying loop, with no sustained
heating, on the other ($\zeta=0.32$), to a sequence of quasi-static
states for the loop, in which the heating timescale is so long as to
mask the loop's intrinsic decay. $F(\zeta)$ and the relationship
between $T_{\rm max}$ and the best-fit peak temperature $T_{\rm obs}$
are both functions that need to be separately determined for each
X-ray detector, depending on its spectral response. For ACIS,
\begin{equation}
F(\zeta) = \frac{0.63}{\zeta - 0.32} + 1.41
\label{eq:fchi}
\end{equation}
 and
\begin{equation}
T_{\rm max} = 0.068 T_{\rm obs}^{1.20}.
\end{equation}

\begin{table*}[thbp]
  \begin{center}

  \caption{Best-fit spectral parameters for the quiescent phase and
    flaring phases of source 79; units are $N_{22} =
  10^{22}~{\rm cm^{-2}}$ and $E\!M_{53} =
    10^{53}$~cm$^{-3}$. Metal abundance was fixed at $Z = 0.3\,Z_\odot$}

    \leavevmode
        \footnotesize
    \begin{tabular}{r|cccccc}
Phase & $N({\rm H})$ & $kT$ & $E\!M$ & $\chi^2$ & $P$ & Rate\\
\hline
~ &  $N_{22}$ & keV & $E\!M_{53}$ & ~ & ~ & cts/s\\
\hline
Quiesc. & 0.98 $\pm$ 0.22 & 0.96 $\pm$ 0.18 & 0.92 $\pm$ 0.69 & 1.32 & 0.16 & 4.2 $\pm$ 0.4\\
Rise & 0.61 $\pm$ 0.20 & 5.45 $\pm$ 3.12 & 1.65 $\pm$ 0.65 & 0.97 & 0.52 & 17.6 $\pm$ 1.2\\
Peak & 0.60 $\pm$ 0.10 & 4.79 $\pm$ 1.22 & 4.95 $\pm$ 1.13 & 0.80 & 0.90 & 49.3 $\pm$ 2.2\\
Decay & 0.44 $\pm$ 0.05 & 3.42 $\pm$ 0.31 & 3.31 $\pm$ 0.45 & 0.93 & 0.72 & 32.9 $\pm$ 0.9\\

    \end{tabular}
    \label{tab:flare}
  \end{center}
\end{table*}

\begin{figure}[!tbp]
        \begin{center} \leavevmode \epsfig{file=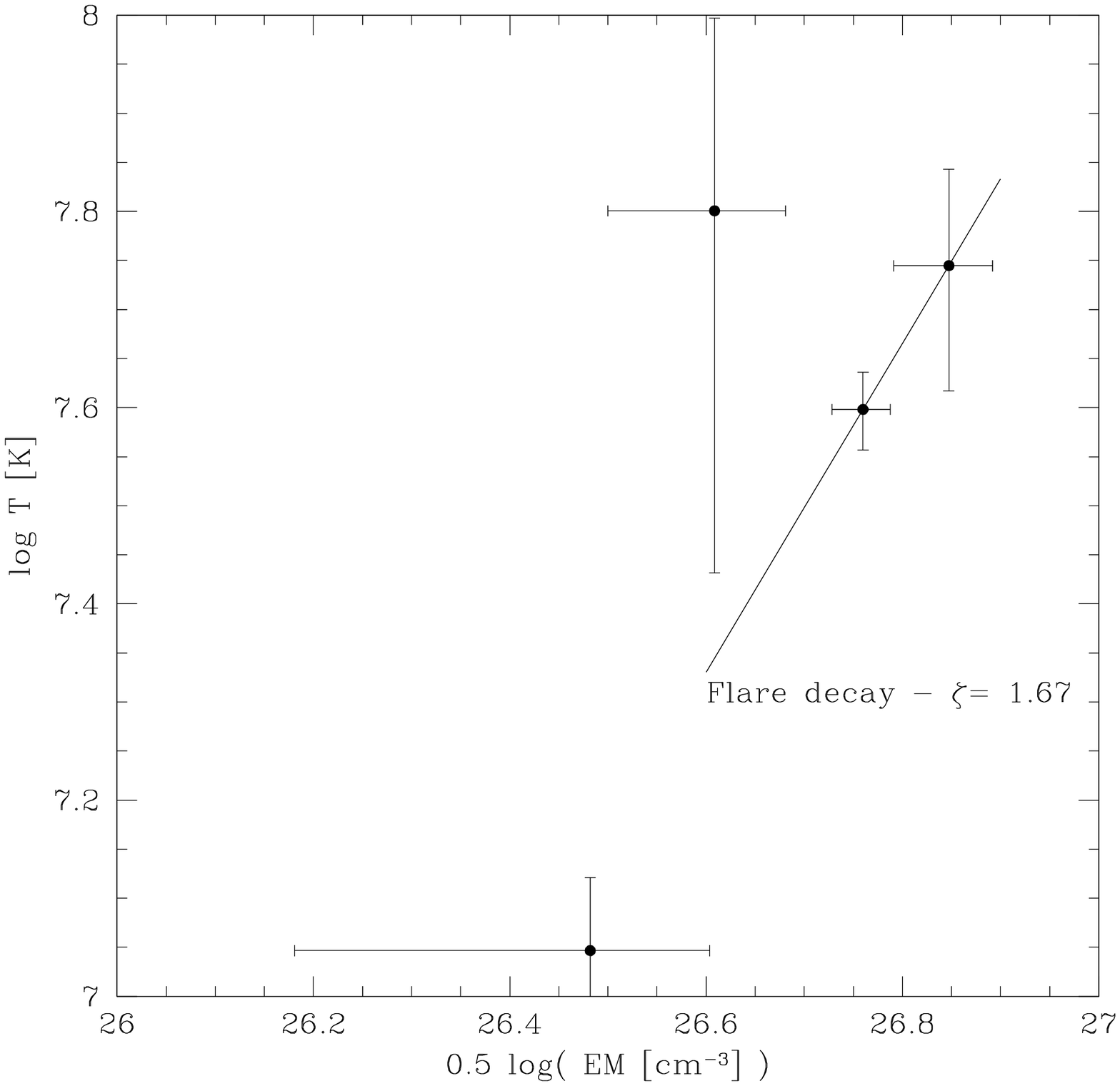, width=9.0cm}
        \caption{The evolution of the flare of source 79 in the
            $\log T$ vs.\ $\log \sqrt{E\!M}$ plane.}
        \label{fig:zeta}
    \end{center}
\end{figure}

The $\log T$ versus $\log \sqrt{E\!M}$ diagram for the flare of source
79 is shown in Fig.~\ref{fig:zeta}.  The lowest point refers to the
quiescent phase, the highest point corresponds to the rise phase of
the flare, and the line joins the points relatives to the flare peak
and its decay. The slope of this line is $\zeta = 1.67$, which is
outside the limit of applicability of Eq.\,\ref{eq:loop}, however the
error bars of the points are compatible with $\zeta = 1.0 - 1.5$. Given
a flare decay $1/e$-folding time $\tau_{\rm LC} = 53$~ks,
Eqs.~\ref{eq:fchi} and~\ref{eq:loop} result in a loop semi-length of
$L = 10-12$ $R_\odot$. 

Source 79 appears to be a very-low-mass star; its position in the
color-magnitude diagram of Fig.\,\ref{fig:cmd_2myr} implyies a mass $M
< 0.1~M_{\sun}$. According to the evolutionary model of \citet{sdf00},
at 2 Myr of age, such a star would have a radius $R < R_{\sun}$,
hence, the flaring loop of source 79 extends to a dinstance of more
than 10 times the stellar radius.  Assuming for the loop a radius $r =
0.1~L$, typical of solar events, from the emission measure at the
flare maximum, one derives an electron density $n_e \simeq 3.6 \times
10^{9}$~cm$^{-3}$ and a corresponding equipartition magnetic field
strength $B \simeq 60$ G.

Very long flaring structures in YSOs have been recently reported for
half a dozen sources in the ONC by \citet{ffr+2005} and, more
tentatively, for HL Tau in L1551 by \citet{gfs+06}. The flare in
source 79 is similar, even though the star is significantly less
massive than HL Tau and the objects in the ONC that exhibited evidence
of very long flaring structures. This indicates that these long
flaring structures develop in stars across a wide mass range (from
$M_{\star}$ apparently well below 0.1 $M_{\sun}$ to $M_{\star} \sim
2.4~M_{\sun}$, the most massive object in the COUP flaring sample).

As the majority of YSOs is surrounded by disks, \citet{ffr+2005}
speculated that the large magnetic structures that confine the flaring
plasma are actually the same type of structures that connect the star
to the circumstellar disk, in the magnetospheric accretion paradigm.
As described below, we detect, in the spectrum of this star, 6.4 keV
Fe fluorescent emission with a large equivalent width, compatible with
it being due to fluorescence from a centrally irradiated circumstellar
disk.

\begin{figure}[!htbp]
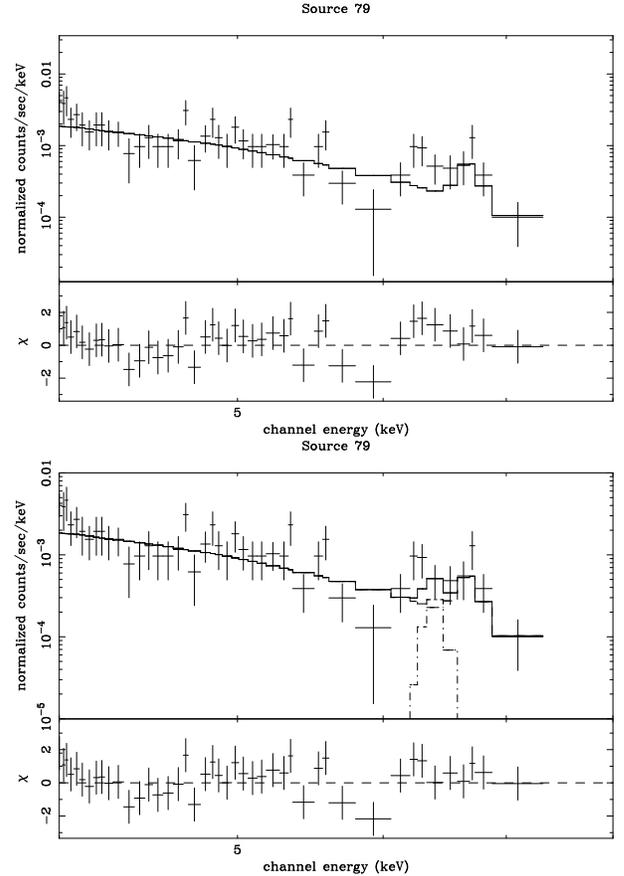

  \begin{center} \leavevmode 

        \epsfig{file=6424fg24.ps, height=8.0cm, angle=270}
        \epsfig{file=6424fg25.ps, height=8.0cm, angle=270}

\caption{Spectrum of source 79 in the $4.0-8.0$ keV region and
spectral fit using an absorbed thermal model ({\em top}) and with the
addition of a narrow Gaussian component ({\em bottom}) at 6.4 keV (Fe
fluorescent line)}.

\label{fig:fluo}
  \end{center}
\end{figure}

The spectrum (integrated over the entire observation) of source 79 is
shown in Fig.\,\ref{fig:lc} together with its spectral fit with an
absorbed 1T plasma model. There appears to be an excess of
emission redward of the Fe K complex at $\simeq 6.7$ keV, at the
expected energy of the 6.4 keV Fe fluorescent line. To determine the
significance of this excess, we fitted the spectrum of source 79 again
with an additional Gaussian line component at 6.4 keV, which was
constrained to be narrow (10 eV), while its normalisation was
unconstrained. The other parameters of the absorbed 1T plasma model
(absorbing column density, temperature and normalisation of the
thermal spectrum) were also unconstrained. The result of this new
fit can be compared with the original fit in Fig.\,\ref{fig:fluo}: the
excess at 6.4 keV is well accounted for by a narrow Gaussian line with
fitted intensity $I = (3.79 \pm 5.78) \times 10^{-7}$ photons
cm$^{-2}$ s$^{-1}$.  The fitted values of the absorbed 1T plasma model
do not change significantly with the addition of the Gaussian line at
6.4 keV (cf.\ Table\,\ref{tab:psfit}), while the null-hypothesis
probability of the fit increases marginally from 50\% to 54\%.

The equivalenth width of the fitted line is $W_{\alpha} = 149$
eV\footnote{In computing the line equivalent width \textsc{xspec} does
not provide an error, as this would involve an estimate of the error
of the fitted continuum. We note, however, that given the large
uncertainties in the fitted line intensity the uncertainties on
$W_{\alpha}$ are also large: i.e. of the order of 150\%.}. As
discussed in \citet{fms+05}, such a large equivalenth width cannot be
explained by fluorescent emission from colder, diffuse (optically
thin) circumstellar material: the absorbing column density required
would in fact be of the order of $10^{24}$~cm$^{-2}$ in conflict with
the value of $0.5 \times 10^{22}$~cm$^{-2}$ derived from the spectral
fit.  Under optically thick conditions, the computation of the
equivalent width of the fluorescence line requires a detailed
radiative transfer treatment and the assumption of a well defined
geometry. Using the computation by \citet{gf91}, for a spectral index
$\gamma = 2 - 2.3$, a centrally irradiated accretion disk (viewed
face-on) would entail $W_{\alpha} = 120 - 150$ eV. The X-ray spectrum
of source 79 is similar to the model spectra that were used by
\citet{gf91}\footnote{The data of source 79
can also be fitted ($P = 0.60$) by an absorbed power law with
spectral index $\gamma = 2.3 \pm 0.09$}, hence, the equivalent width
of the 6.4 keV line is compatible with Fe fluorescent emission from a
centrally irradiated disk.

\section{Conclusions}

Our 90 ks \chandra\ observations of the Serpens Cloud Core allowed the
X-ray emission from this embedded young star cluster to be studied at
high sensitivity. One of the goals of the observation was to search
for X-ray emission from confirmed, bona fide Class 0 sources, a still
elusive target. We did not detect X-ray emission from any of the six
well studied Class 0 objects in the cloud core. The sensitivity of our
observation allows us to put an upper limit to the X-ray emission from
these Class 0 sources: assuming that the absorbing column density is
$\la 4 \times 10^{23}$ cm$^{-2}$ ($A_V \la 200$ mag), and that the
coronal temperature is $kT \simeq 2.5$~keV (as observed in the Class I
sources of Serpens), the typical X-ray luminosity of Class 0 sources
must be $L_{\rm X} \la 0.4 \times 10^{30}$ \es. If one assumes a
higher plasma temperature of 3.5 keV -- the average plasma temperature
of Class I sources in $\rho$ Oph (\citealp{ogm05}) -- then the typical
luminosity of Class 0 sources must be $L_{\rm X} \la 0.1\times
10^{30}$ \es or the absorbing column densities must be even
higher. Class I sources in this region and $\rho$ Oph have typical
luminosities $L_{\rm X} \ga 1\times 10^{30}$ \es.

We performed a spectral and timing analysis of all the
sufficiently X-ray bright YSOs. Our data show a clear pattern of
increasing column densities from Class III to Class I sources, and
also hint at evolution in the X-ray emitting plasma temperature of
these YSOs. No evolutionary trend is found in these sources' temporal
variability.

During our observations the low-mass Class II source, source 79,
undergoes a large, long-duration, flare for which we derive a
semi-loop length of $L = 10-12$ $R_\odot$. Such a long flaring loop
($\sim 0.1$ AU) could be explained if the flare is due to a magnetic
reconnection event of a flux tube linking the star's photosphere with
the inner rim of a circumstellar disk, as in the magnetospheric
accretion model.  Interestingly, we detect, in the spectrum of this
star, 6.4 keV Fe fluorescent emission with a large equivalent width,
compatible with reflection off a circumstellar disk irradiated by the
hard X-ray continuum emission of the flare.

\begin{acknowledgements}

  We thank Ettore Flaccomio for his help in producing the coadded
  image of Class 0 sources, Andrew Thean for helping us improve
  the language of the paper, and Olive Buggy for her prompt help with
  software installation problems. GM and SS acknowledge financial
  contribution from contract ASI-INAF I/023/05/0. This research makes
  use of data products from the Two Micron All Sky Survey, which is a
  joint project of the University of Massachusetts and the Infrared
  Processing and Analysis Center/California Institute of Technology,
  funded by the National Aeronautics and Space Administration and the
  National Science Foundation.

\end{acknowledgements}


\begin{table*}[!thbp]
  \begin{center} \caption{X-ray sources detected with {\sc Pwdetect}
 in the \chandra\ observation of the Serpens star forming region. For
 the sources with a 2MASS counterpart (within a search radius of 3
 arcsec) the values of their magnitude in the $J$, $H$ and $K$ bands
 are also given, together with the radial distance from this
 counterpart ($r$), the associations ($a$) with Tycho 2 or USNOA-2.0
 optical catalog sources, U or T respectively and their optical
 magnitude. The ISO field provides the source number and Class of the
 counterpart in the ISO catalogue by Kaas et al. (2004) (also within a
 search radius of 3 arcsec). The last three fields give the source
 identification numbers and classification in the \spitzer\ data: the
 first id is the source IR id and the second one the YSO id
 number. Class I/II and Class II/III indicate Flat Spectrum and
 Transition Disk objects, respectively. All the \spitzer\ counterparts
 are within 1 arcsec from the X-ray sources, unless their IR id is
 accompanied by $\dag$ or is in between brackets. In the first case,
 although the matching distance is greater than $1''$, we are confident
 in the cross-identification reliability, in the second case some
 doubts remains (these are usually faint IR sources).  }  \leavevmode

\begin{tabular}{r|rrrrrrrcrrr|rrr}
N. & RA(J200) & Dec(J200) & Count rate & $J$ & $H$ & $K$ & $r$ & a &
$B$ & $VR$ & ISO & IR  & YSO  & Class\\ 
~      & deg      & deg       & ks$^{-1}$ & ~ & ~ & ~ & arcsec & ~ & ~ & ~ & ~ & ~\\

\hline
1 & 18 29 10.8 & 01 12 02 & 1.07 $\pm$ 0.20 & 16.5 & 15.4 & 15.2
& 2.4 & 0 & - & - & - & - & - & -\\ 

2 & 18 29 22.7 & 01 10 33 & 3.18 $\pm$ 0.23 & 12.2 & 11.2 & 10.9
& 0.4 & U & 18.0 & 15.1 & - & 13836    &     221  &   	III\\ 

3 & 18 29 27.7 & 01 12 58 & 0.63 $\pm$ 0.11 & 9.7 & 9.5 & 9.4 
& 1.0 & T & 11.5 & 11.1 & - & 8974$^{\dag}$   &    - &          III\\

4 & 18 29 28.2 & 01 15 48 & 0.26 $\pm$ 0.09 & - & - & - & - & - & - &
- & - & 57260    &      - &           -\\

5 & 18 29 29.8 & 01 21 19 & 0.48 $\pm$ 0.14 & - & - & - & - & - & - &
- & - & 44904    &      - &           -\\ 

6 & 18 29 29.8 & 01 21 57 & 0.82 $\pm$ 0.14 & - & - & - & - & - & - &
- & - & 57617    &     119 &            II\\ 

7 & 18 29 31.4 & 01 19 56 & 1.23 $\pm$ 0.15 & - & - & - & - & - & - &
- & - & 44113    &     -  &          -\\ 

8 & 18 29 31.8 & 01 13 39 & 1.09 $\pm$ 0.18 & - & - & - & - & - & - &
- & - & 40525    &     -  &           \\ 

9 & 18 29 32.0 & 01 18 42 & 0.36 $\pm$ 0.11 & 11.6 & 9.8 & 8.6 &
0.3 & 0 & - & - & 159 I/II & 12134 &         36  &        I/II\\
 
10 & 18 29 32.9 & 01 22 40 & 0.69 $\pm$ 0.18 & - & - & - & - & - & - &
- & - &  45772  &        - &           - \\  

11 & 18 29 33.1 & 01 17 16 & 15.91 $\pm$ 0.46 & 11.4 & 10.3 & 10.0 &
0.2 & U & 17.9 & 15.0 & - & 14252 &        225   &         III\\  

12 & 18 29 34.2 & 01 17 03 & 0.08 $\pm$ 0.05 & - & - & - & - & - & - &
- & - & - & - & -\\ 

13 & 18 29 35.2 & 01 23 39 & 4.15 $\pm$ 0.27 & - & - & - & - & - & - &
- & - & 46388  &        50  &        I/II\\ 

14 & 18 29 35.5 & 01 16 54 & 0.23 $\pm$ 0.08 & - & - & - & - & - & - &
- & - & - & - & -\\ 

15 & 18 29 35.6 & 01 23 11 & 0.29 $\pm$ 0.11 & - & - & - & - & - & - &
- & - & -& - & - \\
 
16 & 18 29 38.1 & 01 10 23 & 0.25 $\pm$ 0.09 & - & - & - & - & - & - &
- & - & - & - & -\\ 

17 & 18 29 38.5 & 01 14 50 & 0.11 $\pm$ 0.05 & - & - & - & - & - & - &
- & - & (41344) & - &           -\\
 
18 & 18 29 39.9 & 01 17 56 & 0.71 $\pm$ 0.10 & 14.5 & 12.8 & 11.9 &
0.1 & 0 & - & - & 202 II & 13040   &       94  &          II\\  

19 & 18 29 41.1 & 01 13 43 & 0.08 $\pm$ 0.06 & - & - & - & - & - & - &
- & - & - & - & -\\ 

20 & 18 29 42.1 & 01 13 12 & 0.34 $\pm$ 0.10 & - & - & - & - & - & - &
- & - & -  & - & -\\
 
21 & 18 29 42.3 & 01 12 02 & 1.31 $\pm$ 0.15 & 16.8 & 14.4 & 12.9 &
0.1 & 0 & - & - & - & 13257  &       215    &        III\\  

22 & 18 29 42.7 & 01 14 01 & 0.11 $\pm$ 0.05 & - & - & - & - & - & - &
- & - &  40962   & -   &        -\\ 

23 & 18 29 44.6 & 01 13 11 & 0.92 $\pm$ 0.25 & 13.3 & 12.2 & 11.8 &
0.3 & 0 & - & - & - &  13758$^{\dag}$ &    100 &          II\\
  
24 & 18 29 44.7 & 01 10 03 & 0.18 $\pm$ 0.08 & 16.6 & 14.6 & 13.7 &
2.8 & 0 & - & - & - & (7965)    &      - &            III\\ 

25 & 18 29 45.1 & 01 24 37 & 0.27 $\pm$ 0.10 & - & - & - & - & - & - &
- & - & - & - & -\\ 

26 & 18 29 46.3 & 01 07 47 & 0.54 $\pm$ 0.16 & - & - & - & - & - & - &
- & - & (37471) &         - &           -\\ 

27 & 18 29 47.2 & 01 22 34 & 3.14 $\pm$ 0.22 & 11.8 & 10.5 & 10.1 &
0.2 & U & 19.8 & 16.0 & - & 10876 &        199 &           III\\ 

28 & 18 29 49.6 & 01 17 06 & 0.42 $\pm$ 0.14 & 16.9 & 15.7 & 12.7 &
0.2 & 0 & - & - & 254 I & 10187   &        7      &      I\\ 

29 & 18 29 50.6 & 01 08 58 & 0.79 $\pm$ 0.12 & 14.4 & 13.0 & 12.5 &
0.3 & 0 & - & - & - & 12228$^{\dag}$ &       209 &           III\\ 

30 & 18 29 51.2 & 01 16 41 & 0.30 $\pm$ 0.09 & 16.7 & 15.1 & 13.0 &
1.2 & 0 & - & - & 265 I & 42625  &        24  &          I\\ 

31 & 18 29 51.2 & 01 13 20 & 0.11 $\pm$ 0.05 & 16.5 & 14.5 & 13.6 &
0.2 & 0 & - & - & - & 13144    &      95   &         II\\ 

32 & 18 29 52.2 & 01 15 59 & 1.01 $\pm$ 0.19 & - & - & - & - & - & - &
- & - & 42231   &       22     &       I\\
 
33 & 18 29 52.9 & 01 14 56 & 0.12 $\pm$ 0.06 & - & - & - & - & - & - &
- & 276 I & 41692    &      20  &         I\\ 

34 & 18 29 53.6 & 01 17 02 & 6.11 $\pm$ 0.28 & 12.6 & 11.2 & 10.4 &
0.2 & U & 19.7 & 16.6 & 283 II & 10801   &       81  &          II\\

35 & 18 29 54.5 & 01 14 47 & 0.29 $\pm$ 0.09 & 14.2 & 13.3 & 12.8 &
0.2 & 0 & - & - & - & 11909   &       87  &          II\\ 

36 & 18 29 55.4 & 01 10 34 & 0.35 $\pm$ 0.07 & 14.0 & 13.4 & 12.9 &
0.1 & 0 & - & - & - &  11296  &       201 &           III\\ 

37 & 18 29 55.7 & 01 14 31 & 0.50 $\pm$ 0.08 & 15.3 & 12.5 & 10.7 &
0.2 & 0 & - & - & 298 II & 12547  &        38  &        I/II\\ 

38 & 18 29 56.2 & 01 10 57 & 4.14 $\pm$ 0.25 & 12.5 & 11.7 & 11.4 &
0.2 & 0 & - & - & - & 11728  &       205  &          III \\ 

39 & 18 29 56.4 & 01 12 18 & 0.11 $\pm$ 0.04 & 15.3 & 14.3 & 13.6 &
0.1 & 0 & - & - & - & (11699)  &       204  &          III\\
 
40 & 18 29 56.6 & 01 12 59 & 0.16 $\pm$ 0.06 & 15.0 & 12.8 & 11.5 &
0.3 & 0 & - & - & 304 II & 10636 &          80  &          II\\ 

41 & 18 29 56.7 & 01 19 54 & 0.18 $\pm$ 0.06 & - & - & - & - & - & - &
- & - & (57555)   &       - &           -\\ 

42 & 18 29 56.7 & 01 12 39 & 0.40 $\pm$ 0.08 & 18.1 & 16.5 & 14.1 &
0.2 & 0 & - & - & 306 I &  8797   &        61   &        II\\ 

43 & 18 29 56.9 & 01 14 46 & 0.28 $\pm$ 0.08 & 12.0 & 10.8 & 9.3 & 0.2
& 0 & - & - & 307 I/II & 13114   &        9  &          I \\ 

44 & 18 29 57.0 & 01 12 48 & 2.93 $\pm$ 0.20 & 15.2 & 12.5 & 11.1 & 0.1
& 0 & - & - & 309 II & 11376   &      85   &         II\\ 

45 & 18 29 57.4 & 01 14 50 & 1.28 $\pm$ 0.12 & 13.3 & 11.7 & 10.9 &
0.2 & 0 & - & - & - & 8506   &       190   &         III\\ 

46 & 18 29 57.6 & 01 13 00 & 0.17 $\pm$ 0.06 & 16.5 & 14.1 & 13.1 &
0.2 & 0 & - & - & 312 I & 11670  &         8    &        I\\ 

47 & 18 29 57.7 & 01 10 53 & 2.41 $\pm$ 0.21 & 8.1 & 8.1 & 8.2 & 0.1 &
T & 8.6 & 8.5 & - & 9456   &       192    &        III\\ 

48 & 18 29 57.7 & 01 14 05 & 6.61 $\pm$ 0.36 & 12.2 & 9.2 & 7.2 & 0.4
& 0 & - & - & 314$^{*}$ I/II & 10172$^{\dag}$ &          35   &       I/II\\
 
49$^{\ddag}$ & 18 29 57.8 & 01 14 07 & 0.27 $\pm$ 0.13 & - & - & - & 1.4
& 0 & - & - & - & - & - &- \\ 

50 & 18 29 57.8 & 01 15 32 & 1.64 $\pm$ 0.26 & 16.2 & 12.8 & 10.8 &
0.2 & 0 & - & - & 319 II & 10942     &     83    &        II\\
 
51$^{\ddag}$ & 18 29 57.8 & 01 14 08 & 0.11 $\pm$ 0.05 & - & - & -  & 2.4
& 0 & - & - & - & - & - &-  \\ 

\end{tabular}
    \label{tab:src}
  \end{center}
$\ddag$ sources associated with the elongated PSF of source 48 (SVS 20)
$^{*}$  distance of ISO source greater than 3 arcsec
\end{table*}

\addtocounter{table}{-1}
\begin{table*}[!thbp]
  \begin{center}
    \caption{{\em (continued)} X-ray sources detected with {\sc
        Pwdetect} in the \chandra\ observation of the Serpens star
      forming region.} 
    \leavevmode
   \begin{tabular}{r|rrrrrrrcrrr|rrr}
N. & RA(J200) & Dec(J200) & Count rate & $J$ & $H$ & $K$ & $r$ & a &
$B$ & $VR$ & ISO & IR  & YSO  & Class\\ 
~      & deg      & deg       & ks$^{-1}$ & ~ & ~ & ~ & arcsec & ~ & ~ & ~ & ~ & ~\\

\hline

52 & 18 29 57.8 & 01 12 38 & 0.29 $\pm$ 0.09 & 17.2 & 14.4 & 11.7 &
0.2 & 0 & - & - & 318 I/II & 12355    &      37  &        I/II \\ 

53 & 18 29 57.9 & 01 12 51 & 6.7 $\pm$ 0.50 & 15.8 & 12.5 & 10.5 & 0.2
& 0 & - & - & 317 I/II & 8173$^{\dag}$   &        2  &          I \\ 

54 & 18 29 57.9 & 01 12 46 & 44.99 $\pm$ 0.76 & 16.6 & 12.5 & 10.0 &
0.2 & 0 & - & - & - &  14647   &      105  &          II \\ 

55 & 18 29 58.2 & 01 15 21 & 1.78 $\pm$ 0.18 & 13.1 & 11.2 & 9.9 & 0.2
& 0 & - & - & 321 II & 7685   &        27  &        I/II\\ 

56 & 18 29 58.5 & 01 12 50 & 0.19 $\pm$ 0.06 & 18.1 & 14.8 & 12.3 &
0.3 & 0 & - & - & 322 I/II & 12962$^{\dag}$         &   - &        II \\ 

57 & 18 29 59.1 & 01 11 14 & 0.56 $\pm$ 0.10 & 16.9 & 16.2 & 14.4 & 0.2
& 0 & - & - & - & 10535    &     198  &          II\\ 

58 & 18 29 59.1 & 01 11 20 & 0.31 $\pm$ 0.11 & 14.2 & 12.6 & 11.9 &
0.3 & 0 & - & - & - & 10161 &        196 &           III\\ 

59 & 18 29 59.2 & 01 14 08 & 1.29 $\pm$ 0.20 & 11.8 & 10.4 & 9.5 & 0.2
& U & 19.7 & 16.9 & 328 II & 8450   &        59 &           II\\ 

60 & 18 29 59.6 & 01 11 58 & 0.71 $\pm$ 0.11 & 17.9 & 17.4 & 14.7 &
0.4 & 0 & - & - & 330 I & 8176    &        3   &         I \\ 

61 & 18 29 59.6 & 01 14 12 & 1.32 $\pm$ 0.14 & - & - & - & - & - & - &
- & - & 41458    &      19    &       I \\ 

62 & 18 30 00.2 & 01 14 03 & 0.15 $\pm$ 0.05 & 15.7 & 13.9 & 12.8 &
0.2 & 0 & - & - & - & 8029  &         28 &         I/II\\ 

63 & 18 30 00.3 & 01 09 44 & 0.53 $\pm$ 0.14 & - & - & - & - & - & - &
- & - & 38922    &     110   &         II\\ 

64 & 18 30 00.5 & 01 23 45 & 0.89 $\pm$ 0.24 & - & - & - & - & - & - &
- & - & (57779)   &       - &           -   \\ 

65 & 18 30 00.7 & 01 13 40 & 6.04 $\pm$ 0.28 & 13.4 & 11.2 & 10.1 &
0.2 & 0 & - & - & 338 II & 13265$^{\dag}$   &      216    &   III  \\ 

66 & 18 30 01.1 & 01 13 24 & 0.83 $\pm$ 0.11 & 17.9 & 15.4 & 13.2 &
0.2 & 0 & - & - & 341 I/II & 8794  &         30    &      I/II  \\ 

67 & 18 30 01.2 & 01 15 03 & 5.58 $\pm$ 0.34 & 15.4 & 13.1 & 12.0 &
0.2 & 0 & - & - & - & 10085  &       195 &        III\\ 

68 & 18 30 01.4 & 01 18 08 & 0.16 $\pm$ 0.06 & - & - & - & - & - & - &
- & - & - & - & -\\ 

69 & 18 30 03.4 & 01 16 19 & 2.32 $\pm$ 0.21 & 12.3 & 11.1 & 10.4 &
0.1 & U & 18.9 & 15.0 & - & 10468  &        78  &          II \\ 

70 & 18 30 04.6 & 01 22 33 & 2.21 $\pm$ 0.23 & - & - & - & - & - & - &
- & - &  57713   &       - &           -\\
 
71 & 18 30 04.9 & 01 14 39 & 0.15 $\pm$ 0.06 & 13.7 & 12.7 & 12.2 &
0.4 & 0 & - & - & - & 13700    &      98 &           II\\ 

72 & 18 30 05.5 & 01 14 25 & 0.20 $\pm$ 0.07 & 15.2 & 14.4 & 14.1 &
0.5 & 0 & - & - & - &  11452   &      203   &        III \\ 

73 & 18 30 06.1 & 01 06 17 & 2.35 $\pm$ 0.22 & 12.9 & 11.8 & 11.2 &
0.5 & U & 19.0 & 16.5 & - & 10946  &        84   &      II\\ 

74 & 18 30 06.7 & 01 12 16 & 1.02 $\pm$ 0.13 & - & - & - & - & - & - &
- & - &  (40517) &       - &           - \\ 

75 & 18 30 07.7 & 01 12 04 & 2.95 $\pm$ 0.28 & 12.3 & 10.8 & 10.1 &
0.1 & 0 & - & - & 366 II & 9882  &         73   &         II\\ 

76 & 18 30 08.2 & 01 10 55 & 0.41 $\pm$ 0.12 & - & - & - & - & - & - &
- & - & (39762)   &       - &           - \\ 

77 & 18 30 09.9 & 01 17 07 & 0.16 $\pm$ 0.07 & - & - & - & - & - & - &
- & - & - & - & -\\ 

78 & 18 30 10.4 & 01 10 06 & 0.31 $\pm$ 0.10 & - & - & - & - & - & - &
- & - &  39349    &       - &           -\\ 

79 & 18 30 11.1 & 01 12 38 & 22.70 $\pm$ 0.57 & 13.3 & 12.5 & 12.0 &
0.1 & 0 & - & - & 393 II & 13765    &     101   &         II\\ 

80 & 18 30 11.1 & 01 17 52 & 0.24 $\pm$ 0.08 & - & - & - & - & - & - &
- & - & (43769)  &        - &           -\\ 

81 & 18 30 12.6 & 01 12 27 & 0.84 $\pm$ 0.12 & - & - & - & - & - & - &
- & - &  57165   &       - &           -\\ 

82 & 18 30 16.2 & 01 17 54 & 0.22 $\pm$ 0.07 & - & - & - & - & - & - &
- & - & - & -& -\\ 

83 & 18 30 18.2 & 01 14 17 & 0.68 $\pm$ 0.12 & 13.3 & 11.8 & 10.9 &
0.7 & 0 & - & - & - & 14137  &        40   &       I/II\\ 

84 & 18 30 22.4 & 01 20 44 & 1.46 $\pm$ 0.19 & 13.1 & 12.2 & 11.9 &
0.5 & U & 18.8 & 16.4 & - &  12371    &     157   &       II/III \\ 

85 & 18 30 23.1 & 01 20 09 & 0.85 $\pm$ 0.14 & 13.4 & 12.6 & 12.2 &
1.0 & U & 19.50 & 16.80 & - &  8080   &       128 &        II/III\\ 
\hline
   \end{tabular}
  \end{center}
\end{table*}

\begin{table*}[!thbp]
  \begin{center}

\caption{The cross-identification of the sources below was obtained
from the Simbad database, with a search radius of 5 arcsec. Identifier
are as follow: [B96] Bontemps (1996), [CK86] Churchwell \& Koornneef
(1986), [EC92] Eiroa \& Casali (1992), [GCN98] Giovannetti et
al. (1998), [HB96] Hurt \& Barsony (1996), [KOB2004] Kaas et
al. (2004), [KCM2004] Klotz et al. (2004), [P2003] Preibisch (2003),
[SVS76] Strom, Vrba \& Strom (1976).}

\leavevmode

\begin{tabular}{r|cc|l}
2  & 18 29 22.7  &  +01 10 33 &  [P2003] J182922.8+011032\\
3  & 18 29 27.7  &  +01 12 58 &   BD+01 3686\\
9  & 18 29 32.0  &  +01 18 42 &  IRAS 18269+0116 $-$ ESO-HA 279\\
11 & 18 29 33.1  &  +01 17 16 &  [P2003] J182933.1+011716 \\
13 & 18 29 35.2  &  +01 23 39 &  [P2003] J182935.1+012338 \\
18 & 18 29 39.9  &  +01 17 56 &  [P2003] J182939.7+011754 \\
21 & 18 29 42.3  &  +01 12 02 &  [KCM2004] J182942.36+011201.9\\
23 & 18 29 44.6  &  +01 13 11 &  [EC92] 11\\
27 & 18 29 47.2  &  +01 22 34 &  [P2003] J182947.3+012234\\
28 & 18 29 49.6  &  +01 17 06 &  [EC92] 38\\
30 & 18 29 51.2  &  +01 16 41 &  [EC92] 53\\
31 & 18 29 51.2  &  +01 13 20 &  [EC92] 51 $-$ [GCN98] 37\\
33 & 18 29 52.9  &  +01 14 56 &  [GCN98] 53\\
34 & 18 29 53.6  &  +01 17 02 &  [EC92] 67\\
35 & 18 29 54.5  &  +01 14 47 &  [EC92] 70\\
36 & 18 29 55.4  &  +01 10 34 &  [P2003] J182955.3+011034\\
37 & 18 29 55.7  &  +01 14 31 &  [CK86] 9 $-$ [EC92] 74\\
38 & 18 29 56.2  &  +01 10 57 &  [P2003] J182956.3+011056\\
39 & 18 29 56.4  &  +01 12 18 &  [EC92] 77 $-$ [SVS76] 1\\
40 & 18 29 56.6  &  +01 12 59 &  [EC92] 79\\
42 & 18 29 56.7  &  +01 12 39 &  [EC92] 80\\
43 & 18 29 56.9  &  +01 14 46 &  [SVS76] 2 $-$ [CK86] 3 $-$ [EC92] 82\\
44 & 18 29 57.0  &  +01 12 48 &  [EC92] 84 $-$ [B96] Serpens 5\\
45 & 18 29 57.4  &  +01 14 50 &  [EC92] 86\\
46 & 18 29 57.6  &  +01 13 00 &  [B96] Serpens 6 $-$ [EC92] 88\\
47 & 18 29 57.7  &  +01 10 53 &  [SVS76] 19 \\
48 & 18 29 57.7  &  +01 14 05 &  [SVS76] 20 $-$ [GCN98] 98 $-$ [GCN98] 99 $-$ [B96] Serpens 7 $-$ [EC92] 90 $-$ [KOB2004] 314\\
50 & 18 29 57.8  &  +01 15 32 &  [CK86] 13 $-$ EC93 \\
52 & 18 29 57.8  &  +01 12 38 &  [EC92] 94 \\
53 & 18 29 57.9  &  +01 12 51 &  [EC92] 95 $-$ [B96] Serpens 8\\
54 & 18 29 57.9  &  +01 12 46 &  [B96] Serpens 8\\
55 & 18 29 58.2  &  +01 15 21 &  [CK86] 4 $-$ [EC92] 97\\
56 & 18 29 58.5  &  +01 12 50 &  [EC92] 98\\
57 & 18 29 59.1  &  +01 11 14 &  HH 456 : HH\\
59 & 18 29 59.2  &  +01 14 08 &  [CK86] 8 $-$ [EC92] 105\\
60 & 18 29 59.6  &  +01 11 58 &  [HB96] PS 1\\
61 & 18 29 59.6  &  +01 14 12 &  [GCN98] 122 $-$ [B96] Serpens 10\\
62 & 18 30 00.2  &  +01 14 03 &  [EC92] 114\\
63 & 18 30 00.3  &  +01 09 44 &  [KOB2004] 332\\
65 & 18 30 00.7  &  +01 13 40 &  [EC92] 117 $-$ [CK86] 6 $-$ [B96] Serpens 13\\
66 & 18 30 01.1  &  +01 13 24 &  [EC92] 121 \\
67 & 18 30 01.2  &  +01 15 03 &  [CK86] 12 $-$ [GCN98] 147\\
69 & 18 30 03.4  &  +01 16 19 &  [EC92] 135 $-$ [EC92] 138\\
70 & 18 30 04.6  &  +01 22 33 &  [P2003] J183004.7+012232\\
71 & 18 30 04.9  &  +01 14 39 &  [EC92] 149\\
72 & 18 30 05.5  &  +01 14 25 &  [EC92] 156\\
73 & 18 30 06.1  &  +01 06 17 &  [P2003] J183006.4+010616\\
74 & 18 30 06.7  &  +01 12 16 &  [P2003] J183006.7+011216\\
75 & 18 30 07.7  &  +01 12 04 &  [KOB2004] 366\\
79 & 18 30 11.1  &  +01 12 38 &  [KOB2004] 393\\
83 & 18 30 18.2  &  +01 14 17 &  [P2003] J183018.3+011416\\
84 & 18 30 22.4  &  +01 20 44 &  [P2003] J183022.4+012044\\

\end{tabular}
    \label{tab:cross-id}
  \end{center}
\end{table*}

\textbf{All this to go to ``Electronic material''}

\section{Notes on individual sources}

We extracted the light curves and spectra of all the sources with
count rate $\ge 2$ cts/ks. These are shown in Fig.~\ref{fig:lc} and
Fig.~\ref{fig:lc2} and the results of the spectral fits for these
sources are summarised in Table~\ref{tab:psfit}. Below we comment on
the brighter sources that were not discussed in
Sect.\,\ref{sec:peculiar}.

\subsection{Source 2}

This X-ray source, already detected by \citet{pre2003}, has an optical
and (bright) IR counterpart and it is classified as Class III by its
\spitzer\ colors. It has no counterpart in the ISO catalogue by
\cite{kob+04}. The source displays a variability of a factor of 2
in its X-ray counts and its X-ray spectrum (Table~\ref{tab:psfit}) is
well fit by a 1-$T$ model, with $\nh = 4.1\times 10^{21}$ cm$^{-2}$,
$kT = 1.3$~keV, and $Z=0.11 Z_{\sun}$, typical of a Class III source.

\subsection{Source 11}

One of the brightest X-ray sources in the region, and bright in 2MASS,
its \spitzer\ colors identify it as a Class III source; it has no
counterpart in the ISO catalogue. A fit to its X-ray spectrum gives
$kT = 1.7$ keV and $\nh = 4.7\times 10^{21}$ cm$^{-2}$
(Table~\ref{tab:psfit}), also typical for a Class III source.

\subsection{Source 27}

A bright 2MASS source, classified on the basis of the \spitzer\ colors
as Class III; it has no counterpart in the ISO catalogue. Its
fitted absorbing column density, $\nh = 7.2\times 10^{21}$ cm$^{-2}$,
and plasma temperature, $kT = 1.1$~keV are typical for a Class III
source.

\subsection{Source 34}

A bright 2MASS source already identified as a member of the Serpens
cloud by \citet{ec92} ([EC92] 67), both the ISO and \spitzer\ colors
indicate a Class II source. The X-ray light curve shows variability
within a factor of $\simeq 2$, while the X-ray spectrum is relatively
hot, with $kT = 3.2$ keV, and $\nh = 7.6\times 10^{21}$ cm$^{-2}$. 

\subsection{Source 38}

An object with Class III \spitzer\ colors, which has no counterpart in
the ISO catalogue. The X-ray light curve shows a short-lasting,
impulsive flare. The spectrum has a best fit temperature of $kT = 2.5$
keV and a low absorption, $\nh = 0.5 \times 10^{21}$ cm$^{-2}$. The
low \nh\ value makes it likely that the source is at the surface of
the Serpens cloud.

\subsection{Source 47}


The only optically bright star in our sample, HD 170545, has
photospheric colors compatible with an A0 spectral type
(\citealp{scb96}), and photometrically determined extinction $A_V =
0.53$, compatible with Serpens membership. The source was already
identified in the IR study by \citet{svs76}. \spitzer\ colors indicate
a Class III sources, consistent with a thin disk. \cite{scb96}
indicate that a fainter companion at 7 arcsec is present. The X-ray
spectrum has a moderate temperature ($kT = 0.6$ keV) and a low
absorption $\nh \le 3\times 10^{20}$, compatible with the optical
absorption.

\subsection{Source 53}

This source is classified as a flat-spectrum YSO in the ISO study and
by its \spitzer\ colors. It has a radio counterpart (\citealp{bont96})
and was classified as a YSO by \citet{ec89} and \citet{kob+04}. The
light curve shows significant variability: a factor of three on
the 20 ks time scale. Its best fit plasma temperature ($kT = 3.8$ keV)
and absorbing column density ($\nh = 4.2 \times 10^{22}$ cm$^{-2}$)
indicate an other relatively hot and deeply embedded object.

\subsection{Source 65}

Classified as Class II in ISO study and as Class III by its \spitzer\
colors, this star already identified by \citet{bont96}, has a low best
fit absorbing column density indicating that is probably in the surface
of the cloud. The star shows significant variability (a factor of
3 over 40 ks) and a relatively high plasma temperature ($kT = 2.95$
keV).

\subsection{Source 67}

This source has no counterpart in the ISO catalogue but it is seen in
the \spitzer\ observation and classified as Class III. It was first
observed in the IR by \citet{ck86}. A spectral fit with an absorbed 1T
plasma gives $N({\rm H}) = 0.8\times 10^{22}$ cm$^{-2}$ and $kT = 2.6$
keV.

\subsection{Source 69}

Also this source has no counterpart in the ISO catalogue, but it
is classified as a Class II from its \spitzer\ colors. Its best-fit
absorbing column density ($N({\rm H}) = 1.2\times 10^{22}$ cm$^{-2}$)
is typical for the Serpens cloud. The best-fit plasma temperature is
$kT = 1.0$ keV.

\subsection{Source 73}

An object with \spitzer\ colors compatible with being Class II, 
with no counterpart in the ISO catalogue. The X-ray light curve shows
a short-lasting, impulsive flare. The spectrum has a best fit
temperature of $kT = 1.0$ keV and and a typical absorption value of
$\nh = 0.8 \times 10^{22}$ cm$^{-2}$.

\subsection{Source 75}

A Class II object identified both in the ISO and \spitzer\ data. It
has an absorption value below the average ($\nh = 0.5 \times 10^{22}$
cm$^{-2}$) for this sample and a relatively high plasma temperature
($kT = 2.8$ keV).

\begin{figure*}[!htbp]
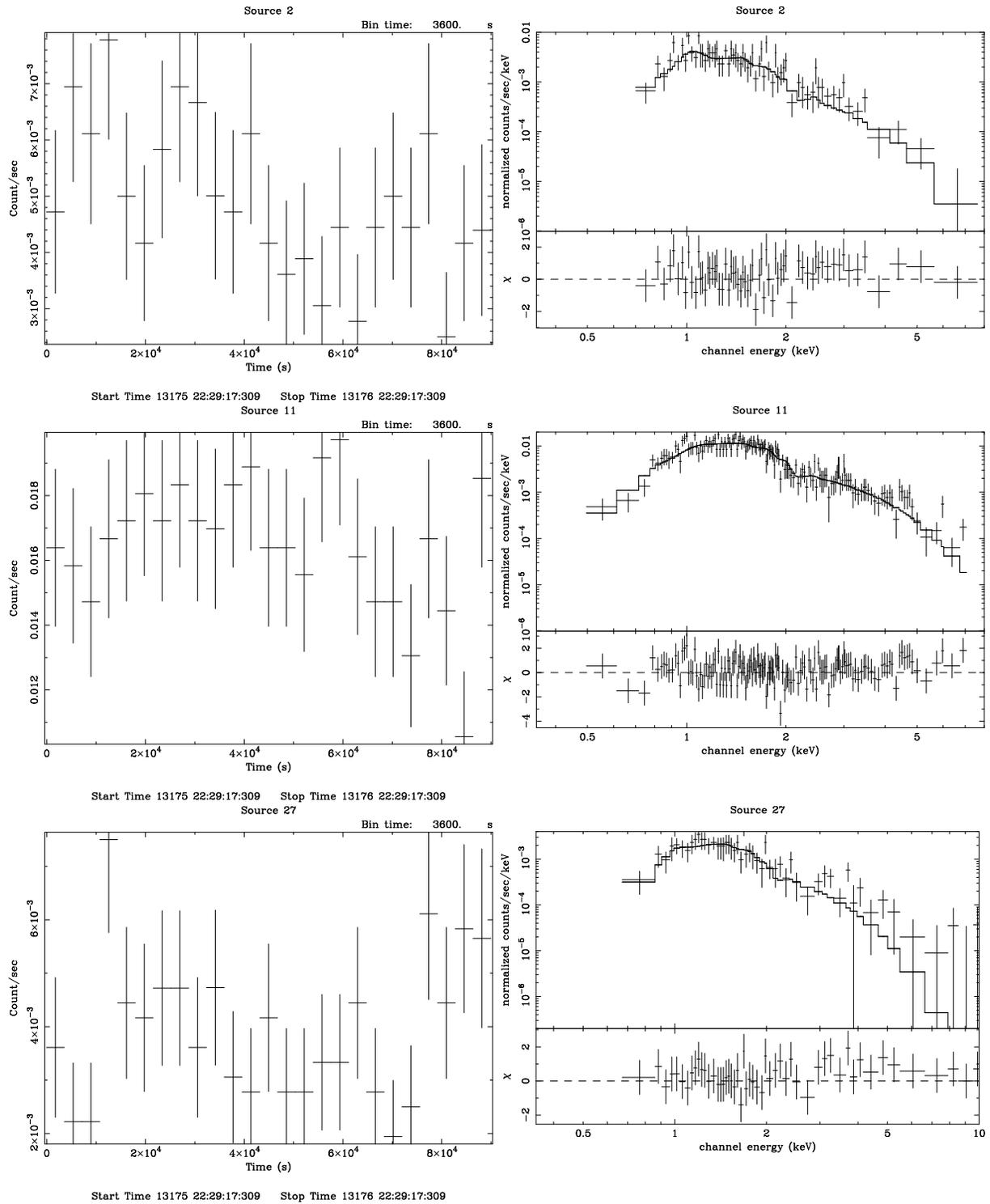

  \begin{center} \leavevmode 

        \epsfig{file=6424fg26.ps, height=8.0cm, angle=270}
        \epsfig{file=6424fg27.ps, height=8.0cm, angle=270} 
        \epsfig{file=6424fg28.ps, height=8.0cm, angle=270} 
        \epsfig{file=6424fg29.ps, height=8.0cm, angle=270}
       	\epsfig{file=6424fg30.ps, height=8.0cm, angle=270}
        \epsfig{file=6424fg31.ps, height=8.0cm, angle=270}

\caption{ Light curves and spectra with spectral fits of sources with more than 2 cts~ks$^{-1}$}.

    \label{fig:lc2}
  \end{center}
\end{figure*}

\addtocounter{figure}{-1}
\begin{figure*}[!htbp]
  \begin{center} \leavevmode

        \epsfig{file=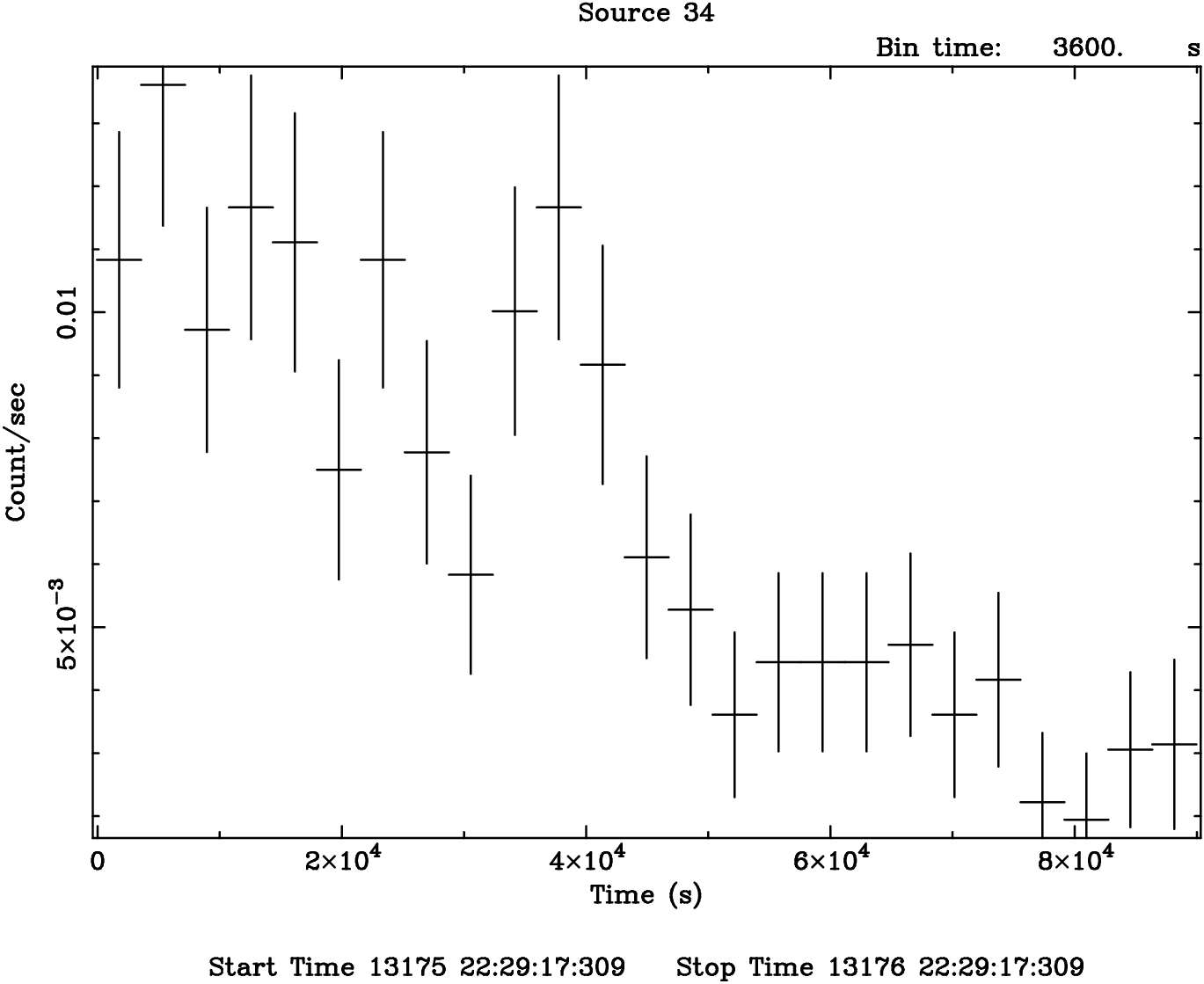, height=8.0cm, angle=270}
        \epsfig{file=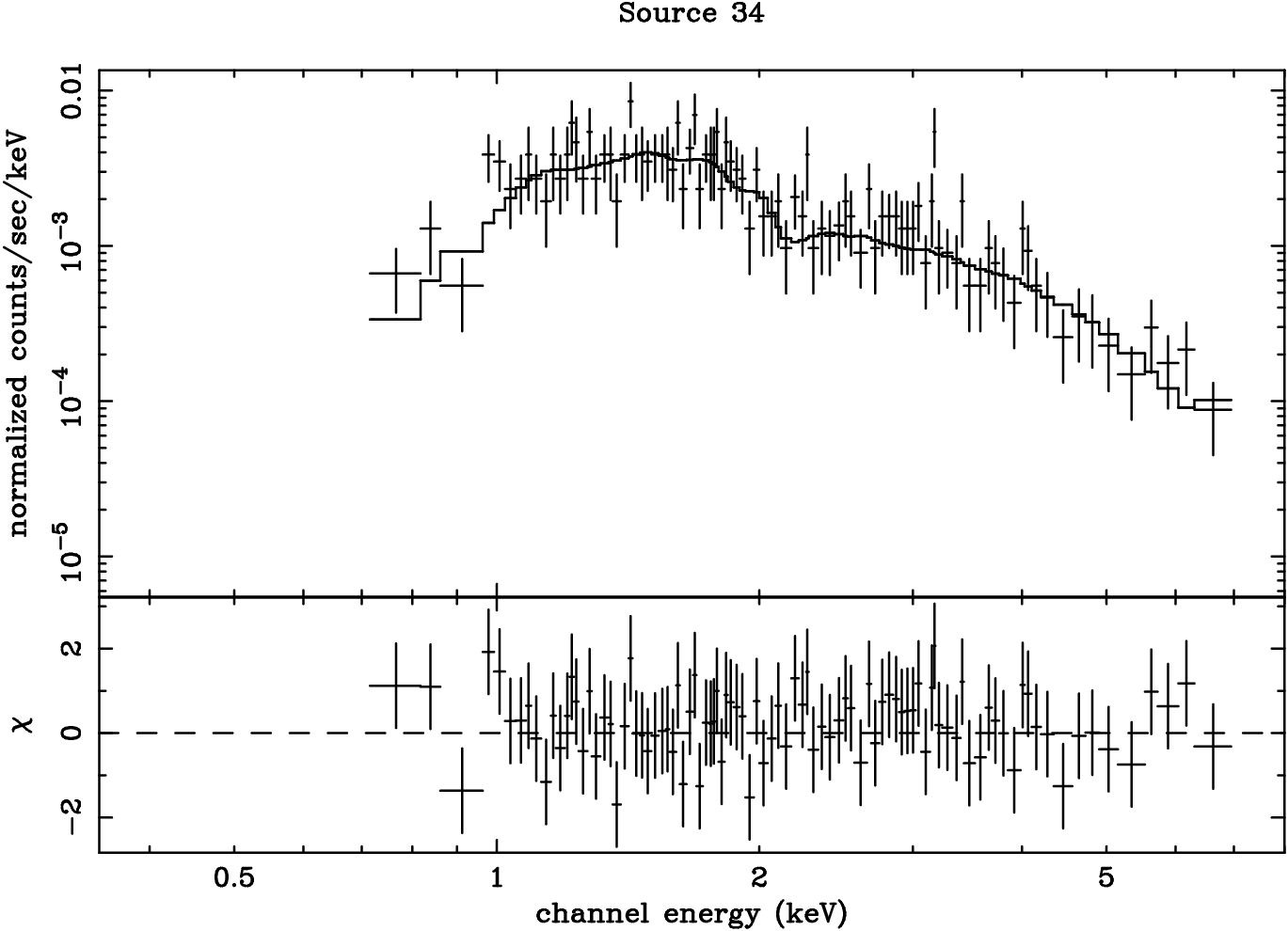, height=8.0cm, angle=270}
        \epsfig{file=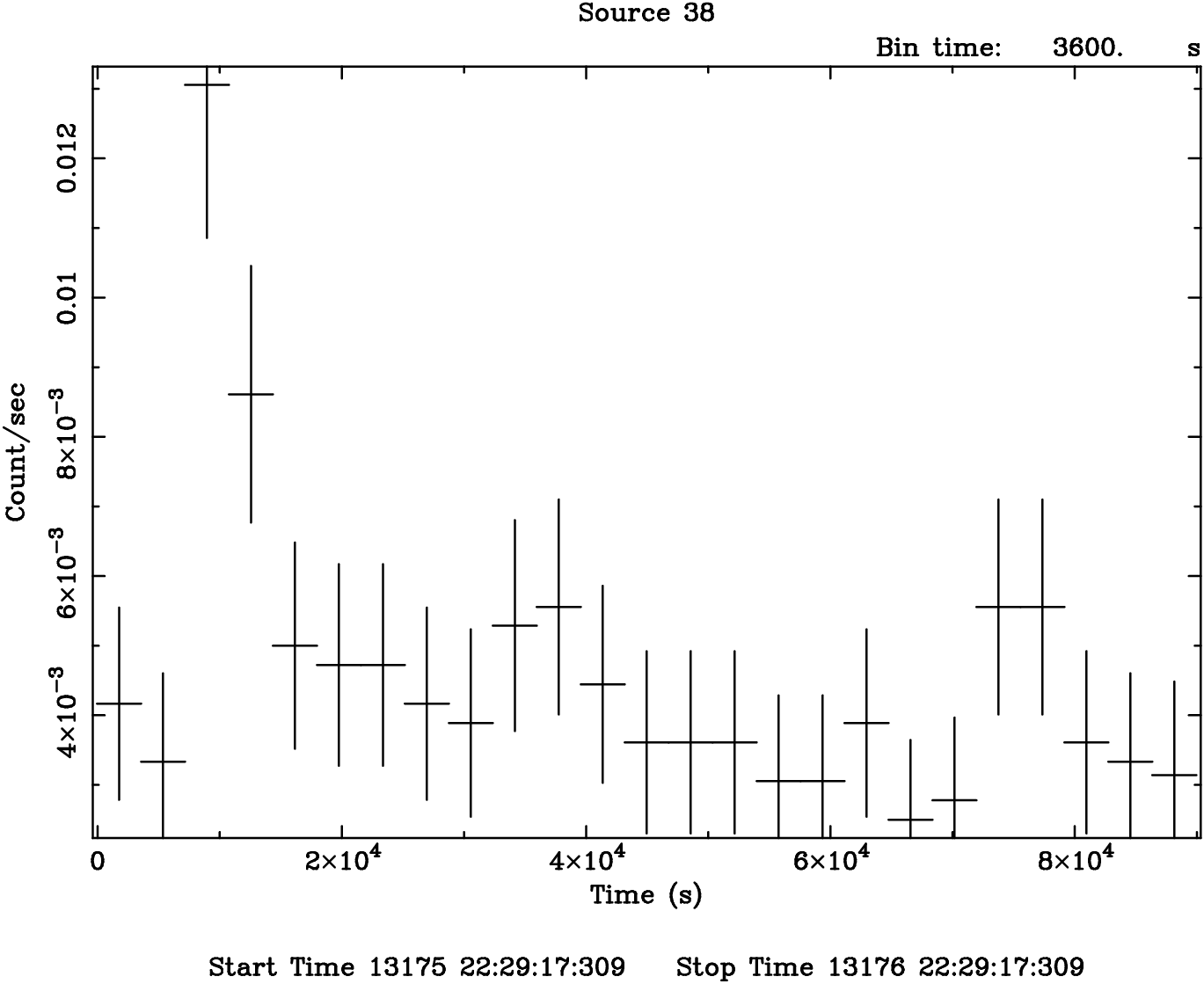, height=8.0cm, angle=270}
        \epsfig{file=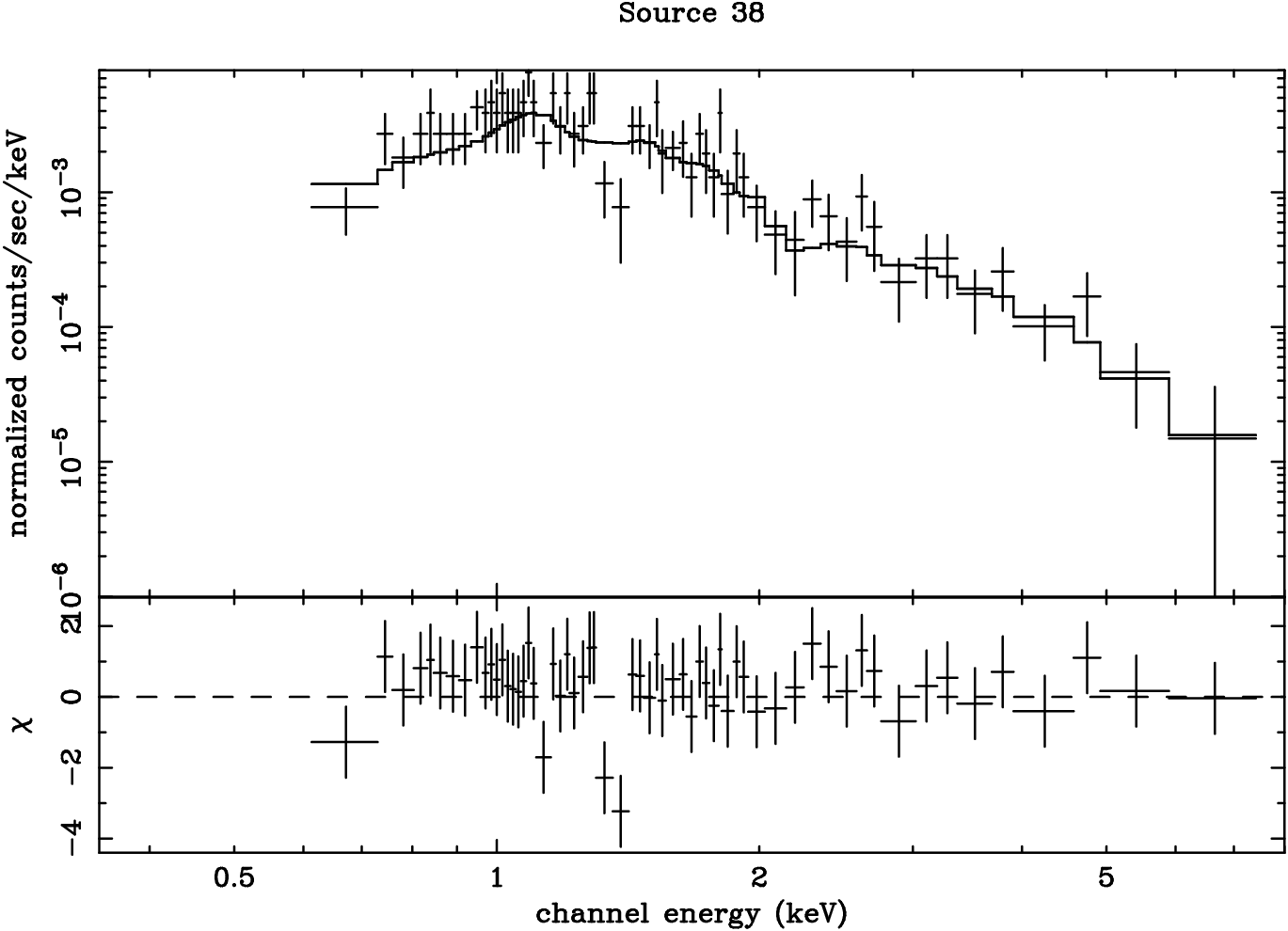, height=8.0cm, angle=270}
	\epsfig{file=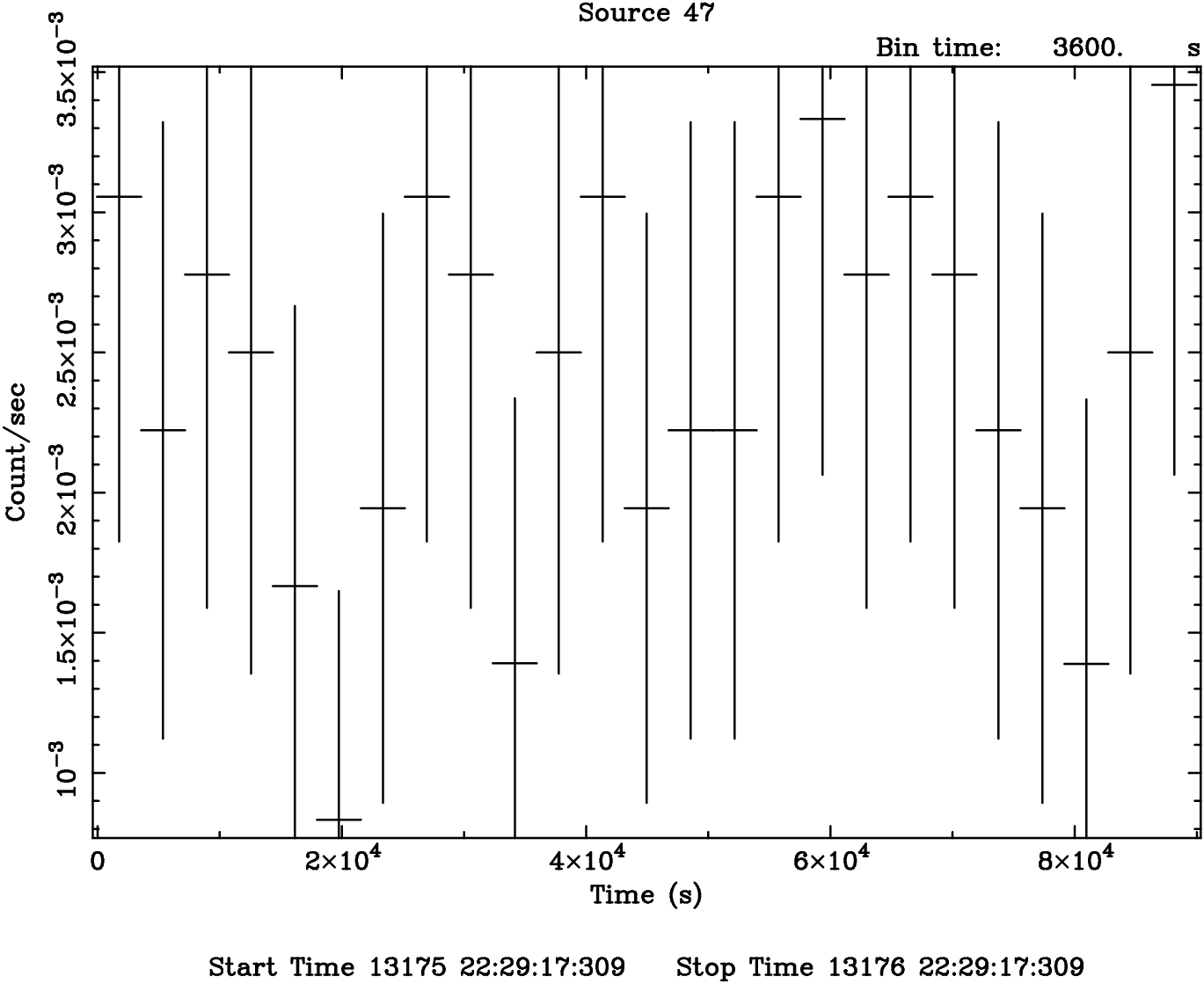, height=8.0cm, angle=270}	
        \epsfig{file=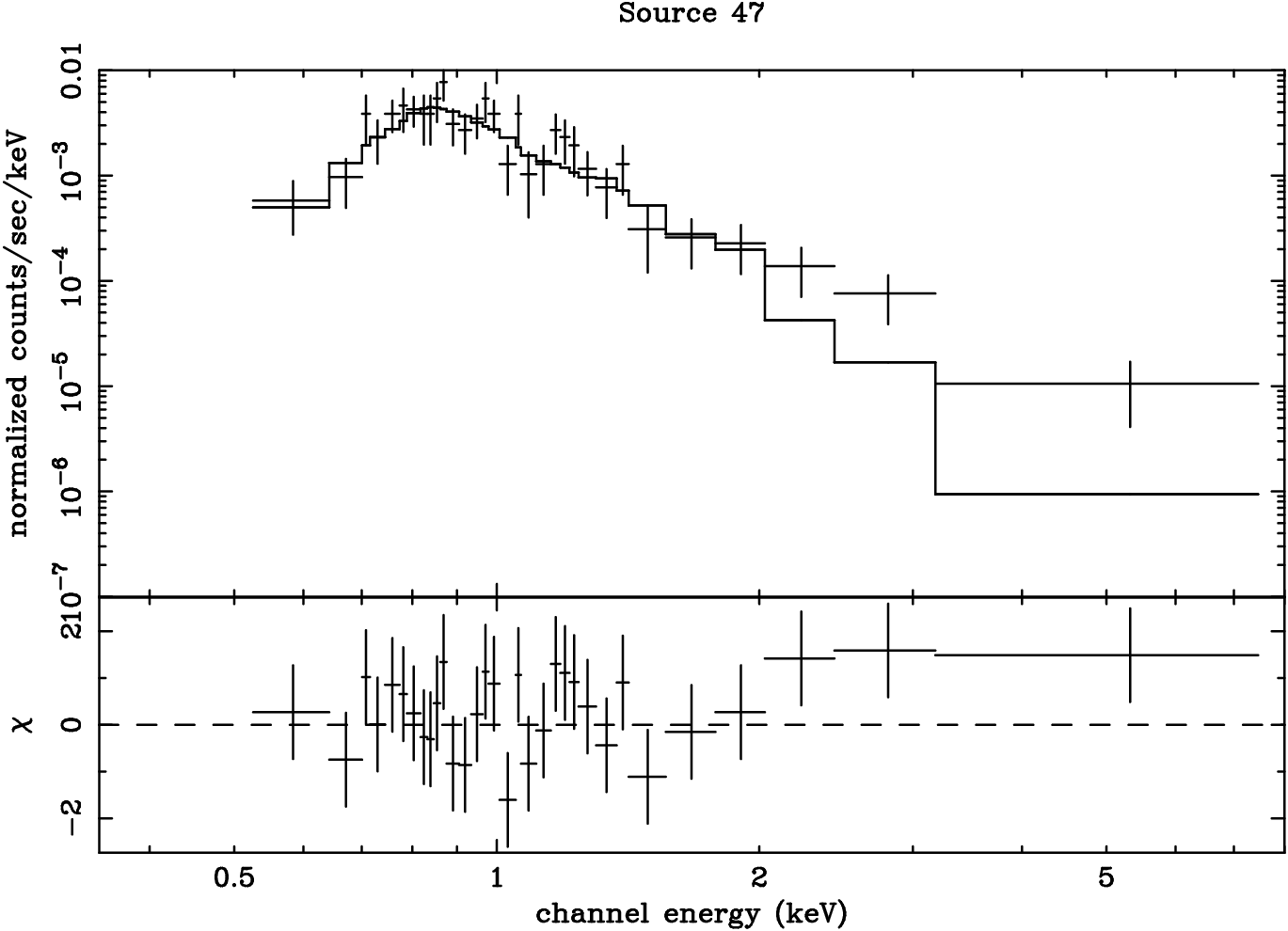, height=8.0cm, angle=270}

\caption{({\it Continued.}) Light curves and spectra with spectral fits of sources with more than 2 cts~ks$^{-1}$}.

  \end{center}
\end{figure*}

\addtocounter{figure}{-1}
\begin{figure*}[!htbp]
  \begin{center} \leavevmode 

	\epsfig{file=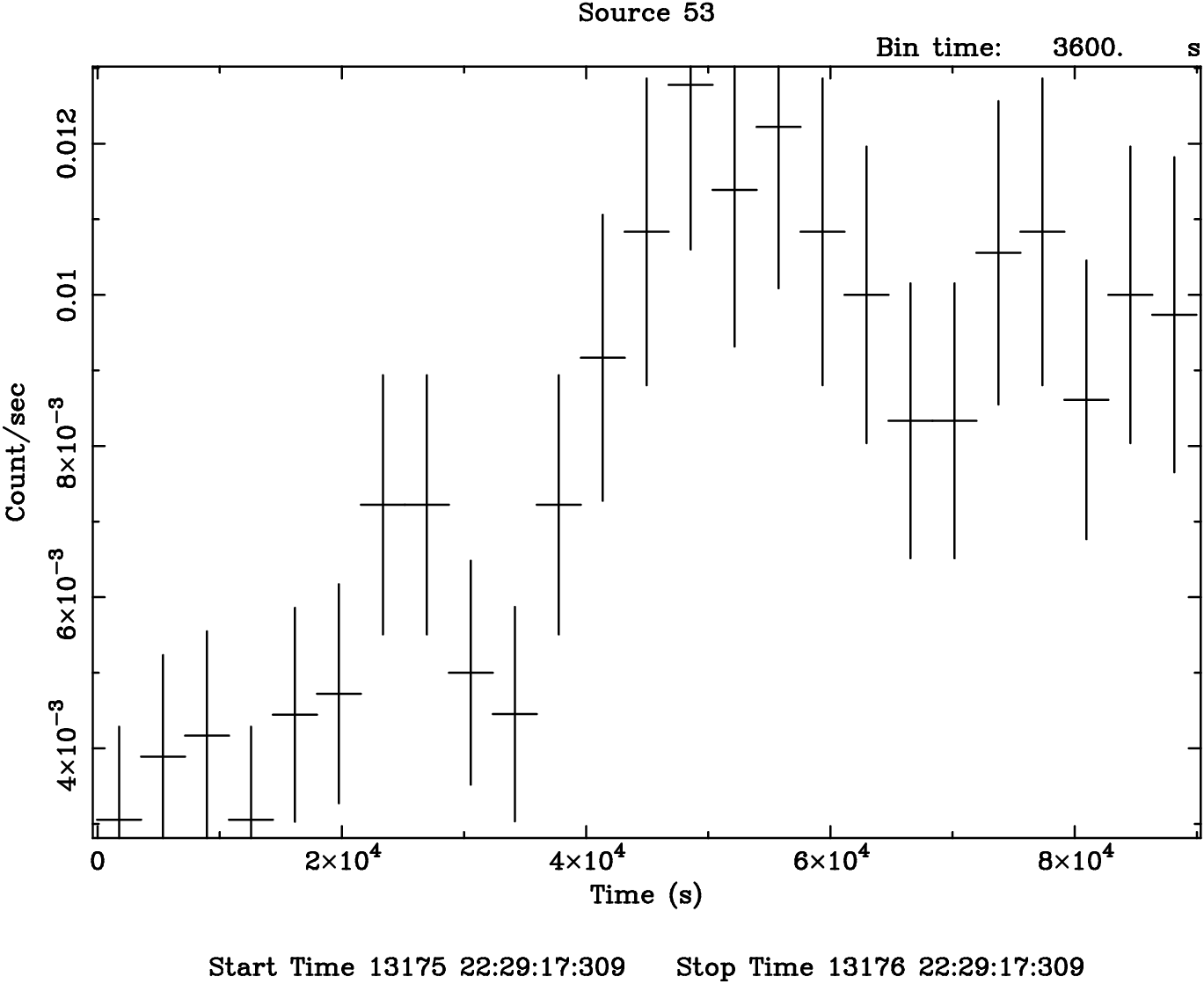, height=8.0cm, angle=270}	
        \epsfig{file=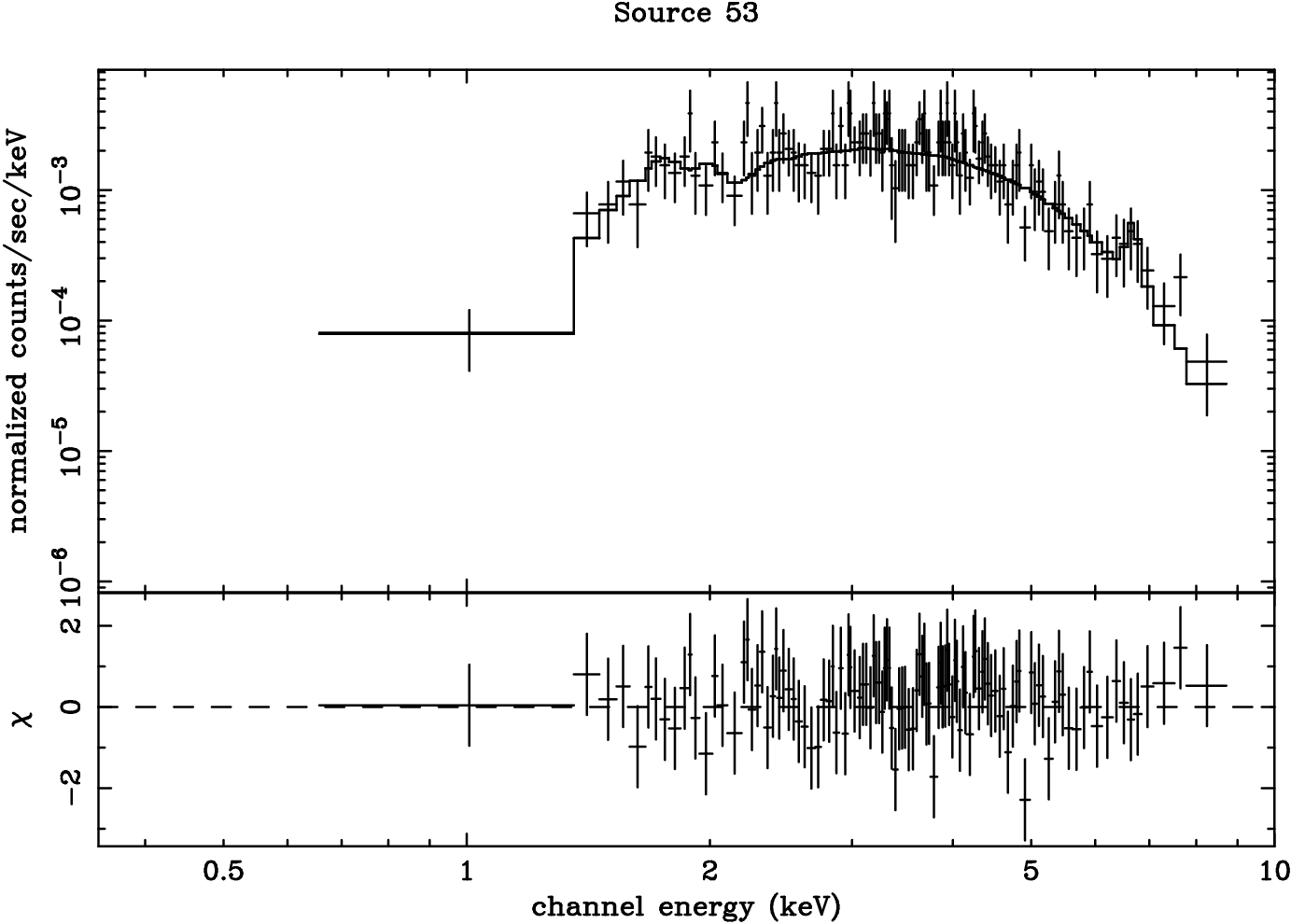, height=8.0cm, angle=270}
	\epsfig{file=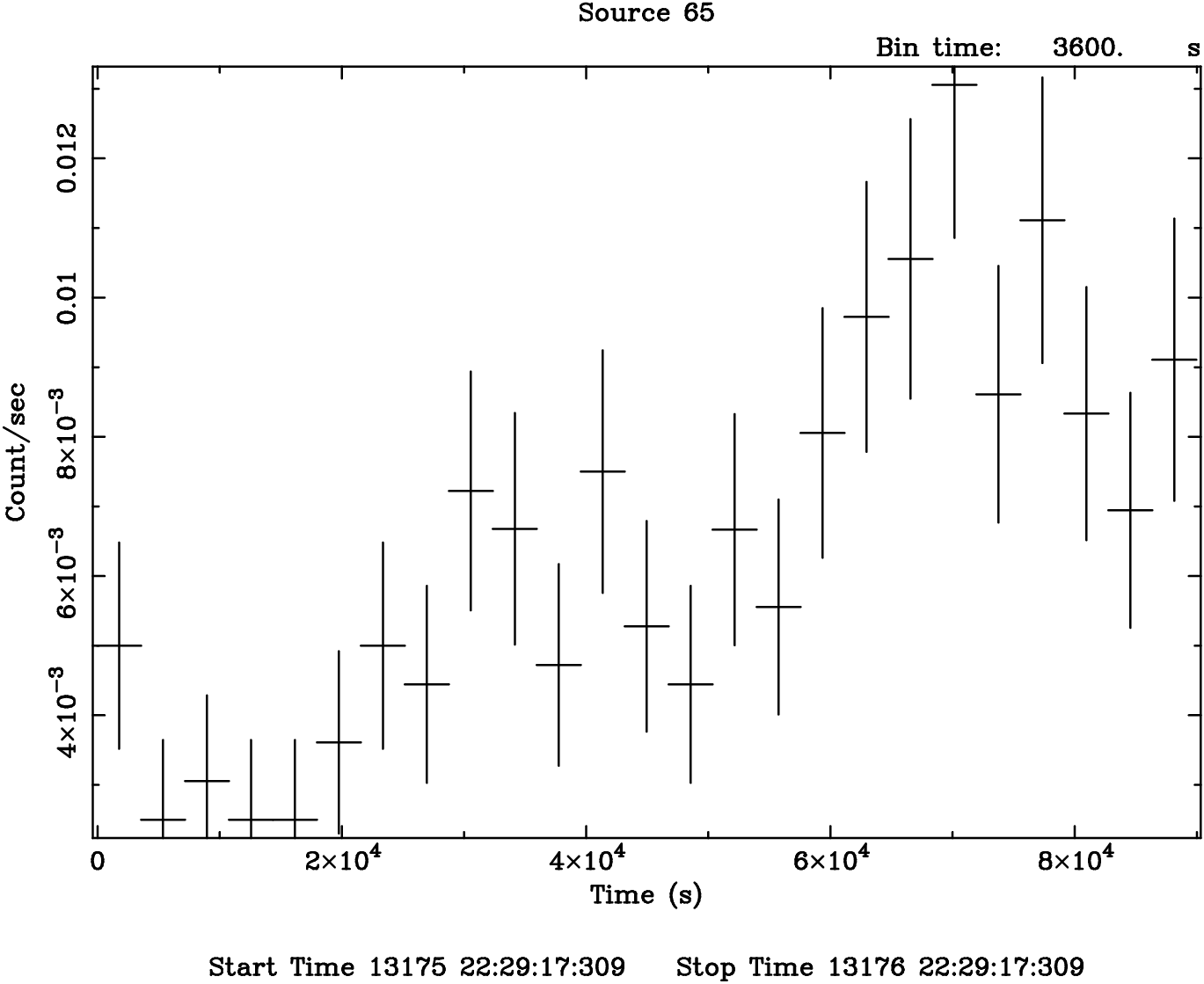, height=8.0cm, angle=270}
        \epsfig{file=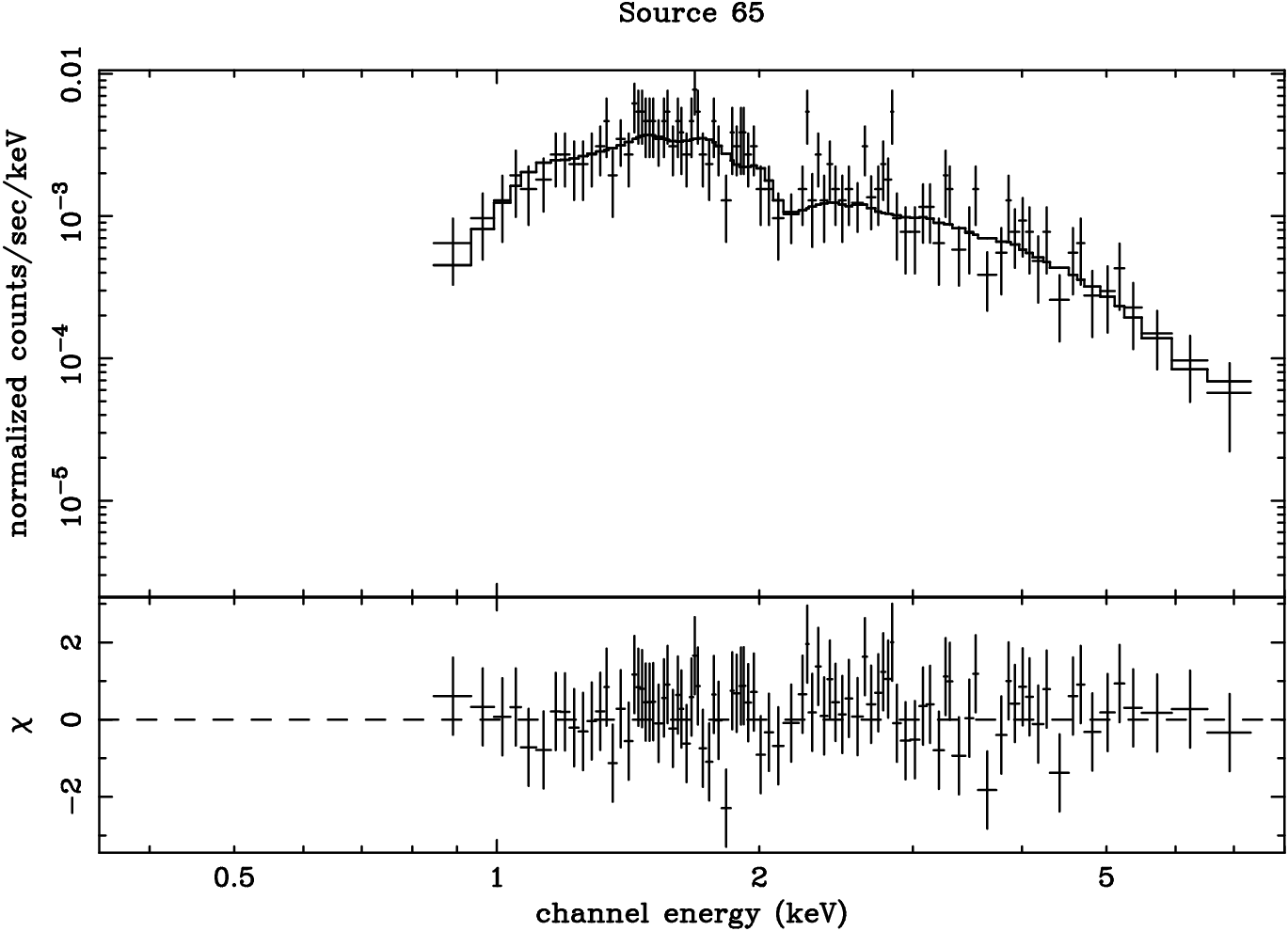, height=8.0cm, angle=270}
        \epsfig{file=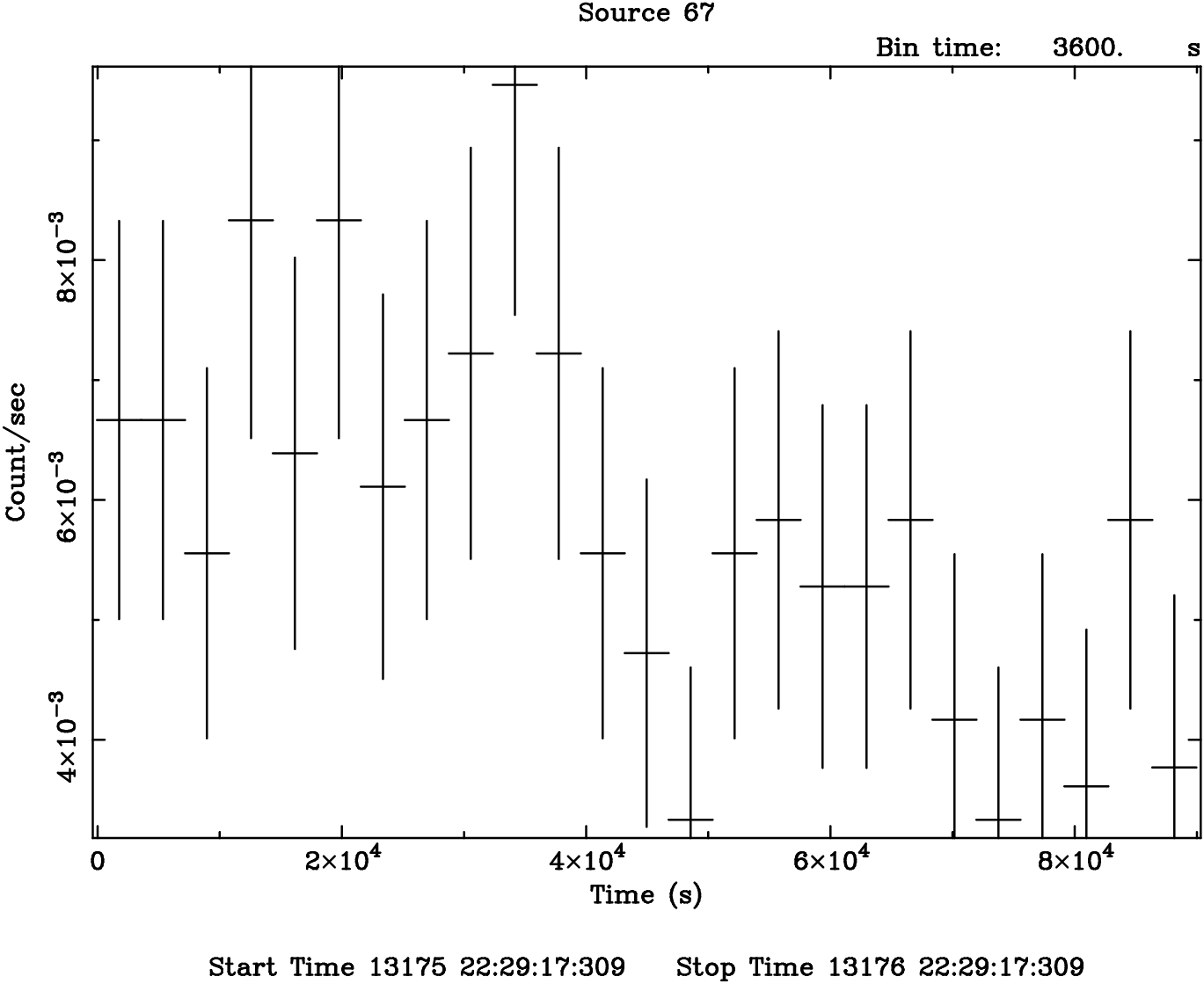, height=8.0cm, angle=270}
        \epsfig{file=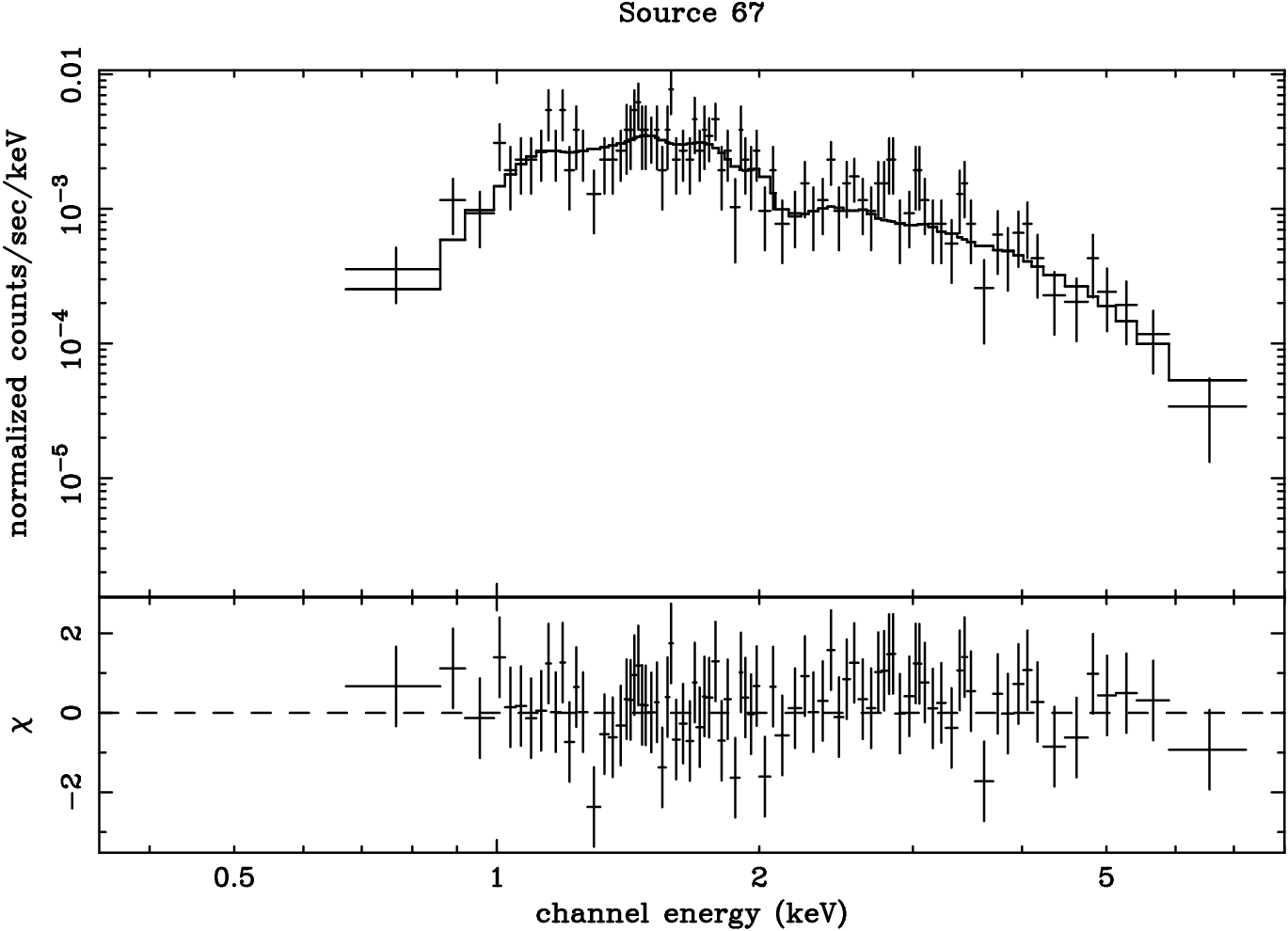, height=8.0cm, angle=270}	

    \caption{({\it Continued.}) Light curves and spectra with spectral fits of sources with more than 2 cts~ks$^{-1}$}.

  \end{center}
\end{figure*}

\addtocounter{figure}{-1}
\begin{figure*}[!htbp]
  \begin{center} \leavevmode 
      
	\epsfig{file=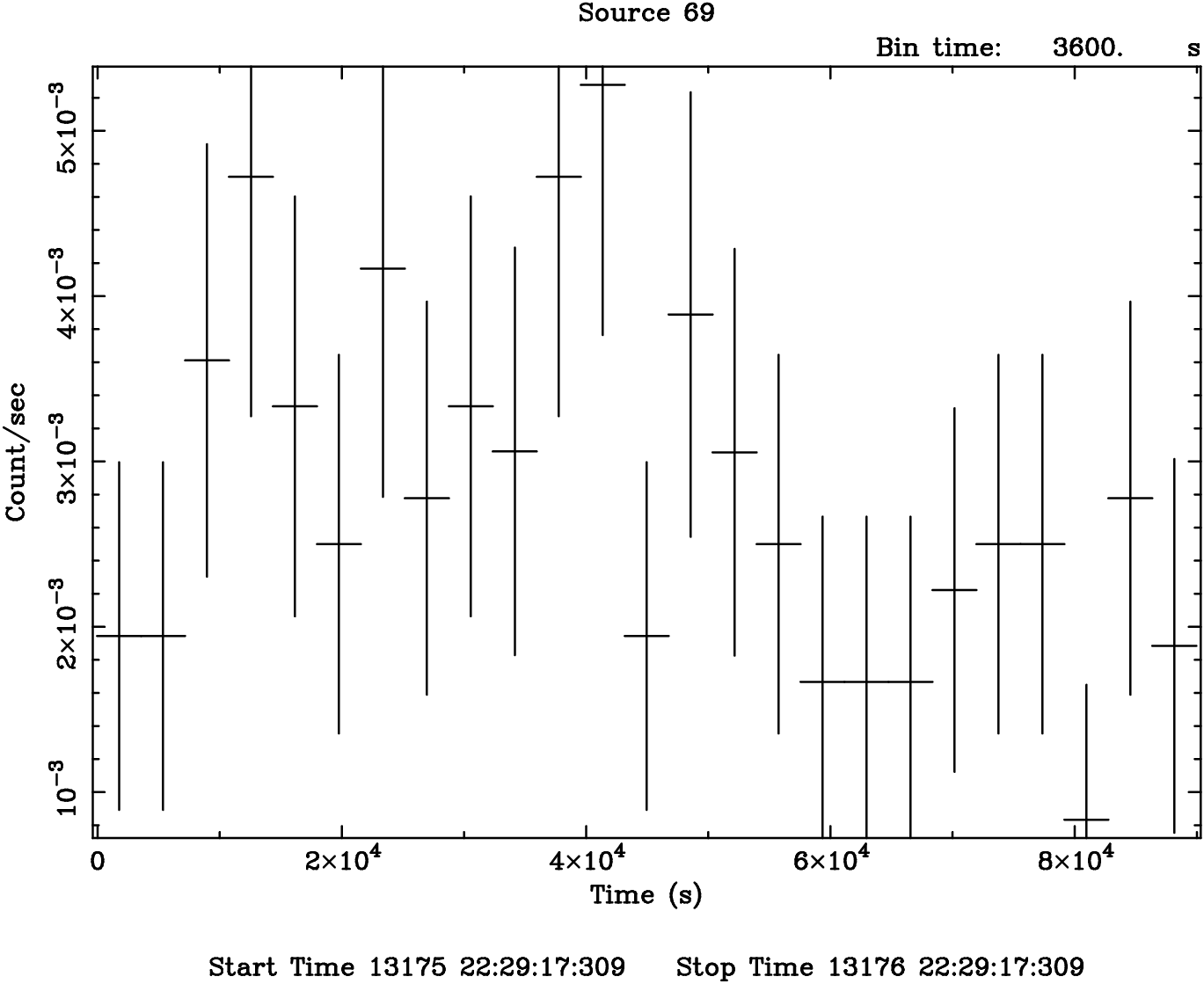, height=8.0cm, angle=270}
        \epsfig{file=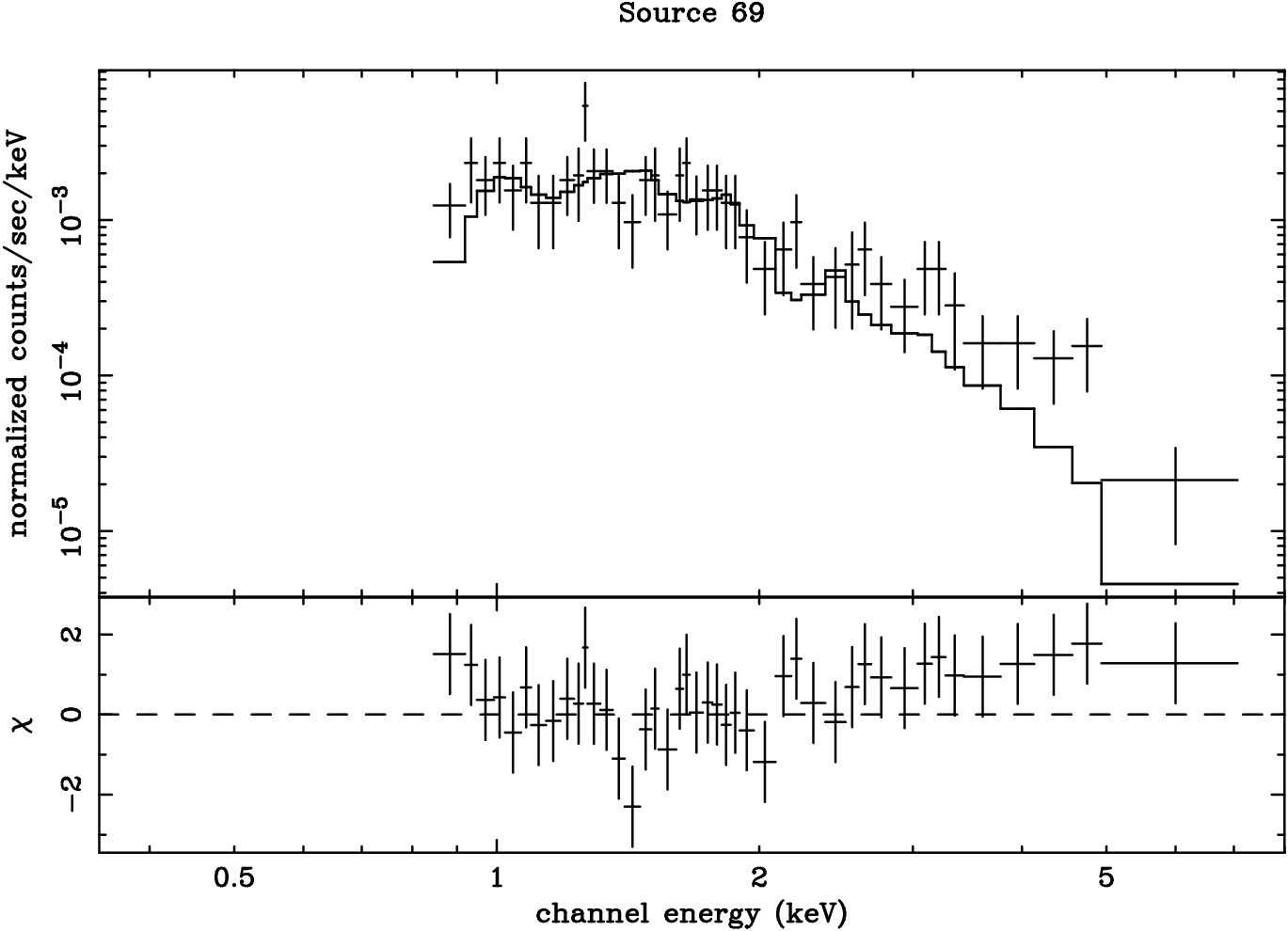, height=8.0cm, angle=270}
       	\epsfig{file=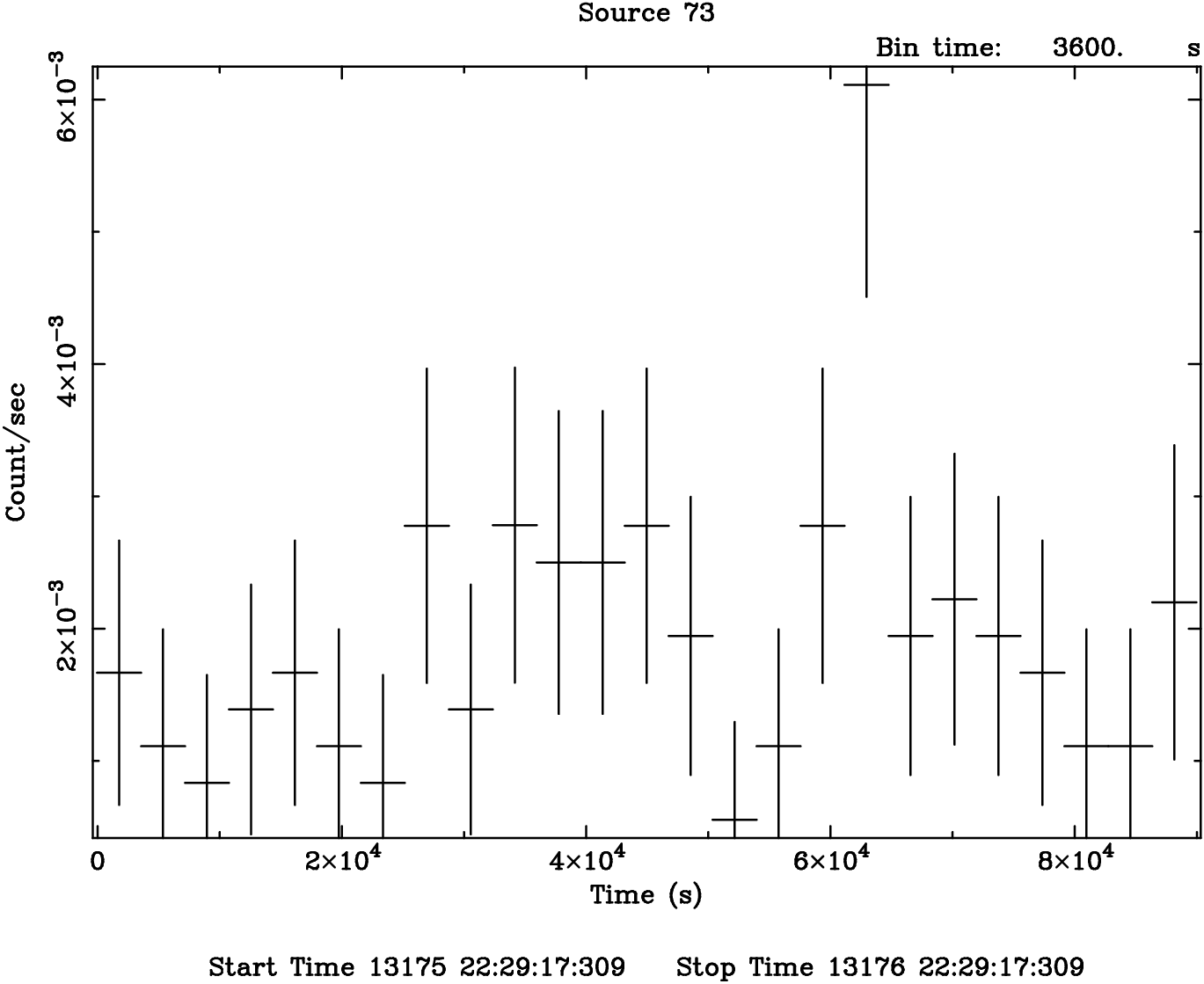, height=8.0cm, angle=270}
        \epsfig{file=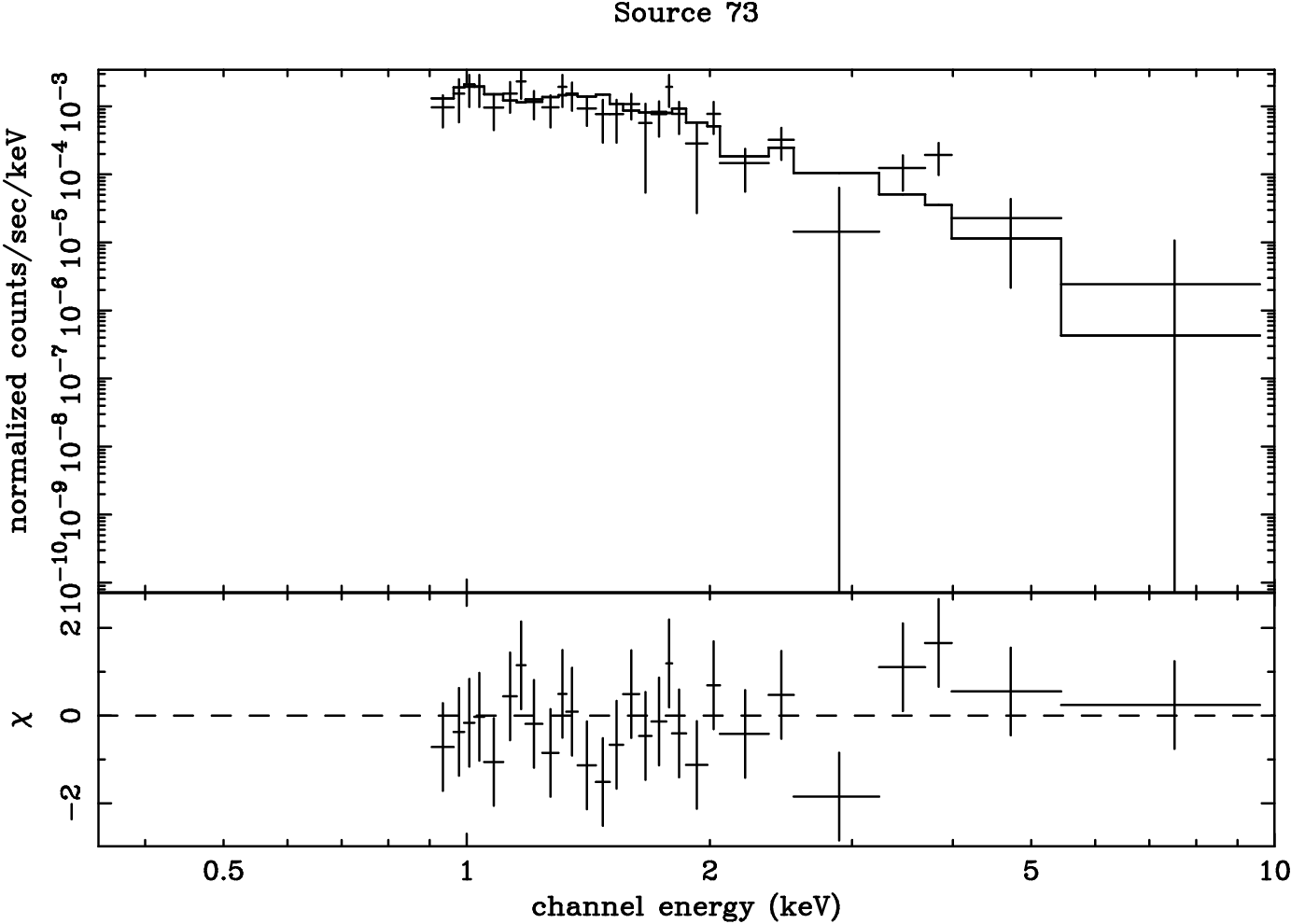, height=8.0cm, angle=270}   
	\epsfig{file=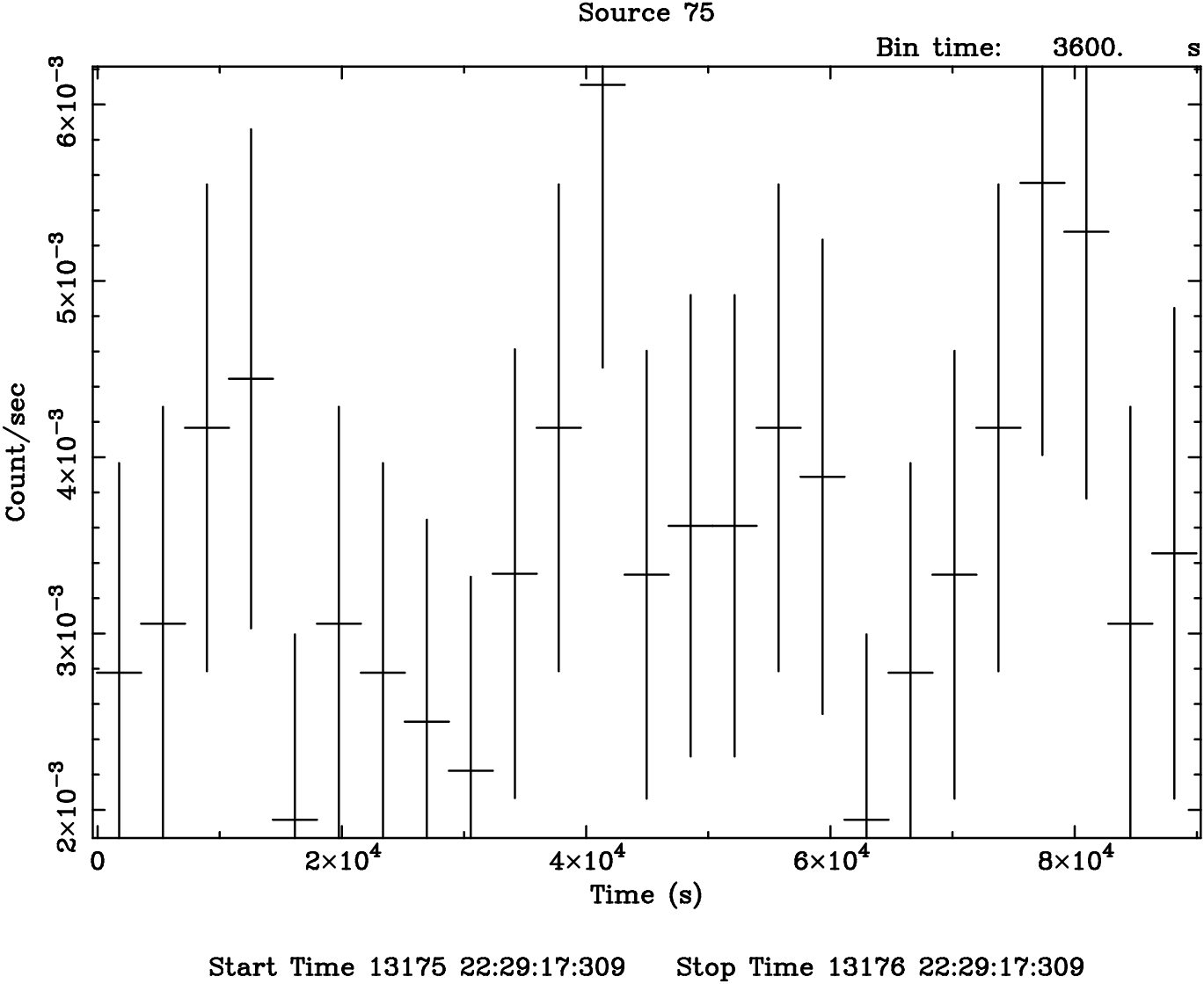, height=8.0cm, angle=270}	
        \epsfig{file=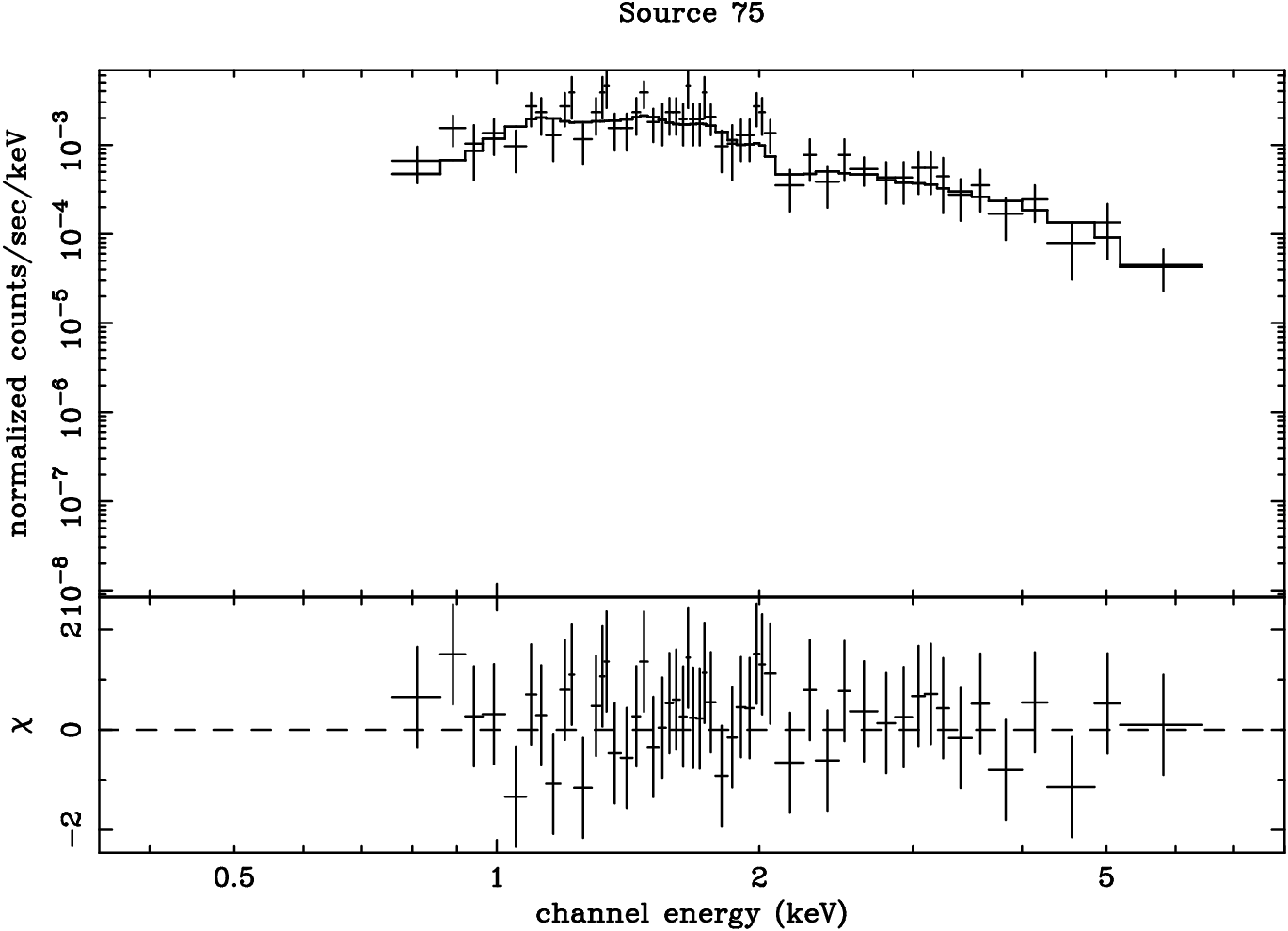, height=8.0cm, angle=270}  
        
    \caption{({\it Continued.}) Light curves and spectra with spectral fits of sources with more than 2 cts~ks$^{-1}$}.

  \end{center}
\end{figure*}

\end{document}